\documentclass[letterpaper,11pt]{article}
\pdfoutput=1

\usepackage{jheppub}
\usepackage{multirow}
\usepackage{placeins}
\usepackage{subfig}
\usepackage{xspace}
\usepackage{xcolor}
\usepackage[countmax]{subfloat}
\usepackage{amsmath}
\usepackage{mathtools}
\usepackage{graphicx}
\usepackage{slashed}

\usepackage{tikz-feynman,contour}
\usepackage{tikz,pgf}
\usepackage{braket}
\usepackage{amsfonts}
\usepackage{amsthm,scrextend,bbold}
\usepackage{comment}
\usepackage{subfiles}
\usepackage{subcaption}
\usepackage{todonotes}

\newcommand{\as}{\alpha_\text{s}}
\newcommand{\cf}{C_{\text{F}}}
\newcommand{\tr}{T_{\text{R}}}
\newcommand{\ca}{C_{\text{A}}}
\newcommand{\nc}{N_{\text{C}}}

\newcommand{\order}[1]{{\cal O}\left(#1\right)}

\DeclareMathOperator{\De}{d}
\newcommand{\de}{\De\!}

\newcommand{\kt}{k_t}

\newcommand{\muf}{\mu_{\text{F}}}

\newcommand{\mur}{\mu_{\text{R}}}

\newcommand{\yjet}{y_{\text{jet}}}
\newcommand{\phijet}{\phi_{\text{jet}}}
\newcommand{\zc}{z_{\text{cut}}}
\newcommand{\ptjet}{p_{T,{\text{jet}}}}
\newcommand{\ptZ}{p_{T,{\text{Z}}}}

\newcommand{\ord}[1]{{\cal O}\left(#1\right)}

\newcommand{\NLO}{\text{NLO}\xspace}    
\newcommand{\NLL}{\text{NLL}\xspace}
\newcommand{\NLLp}{\ensuremath{\text{NLL}^\prime}\xspace}
\newcommand{\NLOpNLL}{\ensuremath{\NLO+\NLL}\xspace}
    
\newcommand{\NLOpNLLp}{\ensuremath{\NLOpNLL^\prime}\xspace}

\usepackage{color}
\definecolor{darkblue}{rgb}{0,0,0.5}
\definecolor{darkgreen}{rgb}{0,0.5,0}
\definecolor{darkorange}{rgb}{0.8,0.3,0}

\newcommand{\sherpa}{\textsc{Sherpa\xspace}}

\newcommand{\softdrop}{\texttt{SoftDrop}\xspace}
\newcommand{\mmd}[1]{\texttt{modified-MassDrop\xspace #1}}
\newcommand{\caesar}{\textsc{Caesar\xspace}}

\title{
Phenomenology of heavy-flavour jet angularities at hadron colliders
}

\author[1]{Andrea Ghira,}
\author[2]{Lorenzo Mai,}
\author[2]{Simone Marzani,}
\author[3]{Daniel Reichelt,}
\author[4]{Steffen Schumann,}
\author[4]{Leon Stöcker}

\affiliation[1]{Technical University of Munich, TUM School of Natural Sciences, Physics Department, James-Franck-Straße 1, 85748 Garching, Germany}
\affiliation[2]{Dipartimento di Fisica, Universit\`a di Genova and INFN, Sezione di Genova, Via Dodecaneso 33, 16146, Italy}
\affiliation[3]{CERN, Theoretical Physics Department, CH-1211 Geneva 23, Switzerland}
\affiliation[4]{Institut für Theoretische Physik, Georg-August-Universität Göttingen, Friedrich-Hund-Platz 1, 37077 Göttingen, Germany}
\emailAdd{andrea.ghira@tum.de}
\emailAdd{lorenzo.mai@ge.infn.it}
\emailAdd{simone.marzani@ge.infn.it}
\emailAdd{d.reichelt@cern.ch}
\emailAdd{steffen.schumann@phys.uni-goettingen.de}
\emailAdd{leon.stoecker@uni-goettingen.de}

\abstract{
We compute resummed and matched predictions for jet angularities in hadronic $Z+$jet events, where the jet is initiated by a $b$-quark. The analysis is performed both with and without grooming the candidate jets using the \texttt{SoftDrop} algorithm. Mass effects are consistently included at both fixed-order and resummed levels. 
Our theoretical predictions also incorporate non-perturbative corrections from the underlying event and hadronization, implemented through parton-to-hadron transfer matrices extracted from dedicated Monte Carlo simulations with \textsc{Sherpa}. Finally, we compare results for $b$-jets with the ones from light-flavour jets, in order to quantify the impact of finite-mass effects.}

\begin{document}
\preprint{CERN-TH-2026-137, MCNET-26-14}
\maketitle

\section{Introduction}

The Large Hadron Collider (LHC) has opened a new era in the exploration of particle physics, providing unprecedented collision energies and luminosities. The vast samples of high-quality data collected by the various experiments have allowed for stringent and precise tests of the Standard Model (SM) and among the huge variety of available observables, jet-substructure ones occupy a privilege spot~\cite{Marzani:2019hun}. Indeed, the ubiquity of final-state jets at hadron colliders qualifies them to be used as a probe of strong-interaction phenomena and dynamics, ranging over several orders of magnitude in the relevant energy scale: jet data are used for the determination of key parameters such as the strong coupling constant \cite{Britzger:2017maj,CMS:2013vbb,ATLAS:2017qir, ATLAS:2015yaa,CMS:2014mna} as well as to constrain parton distribution functions (PDFs) \cite{ATLAS:2021qnl,ATLAS:2013pbc,CMS:2014qtp,CMS:2016lna,NNPDF:2021njg,Bailey:2020ooq,AbdulKhalek:2020jut,Harland-Lang:2017ytb,Pumplin:2009nk,Watt:2013oha}. At the same time, jets are sensitive to potential signals of physics beyond the SM \cite{Soper:2010xk,Godbole:2014cfa,Chen:2014dma, Adams:2015hiv} and determining differences between jets originating from quarks versus gluons helps to enhance sensitivity in searches for new physics. 

It is thus not surprising that in recent years jet-substructure physics have been in the spotlight, creating a dynamic interface between theory and experiment and enabling increasingly refined analyses that push the precision frontier. A key outcome of these studies is the effectiveness of grooming techniques -- jet-substructure algorithms that remove soft, wide-angle radiation from 
jets -- in improving the perturbative description of jet observables. Groomers can mitigate non-perturbative effects from hadronization 
and the underlying event (UE), therefore, improving the description of jet physics by means of perturbation theory, leading to a variety of high-precision calculations, see for instance~\cite{Dasgupta:2013ihk,Dasgupta:2013via,Larkoski:2014wba,Larkoski:2017cqq,Larkoski:2017iuy,Baron:2018nfz,Baron:2020xoi,Marzani:2017kqd,Chien:2024uax,Kang:2018vgn,Dasgupta:2022fim,Frye:2016aiz,Kardos:2020gty,Anderle:2020mxj}.
Furthermore, the all-order structure of the perturbative result is simplified because groomers, such as \mmd/\softdrop~\cite{Dasgupta:2013ihk,Larkoski:2014wba}, can remove the logarithmic enhancement due to soft gluons at wide angles, including the intricate structure of non-global logarithms (NGLs), by turning logarithms of the observable under consideration into logarithms of an external parameter (e.g.\ $\zc$ for \softdrop).

In Ref.~\cite{Caletti:2021oor} a detailed analysis for jet angularity observables~\cite{Larkoski:2014pca}, which probe the angular and transverse-momentum distribution within a jet, was first presented for the hadroproduction of a $Z$-boson (decaying into leptons) in association with jets.  
In this analysis theoretical predictions at next-to-leading logarithmic (NLL) accuracy matched to 
next-to-leading order (NLO) were obtained for jets with and without \softdrop~grooming. To be precise, for both groomed and ungroomed distributions the calculation achieved \NLLp accuracy by keeping track of the jet flavour in the matching procedure. This includes single-logarithmic effects from soft wide-angle emissions -- treated as an expansion in the jet radius -- and leading NGLs in the large-$N_c$ limit. The perturbative \NLOpNLLp predictions were supplemented with non-perturbative (NP) corrections, extracted from hadron-level Monte Carlo (MC) simulations, which are required for a realistic analysis as jet angularities probe QCD dynamics deep into the infrared region. In Ref.~\cite{Reichelt:2021svh} the extraction method for NP corrections was improved based on a transfer-matrix approach and the full \NLOpNLLp+NP predictions were compared with data 
presented by the CMS experiment in \cite{CMS:2021iwu}.

In recent years, precision studies of heavy-flavour jets have attracted growing interest within the jet-substructure community.
Jets containing heavy flavours, namely charm ($c$) and beauty ($b$), are relevant for a wide range of analyses at the LHC.
On the one hand, they play a central role in Higgs-boson studies.
On the other hand, measurements of heavy-flavour jets produced in association with electroweak bosons provide a valuable handle to probe the heavy-quark content of the PDFs.
In this context, the recent development of Infra-Red and Collinear (IRC) safe flavour-jet algorithms~\cite{Banfi:2006hf,Caletti:2022hnc,Czakon:2022wam,Gauld:2022lem,Caola:2023wpj} opens the door to a largely unexplored flavour-jet substructure programme at the LHC.
The first resummed calculations for jets initiated by heavy quarks were carried out in the context of studies of $B$-hadrons~\cite{Aglietti:2006wh,Aglietti:2007bp,Aglietti:2008xn,Aglietti:2022rcm,Ghira:2023bxr}
and top-quark jets~\cite{Fleming:2007qr,Fleming:2007xt,Bachu:2020nqn,Jain:2008gb,Hoang:2019fze,Bris:2020uyb}, within the framework of effective field theory.
Over the past few years, the number of investigations devoted to the jet substructure of heavy-flavour jets has increased substantially~\cite{Maltoni:2016ays, Lee:2019lge,Llorente:2014bha,Li:2017wwc,Li:2021gjw, Craft:2022kdo,Cunqueiro:2022svx,Fedkevych:2022mid,Caletti:2023spr,Blok:2023ugf, Zhang:2023jpe,Ghira:2025nym}.
Among the most prominent QCD effects influencing heavy-flavour jet substructure is the so-called dead-cone effect~\cite{Dokshitzer:1991fd,Dokshitzer:1995ev}, namely the suppression of collinear radiation around massive quarks. Its first direct experimental observation was recently reported by the ALICE collaboration~\cite{ALICE:2021aqk}\footnote{For indirect observations of the dead-cone effect, see also Refs.~\cite{DELPHI:1992pnf, OPAL:1994cct, OPAL:1995rqo, SLD:1999cuj, DELPHI:2000edu, ALEPH:2001pfo, ATLAS:2013uet}.}.

The study of heavy-flavour energy-correlation functions (ECFs) was initiated in Ref.~\cite{Lee:2019lge}. More recently, the all-order resummation of several variants of ECFs and jet angularities, both with and without \softdrop grooming, has been carried out in Refs.~\cite{Dhani:2024gtx, Dhani:2025fbk}.
In this work, we incorporate these all-order results into resummed and matched theoretical predictions suitable for phenomenological applications. This is achieved by extending the framework of Refs.~\cite{Caletti:2021oor,Reichelt:2021svh}, retaining logarithmically enhanced quark-mass effects associated with jets initiated by $b$ quarks. To this end, we translate the required modifications with respect to the massless case, as presented in Ref.~\cite{Caletti:2023spr}, into the \textsc{Caesar} formalism~\cite{Banfi:2004yd} and implement them within the \sherpa\ resummation framework~\cite{Gerwick:2014gya}.
Following Refs.~\cite{Caletti:2021oor,Reichelt:2021svh}, we obtain predictions that are fully differential in both jet and lepton kinematics, enabling realistic fiducial cuts and direct comparisons with Monte Carlo simulations or experimental measurements. Finally, we present updated predictions for comparison with the existing CMS measurement~\cite{CMS:2021iwu}, consistently accounting for finite-mass effects both in the running of the strong coupling and explicitly in partonic channels involving $b$ quarks.

This paper is organized as follows. In Sec.~\ref{sec:obs_def} we provide the definition of jet angularities, introduce the fiducial phase space used in our study for $Z+$jet and detail our calculational setups. In Sec.~\ref{sec:caesar_review} we review the theoretical framework employed for making predictions at \NLOpNLLp accuracy, based on the implementation of the \caesar~formalism in \sherpa. Here we also present and discuss the main differences with respect to Refs. \cite{Caletti:2021oor,Reichelt:2021svh} in retaining logarithmically enhanced mass contributions associated with
heavy-flavour quarks. 
Sec.~\ref{sec:HLpredictions} is devoted to particle-level predictions: we perform full particle-level simulations with the \sherpa~event-generator framework, use these to incorporate non-perturbative corrections into our \NLOpNLLp results using the transfer-matrix approach first presented in \cite{Reichelt:2021svh}, and present our final \NLOpNLLp+\;NP results.
We summarize our achievements and discuss future prospects in Sec. \ref{sec:conclusion}.

\section{Setting the scene}\label{sec:obs_def}

\subsection{Observable definition}

The substructure observables we are interested in belong to the family of jet angularities~\cite{Larkoski:2014pca}. These probe both the angular and the transverse-momentum distribution of particles within a given jet. They are defined as follows:

\begin{equation}\label{eq:jet_ang}
    \lambda_\alpha^\kappa = \sum_{i\in {\text{jet}}} \left(    \frac{p_{T_i}}{\ptjet}\right)^\kappa \left(\frac{\Delta R_i}{R_0}\right)^\alpha\,,
\end{equation}
where
\begin{equation}
    \Delta R_i = \sqrt{(\phi_i-\phi_{\text{jet}})^2 +(y_i-y_{\text{jet}})^2}
\end{equation}
is the azimuth-rapidity distance of particle $i$ from the jet axis, which is determined using the Winner-Take-All (WTA) recombination scheme \cite{Larkoski:2014uqa} for $\alpha \leq 1$, and using the default $E$-scheme axis otherwise. 
The quantities $p_{T_i}$, $y_i$, $\phi_i$  denote the transverse momentum, rapidity and azimuth of the $i$-th particle with respect to the colliding beam, while $\ptjet$, $\yjet$, $\phijet$ correspond to the respective variables of the jet.
Finally, $R_0$ denotes the radius parameter used for the anti-$k_t$ \cite{Cacciari:2008gp} jet clustering. We restrict our analysis to the infrared-and-collinear-safe case $\kappa=1$ with $\alpha \in\{1/2,1,2\}$. In previous studies \cite{Larkoski:2014pca,Andersen:2016qtm} these have been dubbed Les Houches Angularity (LHA) $\lambda^1_{1/2}$, Width  $\lambda^1_{1}$, and Thrust  $\lambda^1_{2}$, respectively. We adopt this naming convention here. 
We highlight that eq.~\eqref{eq:jet_ang} is not the only possible definition in the context of jets seeded by a heavy flavour. Alternative definitions have been studied in \cite{Lee:2019lge,Dhani:2024gtx}, however, as found in~\cite{Dhani:2024gtx}, eq. \eqref{eq:jet_ang} is the most suitable to expose the dead-cone effect and is less affected by hadronization corrections.

\subsection{Process definition and phase-space cuts}
\label{subsec: process and phase space cuts}
In this paper, we consider the analysis of jets produced in association with a $Z$-boson in proton--proton collisions. In particular, we are interested in analysing the impact of mass effects on the jet-angularity distribution.  To this end, we inspect two possible subclasses of the process of interest, namely the production of a $Z$-boson together with either a light-flavour 
jet $j$ or a $b$-jet. The phase-space cuts that we apply equal those used in the CMS measurement presented in \cite{CMS:2021iwu}.
For completeness, we here briefly summarize the final settings, and we refer the reader to Ref.~\cite{CMS:2021iwu} for additional details. 

We consider the inclusive production of a pair of oppositely charged muons in proton--proton collisions at a centre-of-mass energy of $13~\text{TeV}$. Both muon candidates need to satisfy
\begin{equation}
  p_{T,\mu} > 26~\mathrm{GeV}\,,\;\;\text{and}\;\;|\eta_\mu|<2.4\,.
\end{equation}
The lepton pair has to pass the additional conditions
\begin{equation}
\quad70~\mathrm{GeV}<m_{\mu^+\mu^-}<110~\mathrm{GeV}\,,\quad  p_{T,\mu^+\mu^-}>30\;\text{GeV}\,, \quad |\eta_{\mu^+ \mu^-}|<5.
\end{equation}
In the following, we refer to the lepton pair collectively as the $Z$-boson, implicitly 
including off-shell effects and $\gamma^*$ exchange. In what follows, we will consider three 
possible hard-process configurations that we evaluate at NLO QCD accuracy:
\begin{enumerate}
\item[(i)] \textbf{heavy-flavour production}: refers to the partonic processes $b+g\to Z+b$ and $\bar{b}+g \to Z+\bar{b}$ with the $b$-quark treated as massive,
\item[(ii)] \textbf{light-flavour production}: refers to the $jj\to Z+j$ process, where $j$ is 
either a light-flavour (anti-)quark, treated as massless, or a gluon,
\item[(iii)] \textbf{inclusive production}: refers to the combination of the former two cases, assuming four massless quark flavours for the $jj\to Z+j$ channel.
\end{enumerate}

Note that for the light-flavour production mode we here assume four massless quarks. Instead, 
in Refs.~\cite{Caletti:2021oor,Reichelt:2021svh} we worked in a five-flavour scheme, i.e. 
treating the $b$-quarks massless as well. 

We select events that exhibit at least one anti-$k_t$ jet \citep{Cacciari:2008gp} with 
\begin{equation}
  |y_{\text{jet}}| < 1.7\,,\quad\text{and}\quad R_0=0.4\,,
\end{equation}
and consider several bins in $p_{T,\text{jet}}$ starting at $50~\text{GeV}$ in close correspondence to~\cite{CMS:2021iwu}.\footnote{Our framework is, of course,
  general and theoretical predictions for different jet radii, angularity exponents and \softdrop
  parameters, as well as different selection cuts and $p_{T,\text{jet}}$ bins,
  can be easily obtained. They can be provided upon request.}
In order to gain better control over NLO QCD corrections for the $Zj$ production process,
that become large when the $Z$-boson transverse momentum is significantly smaller than
$p_{T,\text{jet}}$~\citep{Rubin:2010xp}, we require the transverse momenta of the lepton pair
and the leading jet to be largely balanced. To this end we impose the constraint
\begin{equation}\label{eq:imbalance_cut}
\Delta^{p_T}_{Z, {\rm jet}}  \equiv \left| \frac{p_{T,\rm jet} - p_{T, \mu^+\mu^-} }{ p_{T,\rm jet} + p_{T, \mu^+\mu^-}}  \right| < 0.3\,.
\end{equation}
Finally, we require the $Z$-boson and the leading jet to be well separated in azimuthal angle, i.e.\
\begin{equation}
  \Delta^\phi_{Z, {\rm jet}}  \equiv \left|\phi_Z - \phi_{\rm jet}\right| > 2\,.
\end{equation}

We also consider jets processed with the \softdrop~algorithm. The event-selection criteria are unchanged, and we choose the \softdrop~parameters $\beta = 0$ and $\zc = 0.1$.

\section{NLL resummation and matching to NLO}\label{sec:caesar_review}

In this section we detail our soft-gluon resummation calculations, based on the well-known \caesar~formalism~\cite{Banfi:2004yd,Banfi:2004nk, Banfi:2003je}. After a brief general review, we describe all necessary ingredients to account for massive radiator legs considering, in particular, the case of jet-angularities with and without \softdrop\ grooming. We carefully validate expansions of our resummed expressions against fixed-order calculations of the considered observables at LO and NLO in the strong coupling. 

We consider a well-separated hard Born configuration $\mathcal{B}$, with external legs labeled $l \in \mathcal{B}$. For an observable $\mathcal{V}$ that vanishes at Born level we assume that the contribution from a single additional gluon with momentum $k$ emitted collinear to leg $l$ can be written as  
\begin{equation}\label{eq: CAESAR parametrization}
    \mathcal{V}^{(l)}(k) =
    \left(\frac{\kt^{(l)}}{\mu_Q}\right)^{a}\,
    e^{-b_l\, \eta^{(l)}}\,
    d_l(\mu_Q)\,
    g_l(\phi).
\end{equation}
Here $\kt^{(l)}$ and $\eta^{(l)}$ denote the transverse momentum and pseudorapidity of the emission relative to leg $l$, $\phi$ is the azimuthal angle of the emission. Finally, $\mu_Q$ denotes the resummation scale, which is of the order of the hard scale of the process. In the case of jet angularities, cf. eq.~\eqref{eq:jet_ang}, it is easy to show that their \caesar\ parametrization is given by (see App.~\ref{app: kinematics} for a detailed derivation):
\begin{align}\label{eq:AngParams}
    a=1, \quad b= \alpha-1, \quad d(\mu_Q) g(\phi)= \frac{\mu_Q}{\ptjet R_0}  \left(\frac{2 \cosh \yjet}{R_0}\right)^{\alpha-1}.
\end{align}
For simplicity of notation, here we drop the superscript $(l)$. In what follows, all leg-dependent quantities appearing in the same expression are implicitly understood to be associated with the same leg $l$.

Given the observable parametrization, the resummed cumulative distribution at NLL for a value $v$ of the observable $\mathcal{V}$ can be written in terms of the following generic master formula:
\begin{align}\label{eq:caesar_master}
    &\Sigma_{\text{res}}(v) = \sum_\delta \Sigma^\delta_{\text{res}}(v), 
    \quad \text{with} \nonumber \\
    &\Sigma^\delta_{\text{res}}(v) = \int \de\mathcal{B}_\delta \,
    \frac{\de\sigma_\delta}{\de \mathcal{B}_\delta} \,
    \exp\!\left[-\sum_{l\in \delta} R_l^{\mathcal{B}_\delta}(v)\right]\,
    \mathcal{S}^{\mathcal{B}_\delta}(v)\,
    \mathcal{P}^{\mathcal{B}_\delta}(v)\,
    \mathcal{F}^{\mathcal{B}_\delta}(v)\,
    \mathcal{H}^{\delta}(\mathcal{B}_\delta),
\end{align}
where the sum extends over different partonic channels $\delta$. 

At NLL accuracy, one systematically exponentiates all contributions of type $\as^n L^n$ with $L=-\ln v$ and higher logarithmic powers. This formulation is very general and applies to a wide class of observables. The main ingredients are:
\begin{itemize}
    \item[--] the hard function $\mathcal{H}$, encoding the kinematic cuts on the Born-level configuration $\mathcal{B}$;
    \item[--] the multiple-emission function $\mathcal{F}$;
    \item[--] the soft function $\mathcal{S}$, implementing the non-trivial colour evolution;
    \item[--] the collinear radiators $R_l$ for each hard leg $l$;
    \item[--] the PDF ratio $\mathcal{P}$, which accounts for the true initial-state collinear scale via
    \begin{equation}\label{eq:p}
    \mathcal{P}^{\mathcal{B}_\delta}(L) =
    \prod_{l=1}^2 
    \frac{q_l\!\left(x_l^{\mathcal{B}_\delta},\, e^{-L/(a+b_l)}\muf\right)}
         {q_l\!\left(x_l^{\mathcal{B}_\delta},\, \muf\right)}.
\end{equation}
\end{itemize}

For a detailed discussion of the construction and applicability of the \caesar~approach, we refer the reader to the original literature, in particular~\cite{Banfi:2004yd}. The formalism has been successfully applied to the resummation of several event-shape~\cite{Banfi:2010xy} and jet-shape~\cite{Reichelt:2021svh} observables in hadron--hadron collisions. 
In the following, we present the general framework with and without \softdrop\ for the case of jet angularities initiated by a $b$-quark, as generalization of ~\cite{Reichelt:2021svh}. Our analysis makes use of the implementation of the \caesar~formalism within the \sherpa~framework, originally presented in~\cite{Gerwick:2014gya}, and subsequently employed for resummed predictions of \softdrop\ thrust~\cite{Marzani:2019evv} and multijet resolution scales~\cite{Baberuxki:2019ifp} in $e^+e^-$ collisions, 
as well as event-shape variables in DIS~\cite{Knobbe:2023ehi}, hadronic Higgs-boson decays~\cite{Gehrmann-DeRidder:2024avt}, and hadron--hadron collisions~\cite{Baron:2020xoi}.

For all resummation calculations we use version 3.0.3 of the \sherpa\ event generator~\cite{Sherpa:2024mfk}. 
The required tree-level matrix elements are provided by its built-in matrix-element generators \textsc{Amegic}~\cite{Krauss:2001iv} and \textsc{Comix}~\cite{Gleisberg:2008fv}, one-loop virtual matrix elements we obtain from \textsc{OpenLoops2}~\cite{Buccioni:2019sur} using the \textsc{Collier} library~\cite{Denner:2016kdg}. 
The input value for the strong coupling is set to $\as(m^2_Z)=0.118$, consistent with the PDF4LHC21 set of PDFs~\cite{PDF4LHCWorkingGroup:2022cjn} used for the proton, accessed through 
\textsc{LhaPDF}~\cite{Buckley:2014ana}. For the mass of the $b$-quark we use $m_b=4.92\,\text{GeV}$. 
For the factorization ($\muf$), renormalization ($\mur$), and resummation scale ($\mu_Q$) we use
\begin{equation}\label{eq:scalesResummation}
    \muf=\mur=\ptZ\,,\quad \mu_Q = \ptjet R_0\,.
\end{equation}

To quantify the theoretical uncertainty, we first consider variations of $\muf$ and $\mur$ by factors of $1/2$ and $2$, excluding the combinations where one scale is doubled and the other halved. To estimate the resummation uncertainty, we introduce a parameter $x_L$ varying between $1/2$ and $2$, which rescales the argument of logarithms of the observable, cf. Sec.~\ref{sec:matching}. The total perturbative uncertainty is then estimated by the envelope of all $\muf$, $\mur$, and $x_L$ variations.

\subsection{NLL resummation for ungroomed jet shapes}
\label{sec: NLL}
When considering hard legs $l$ that feature a non-vanishing mass, some elements of the \caesar\ formalism
need to be generalized. We begin by considering the NLL resummation for ungroomed jet angularities for a 
heavy-flavour jet. All-order calculations can be performed exploiting the fact that QCD scattering amplitudes 
with massive partons factorize in the quasi-collinear limit. In this approximation the transverse momentum 
of the emitted radiation, $\kt$, and the mass $m$ of the heavy quark are assumed to be small compared to the 
hard scale of the process~\cite{Cacciari:2001cw}. However, their ratio is assumed to be of order one. Using the standard 
Sudakov decomposition, we obtain in this approximation

\begin{align}\label{eq:qc_fact}
    |\mathcal{M}|^2\simeq |\mathcal{M}_0|^2 \frac{8\pi \as z(1-z)}{\kt^2+z^2 m^2}P_{gb}(z,\kt^2)\,,
\end{align} 
with $\mathcal{M}$ the original scattering amplitude and $\mathcal{M}_0$ the reduced amplitude with one external parton less. In eq.~\eqref{eq:qc_fact}, $z$ denotes the fraction of the momentum transferred in the splitting process $b \to g + b$, and $P_{gb}$ is the leading order time-like massive splitting function:

\begin{align}\label{eq:massive_splitting}
    P_{gb}(z,\kt^2)= \cf\left(\frac{1+(1-z)^2}{z}-\frac{2 m^2 z (1-z)}{\kt^2+z^2 m^2}\right)\,.
\end{align}
In the remainder of this section, we compute the final-state $b$-quark radiators for the \caesar\ formalism, neglecting contributions suppressed by power corrections in the heavy-quark mass. The general phase-space constraints for emissions are the following:  

\begin{enumerate}
    \item A lower bound on the emission rapidity with respect to the heavy-quark leg $l$,
    \begin{equation}
        \eta>\eta_{\min} = \ln\!\left(\frac{2 \cosh y_\text{jet}}{R_0}\right)\,.
    \end{equation}
    \item An upper bound from collinear momentum conservation,
    corresponding to $z < 1$. 
    \item The observable constraint, $\mathcal{V}(k)>\lambda^1_{\alpha}$,
    imposed by the parametrization of eq.~\eqref{eq: CAESAR parametrization}.
\end{enumerate}

The all-order calculation of jet angularities $\lambda^{1}_\alpha$ for massive quarks with $\alpha\in (1/2,1,2)$ was first presented in~\cite{Dhani:2024gtx}.
Within this framework, we are able to control soft and collinear logarithms to NLL accuracy. Depending on the kinematical region, these logarithms can manifest themselves as logarithms of the angularity or logarithms of the heavy-quark mass divided by the jet’s transverse momentum.
The expression for the radiator associated to a $b$-quark reads\footnote{For sake of notation, from now on we omit the label $\mathcal{B}_\delta$.}
\begin{align}\label{eq:radiator_ungromeed_massive}
    R_b=\int^1_0 \de z&\int^{\mu^2_Q}_{z^2 m^2}\frac{\de \kt^2}{\kt^2} \frac{\as^{\text{CMW}}(\kt^2)}{2\pi} P_{gb}(z,\kt^2-z^2m^2)\nonumber\\
 \times&\Theta\left( \frac{\kt}{\ptjet R_0} e^{-(\alpha-1)(\eta-\eta_{\text{min}})}-
 \lambda^{1}_\alpha\right) \Theta(\eta-\eta_{\text{min}}),
\end{align}
\\
where the constraint $\kt>z m$, arises from the dead-cone effect condition (see App.~\ref{app: kinematics} for further details).
Here, CMW refers to the Catani--Marchesini--Webber scheme \cite{Catani:1990rr}, with
\begin{equation} \label{eq:decoupling}
    \as^{\text{CMW}}(\kt^2)= \as(\kt^2)\left(1+\frac{\as(\kt^2) K^{(n_f)}}{2\pi}\right),
\end{equation}
where $K^{(n_f)}= \ca \left(\frac{67}{18}-\zeta_2\right)-\frac{10}{9}\tr n_f$, 
with $n_f$ the number of active flavours at scale $\kt^2$. The expression 
for $\as(\kt^2)$ is given by
\begin{align}
    \as(\kt^2) = \as^{(5)}(\kt^2)\Theta(\kt^2-m_b^2) 
+ \as^{(4)}(\kt^2)\Theta(m_b^2-\kt^2)
\end{align}
with
\begin{align}\label{eq:rc-nf}
    &\as^{(5)}(\kt^2)= \frac{\as}{\ell_5}\left(1-\frac{\beta_1^{(5)}}{\beta_0^{(5)}}\frac{\ln \ell_5}{\ell_5}\right), \nonumber \\
    &\as^{(4)}(\kt^2)= \frac{\as}{\ell_4+\delta_{54}}\left(1-\frac{\beta_1^{(4)}}{\beta_0^{(4)}}\frac{\ln\left(\ell_4+\delta_{54}\right)}{\ell_4+\delta_{54}}-\left(\frac{\beta^{(5)}_1}{\beta^{(5)}_0}-\frac{\beta^{(4)}_1}{\beta^{(4)}_0}\right)\frac{\ln \ell_5^{(m)}}{\ell_4+\delta_{54}}\right), \nonumber \\
    &\ell_{n_f}= 1+\as\beta_0^{(n_f)}\ln{\frac{\kt^2}{\mu^2_Q}}, \quad  \ell_{n_f}^{(m)}= 1+\as\beta_0^{(n_f)}\ln{\frac{m^2}{\mu^2_Q}}, \quad \delta_{54}= \ell_5^{(m)}-\ell_4^{(m)},
\end{align}
where $\as= \as^{(5)}(\mu_Q^2)$. 
Here $\beta_0^{(n_f)}= \frac{11 \ca-2 n_f}{12 \pi}$ and $\beta_1^{(n_f)}= \frac{17 \ca^2- 5n_f \ca-3n_f \cf}{24 \pi^2}$ denote the one- and two-loop coefficients of the QCD $\beta$-function.
\\
The expression of the radiator in eq.~\eqref{eq:radiator_ungromeed_massive} is given by
\begin{align}
\label{eq: massive ungroomed radita}
    R_b =\cf\Bigg[&r\left(L_\alpha\right)+B_1\;t(L_\alpha) \Theta\left(-\alpha \ln \left(\frac{m}{\mu_Q}\right)-L_\alpha\right)    \nonumber\\
    +&\left(B_1\;t\left(-\ln\left(\frac{m}{\mu_Q}\right)\right) + H_1\; t(L_\alpha)\right)\Theta\left(L_\alpha+\alpha \ln\left(\frac{m}{\mu_Q}\right)\right) \nonumber\\
    +& \left(t(L_\alpha)-\alpha r'(L_\alpha) \right)\ln\left(\frac{ \ptjet R_0}{\mu_Q}\right)\Bigg],
\end{align}
where $L_\alpha=-\ln \lambda^1_\alpha$, the function $t$ is defined as
\begin{align}
    t(x)= \int^{\mu^2_Q}_{\mu^2_Q e^{-2x}} \frac{\de \kt^2}{\kt^2}\frac{\as(\kt^2)}{\pi},
\end{align}
while $B_1= -\frac{3}{4},\; H_1=-\frac{1}{2}$.
The contribution $r(L_\alpha)$ is obtained by performing the integrals:
\begin{align}
\label{eq:r_ungroomed}
  r(L_\alpha) = \int_0^{m^2}\frac{\de \kt^2}{\kt^2} \int_{\frac{\kt}{\ptjet R_0}}^{\frac{\kt}{m}} \frac{\de z}{z}  \frac{\as^{\text{CMW}}(\kt^2)}{\pi} \,\Theta_\alpha + \int_{m^2}^{\mu_Q^2}\frac{\de \kt^2}{\kt^2} \int_{\frac{\kt}{\ptjet R_0}}^1  \frac{\de z}{z}  \frac{\as^{\text{CMW}}(\kt^2)}{\pi} \,  \Theta_\alpha\,, 
\end{align}
where $\Theta_\alpha$ denotes the observable constraint and reads
\begin{align}
    \Theta_\alpha=\Theta\left(L_\alpha-\alpha\ln\left(\frac{\ptjet R_0}{\kt}\right)-(\alpha-1)\ln z\right)\,.
\end{align}
Finally, we have defined $r'\equiv \frac{\de r}{\de L_\alpha}$.
In order to highlight the main features of the massive radiator computed in eq.~\eqref{eq: massive ungroomed radita}, it is interesting to study its expansion at order $\order{\as^2}$ in the asymptotic region, i.e. for $\lambda^1_\alpha \ll \text{min}\left(\left(\frac{m}{\mu_Q}\right)^\alpha, \frac{m}{\mu_Q}\right)$:
\begin{align}
\label{eq:exp_res_ug}
    R_b&= \sum^2_{i,j=1} G_{ij} \left(\frac{\as}{2\pi}\right)^i L_\alpha^j + \order{\as^3},
    \end{align}
    with the coefficients $G_{ij}$ given by
    \begin{subequations}
    \begin{align}
    \label{eq: G_ij ungroomed}
G_{10}&=\cf\ln\left(\frac{m}{\mu_Q}\right)\left[
4 \alpha \ln\!\left(\frac{\ptjet R_0}{\mu_Q}\right)
-  2\alpha \ln\!\left(\frac{m}{\mu_Q}\right)
- 4 B_1
+ 4 \alpha H_1
\right], \\
    G_{11}&=4\cf\Bigg[
 \ln\!\left(\frac{ \ptjet R_0}{\mu_Q}\right)
-  \ln\!\left(\frac{m}{\mu_Q}\right)
+ H_1
\Bigg], \\
G_{12}&=0,
\\
G_{20}&=\frac{2}{3} C_F \ln^2{\left(\frac{m}{\mu_Q}\right)} \Bigg[
3 \Big(
K^{(5)} - K^{(4)}(1+\alpha)
+ 4\pi \big(
B_1 \beta^{(5)}_{0}
+ H_1 \alpha(-2\beta^{(5)}_{0} + \beta^{(4)}_{0}\alpha)
\big)\nonumber
\\
&+ 4\pi\ln\left(\frac{p_T R_0}{\mu_Q}\right) \left(
\beta_0^{(4)}-\beta^{(5)}_{0}
+\alpha(\beta_0^{(4)}(1+\alpha)-2 \beta_0^{(5)}) 
\right)
\Big) \nonumber \\
&- 4\pi \ln \left(\frac{m}{\mu_Q}\right) \left(
-\beta^{(5)}_{0}(1+3\alpha)
+ \beta^{(4)}_{0}(1+\alpha+\alpha^2)
\right)
\Bigg] ,\\
G_{21}&= -4 C_F \ln\left(\frac{m}{\mu_Q}\right)
\Big[
K^{(4)}
+ 4  \pi H_1 (\beta^{(5)}_{0} - \beta^{(4)}_{0}\alpha)\nonumber 
\\
&- 4 \pi \left(\beta^{(4)}_{0}(1+\alpha) - \beta_{0}^{(5)}\right)\ln\left(\frac{p_T R_0}{\mu_Q}\right)
+2 \pi \ln\left(\frac{m}{\mu_Q}\right)(\beta^{(4)}_{0}(1+\alpha)- 2\beta^{(5)}_{0} )
\Big],\\
G_{22}&=\; 8\pi \cf \beta_0^{(4)}\left(2  \ln \left(\frac{\ptjet R_0}{\mu_Q}\right)-  \ln\left(\frac{m}{\mu_Q}\right)+H_1\right).
\end{align}
\end{subequations}
As can be seen, there appear at most single-logarithmic contributions, i.e. $\as^n L^n_\alpha$. This is in contrast to the massless case, where the dominant logarithms in the radiator are of the form $\as^n L^{n+1}_\alpha$~\footnote{Note that we here include coefficients not enhanced by logarithms of the observable, i.e. $G_{10}$ and $G_{20}$, which, however, feature logarithms of the heavy-quark mass.}.

To further investigate the role of quark masses, we also consider the radiator for light quark and gluon jets with thresholds in the running coupling. The corresponding radiators read

\begin{figure}
    \centering    \includegraphics[width=0.49\linewidth]{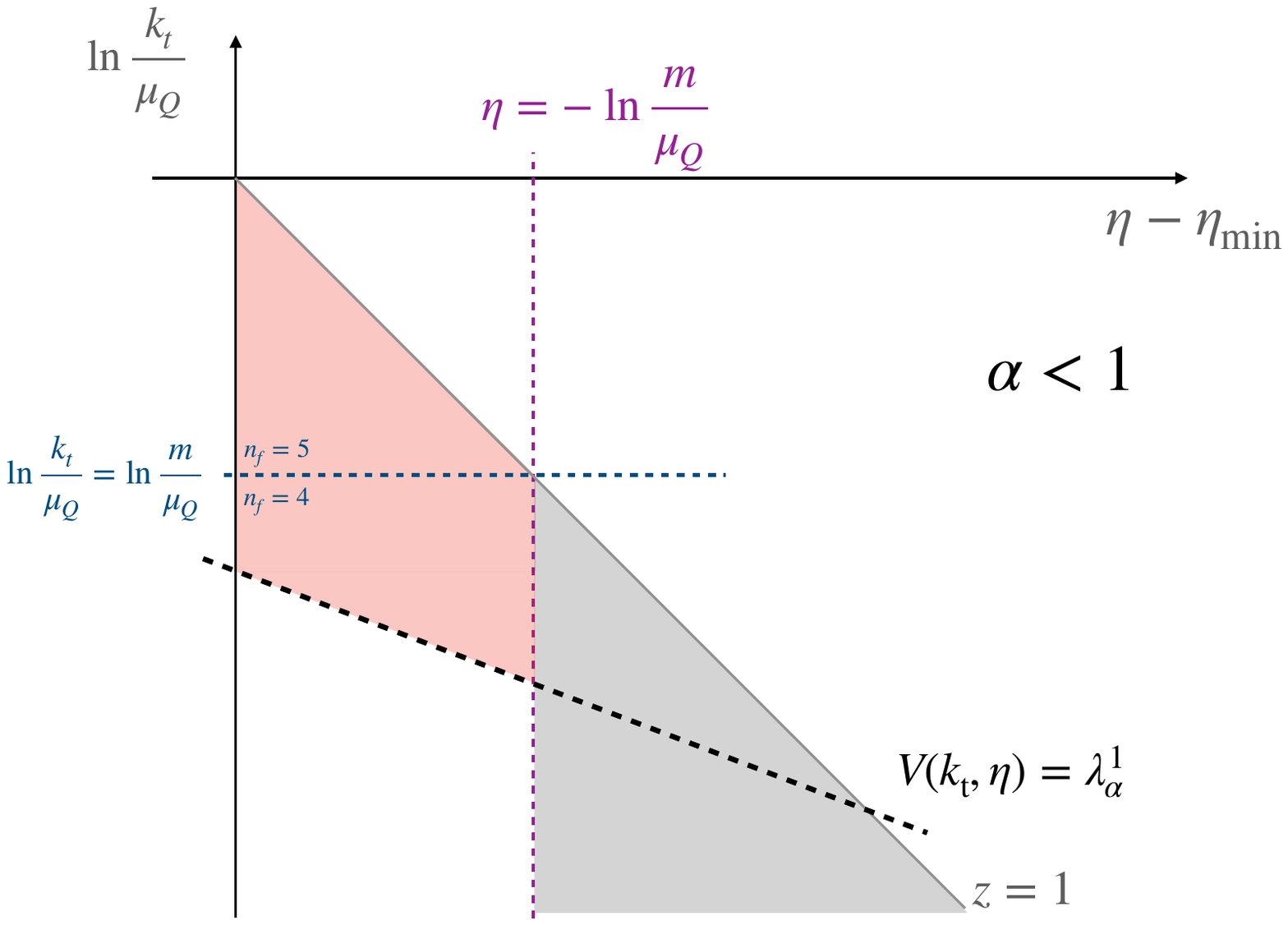}
    \includegraphics[width=0.49\linewidth]{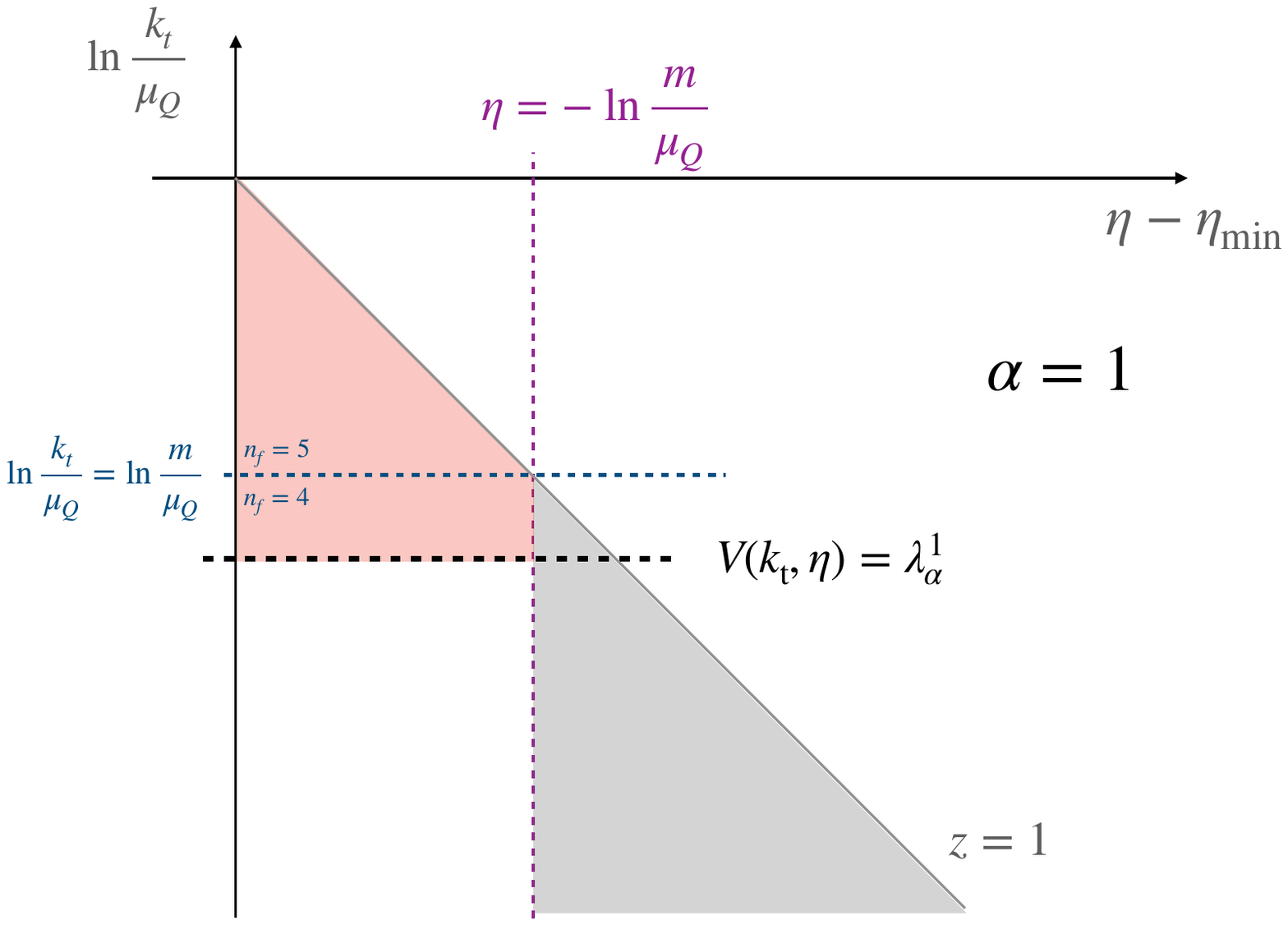}
    \includegraphics[width=0.49\linewidth]{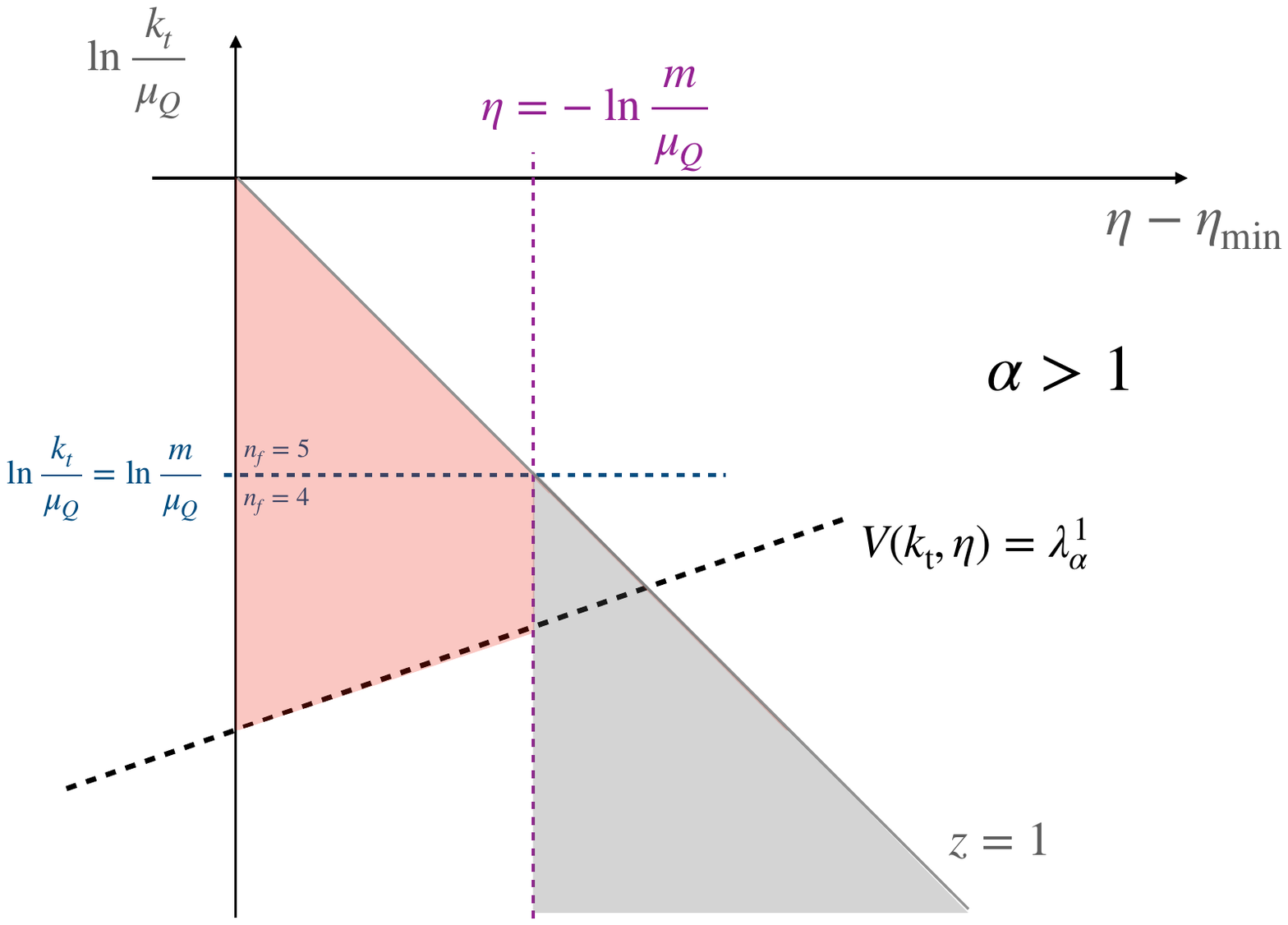}
    \caption{Lund-plane representation of the soft and quasi-collinear phase space for emissions off a $b$ quark for the case of an ungroomed jet. The dead-cone region is indicated in grey. The dashed line in black indicates a measurement of the jet angularity. We explicitly show the cases of $\alpha=1/2$, $\alpha=1$, and $\alpha>1$. The corresponding area in red is the vetoed region, which gives rise to the Sudakov form factor. Finally, the horizontal dashed blue line indicates the boundaries between different flavour numbers for the running coupling.}
    \label{fig:lund_ungroomed}
\end{figure}

\begin{subequations} \label{eq:radiator_ungromeed_light}
\begin{align}
    R_q=&\int^1_0 \de z\int^{\mu^2_Q}_{0}\frac{\de \kt^2}{\kt^2} \frac{\as^{\text{CMW}}(\kt^2)}{2\pi} P_{gq}(z)
 \Theta\left( \frac{\kt}{\ptjet R_0} e^{-(\alpha-1)(\eta-\eta_{\text{min}})}-
 \lambda^{1}_\alpha\right) \Theta(\eta-\eta_{\text{min}}),\\
 R_g=&\int^1_0 \de z\int^{\mu^2_Q}_{0}\frac{\de \kt^2}{\kt^2} \frac{\as^{\text{CMW}}(\kt^2)}{2\pi} P_{xg}(z)
 \Theta\left( \frac{\kt}{\ptjet R_0} e^{-(\alpha-1)(\eta-\eta_{\text{min}})}-
 \lambda^{1}_\alpha\right) \Theta(\eta-\eta_{\text{min}}),
\end{align}
\end{subequations}
with

\begin{align}
    P_{gq}(z)=& \cf \left(\frac{1+(1-z)^2}{z}\right), \quad &&P_{gg}(z)=\ca\left(2\frac{1-z}{z}+z(1-z)\right),\nonumber &&\\
    P_{qg}(z)=& \tr n_f\left(z^2+(1-z)^2\right),\quad &&P_{xg}(z)= P_{gg}(z)+ P_{qg}(z). &&
\end{align}
The calculation of eqs.~\eqref{eq:r_ungroomed} and~\eqref{eq:radiator_ungromeed_light} was carried out in \cite{Dhani:2024gtx} using Lund-plane diagrams~\cite{Andersson:1988gp} as depicted in Figure~\ref{fig:lund_ungroomed}.
In total there are eight different kinematical regions for both the heavy-quark case and light-quark/gluon case, depending on the hierarchy between the observable value $\lambda^1_\alpha$ and $m/\mu_Q$, divided among $\alpha=1$ (2 regions), $\alpha<1$ (3 regions) and $\alpha>1$ (3 regions). Mass effects thereby enter the expression for the light-flavour radiators through flavour thresholds in the running coupling and the gluon splitting functions.

The contributions to the PDF ratio in eq.~\eqref{eq:p} originate from radiation collinear to the initial-state legs; thus, they can be neglected for the jet angularities considered here, and the ratio can accordingly be set to unity, i.e. $\mathcal{P} = 1 $. Moreover, the multiple-emission function receives contributions only from the jet-collinear limit and for additive observables it takes the form
\begin{equation}
    \mathcal{F}^{\mathcal{B}_\delta}(L_\alpha) = \frac{e^{-\gamma_{\text{E}} r' (L_\alpha)}}{\Gamma\left(1+r'(L_\alpha)\right)}.
\end{equation}

 Finally, soft wide-angle emissions can directly contribute to the jet, setting an observable value. In this case, radiation from each dipole is integrated over the phase space constrained by the Born kinematics and the applied cuts. Two distinct contributions enter the soft function $\mathcal{S}$, which we shall refer to as global and non-global effects, i.e.
\begin{align}
    \mathcal{S}^{\mathcal{B}_\delta}(L_\alpha)= \mathcal{S}_{\text{global}}^{\mathcal{B}_\delta}\left(t\left(L_\alpha\right)\right)\mathcal{S}_{\text{non-global}}^{\mathcal{B}_\delta}\left(t\left(L_\alpha\right)\right)\,.
\end{align}
In general, the global part of the soft function can be written as
\begin{align}
\label{eq: Sglobal}
    \mathcal{S}_{\text{global}}^{\mathcal{B}_\delta}\left(t\right)= \text{Tr}\left[H e^{-t \left(\mathbf\Gamma^{\mathcal{B_\delta}}\right)^\dagger}c \,e^{t \mathbf\Gamma^{\mathcal{B_\delta}}} \right]\frac{1}{\text{Tr}\left[c H\right]}\,,
\end{align}
where $c$ and $H$ denote the colour metric and the hard function, respectively (see \cite{Gerwick:2014gya} and \cite{Baberuxki:2019ifp} for details on the notation) and $\mathbf\Gamma$ is given by a sum over all possible dipoles:
\begin{align}
\label{eq: dipoles sum}
    \mathbf\Gamma^{\mathcal{B}_\delta}= \sum_{i> j \in \mathcal{B}_\delta} \mathbf{T}_i \cdot \mathbf{T}_j~I_{ij}\,.
\end{align}
The jet-collinear parts of the dipoles are already included through the radiator $R_i$ and therefore need to be subtracted. As a consequence, the global part of the soft function is proportional to power corrections in the jet radius and the masses of the particles involved in the specific partonic channel. The results for the dipoles that contribute to the process $pp\to Z+b$ are presented in App.~\ref{app:dipoles-Zb}~\footnote{In addition, in view of upcoming measurements by the LHC collaborations, in App.~\ref{app:dipoles-Zbb} we furthermore evaluate the dipoles that contribute to $pp\to Z+b \bar b$, which exhibits a richer colour structure.}. Only three dipoles $(ij)$ contribute to the soft function, namely $(12)$, $(13)$, and $(23)$.
Exploiting colour conservation, all colour operators $\mathbf{T}_i \cdot \mathbf{T}_j$ can be expressed as linear combinations of the $SU(3)$ Casimir invariants $\cf$ and $\ca$. As a result, the matrix structure of eq.~\eqref{eq: dipoles sum} collapses, since the product of the generators becomes diagonal in colour space.

On the other hand, non-global logarithms (NGLs) contribute to the soft function $\mathcal{S}$. 
They first arise at two-loop order from the correlated emission of two gluons at large but comparable angles near the jet boundary. 
Because these effects are confined to the boundary region, mass corrections only enter as power-suppressed terms, as demonstrated in Ref.~\cite{Dhani:2024gtx}. 
Such configurations spoil globalness, i.e.\ the property that guarantees that only flavour combinations with fixed parton multiplicities define the Born configurations $\mathcal{B}_\delta$ around which the resummation is organized. 
As a result, in addition to resumming around $n$-parton final states, one must also resum around $(n+1)$-parton states, 
at least in the region of phase space where the extra parton lies outside the jet. 
The non-global component of $\mathcal{S}^{\mathcal{B}_\delta}(t)$ is computed numerically in the large-$\nc$ limit, using the algorithm of~\cite{Dasgupta:2001sh}. This method can be straightforwardly extended to ungroomed angularities, as done in~\cite{Caletti:2021oor}. We emphasize that the non-global contribution depends only on $\lambda^1_\alpha$ (and on the jet transverse momentum through $\as(\mur^2)$), 
but not on the angularity parameter $\alpha$. 
Similarly, in the groomed case, it depends only on $\zc$ and not on $\beta$. 
The origin of this feature is that, at single-logarithmic accuracy, NGLs originate from configurations that are strongly ordered in energy, with angles of order of the jet radius $R_0$. In practice, $\mathcal{S}_{\text{non-global}}(t)$ is computed separately for each colour dipole, either incoming--incoming or incoming--final. These contributions are independent of the jet rapidity up to power corrections in the mass and jet radius, implying, in particular, that the two incoming--final configurations are identical up to mass-suppressed corrections. 
Overall, since we are neglecting all mass-suppressed power corrections, mass effects in the soft function are included only through the running coupling. The impact of mass power corrections at order $\order{\as}$ will be examined in App. \ref{app:dipoles-Zb}.

\subsection{NLL resummation for groomed jet shapes}
\label{sec: NLL SD}
In this section, we briefly outline the necessary modifications to our resummed formula when considering  \softdrop-groomed jets. \softdrop\ jets are under particularly strong theoretical control, as the grooming procedure suppresses the impact of large-angle soft emissions. This simplifies the colour structure of the resummation.  Groomed jet shapes have been studied extensively, both theoretically and experimentally.
We work in the limit where $\zc$ is small, such that power corrections can be neglected, but where $\ln \zc$ is not too large, so that a systematic resummation of these contributions is not required. 
Moreover, for~\softdrop-groomed distributions, the non-global factor remains the same for $\lambda^1_\alpha \gtrsim \zc$ and saturates at that value.

\softdrop-groomed jets initiated by massive quarks have been studied in Refs.~\cite{Lee:2019lge,Caletti:2023spr, Dhani:2024gtx, Dhani:2025fbk}. The quasi-collinear radiator with grooming is a straightforward generalization of eq.~\eqref{eq:radiator_ungromeed_massive}, and reads

\begin{align}
\label{eq:radiator_gromeed_massive}
    \bar R_b=\int^1_0 \de z\int^{\mu^2_Q}_{z^2 m^2}\frac{\de \kt^2}{\kt^2} &\;\frac{\as^{\text{CMW}}(\kt^2)}{2\pi} P_{gb}(z,\kt^2-z^2m^2)  \Theta(\eta-\eta_{\text{min}})\times \nonumber \\ 
    & \Theta\left( \frac{\kt}{\ptjet R_0} e^{-(\alpha-1)(\eta-\eta_{\text{min}})}-\lambda^1_\alpha\right)\Theta\left(\frac{\kt e^{(1+\beta)\eta}}{2 \ptjet \cosh\yjet}-\zc'\right),
\end{align}
where the last constraint derives from the \softdrop\ condition, and
\begin{align}
\label{eq: zcut prime}
    \zc'= \zc \left(\frac{2 \cosh \yjet}{R_0}\right)^\beta
\end{align}
(see App. \ref{app: kinematics} for further details).
The radiator in eq.~\eqref{eq:radiator_gromeed_massive} evaluates to:
\begin{align}
    \bar R_b=&\;\cf\Bigg[\bar r(L_\alpha)+B_1\;t(L_\alpha) \Theta\left(-\alpha \ln\left(\frac{m}{\mu_Q}\right)-L_\alpha\right)+ B_1\;t\left(-\ln\left(\frac{m}{\mu_Q}\right)\right)\Theta\left(L_\alpha+\alpha \ln\left(\frac{m}{\mu_Q}\right)\right) \nonumber\\
    +&H_1\;t(L_\alpha)\Theta\left(L_\alpha+\alpha \ln\left(\frac{m}{\mu_Q}\right)\right)\Theta\left(-\ln \zc-(\alpha+\beta)\ln\left(\frac{m}{\mu_Q}\right)-L_\alpha\right) \nonumber \\
    +&H_1\;t\left(-\ln \zc+(\alpha+\beta)\ln\left(\frac{m}{\mu_Q}\right)\right)\Theta\left(L_\alpha+\ln \zc+(\alpha+\beta)\ln\left(\frac{m}{\mu_Q}\right)\right)\nonumber \\
    +&\ln\left(\frac{ \ptjet R_0}{\mu_Q}\right) \left(-\alpha \bar{r}'(L_\alpha) +\beta\;\dot r(L_\alpha)+t(-\ln \zc)\right)\Bigg]\,,
\end{align}
where $\dot r(L_\alpha)\equiv-\frac{\de \bar r}{\de \ln \zc}$. 
Additionally, the groomed variant of the contribution $r$ reads
\begin{align}
 \bar r(L_\alpha) &= \int_0^{m^2}\frac{\de \kt^2}{\kt^2} \int_{\frac{\kt}{\ptjet R_0}}^{\frac{\kt}{m}} \frac{\de z}{z}\;  \frac{\as^{\text{CMW}}(\kt^2)}{\pi} \Theta_\alpha \Theta_{\text{SD}}\nonumber \\
  &+ \int_{m^2}^{\mu_Q^2}\frac{\de \kt^2}{\kt^2} \int_{\frac{\kt}{\ptjet R_0}}^1  \frac{\de z}{z}  \;\frac{\as^{\text{CMW}}(\kt^2)}{\pi} \Theta_\alpha \Theta_{\text{SD}},
\end{align}
where $\Theta_{\text{SD}}$ implements the \softdrop\ condition
\begin{align}
    \Theta_{\text{SD}}= \Theta\left((1+\beta)\ln z+\beta \ln\frac{\ptjet R_0}{\kt}-\ln \zc\right)\,.
\end{align}
Analytical results at NLL accuracy for general values of the exponent $\beta \geq 0$ were presented in \cite{Dhani:2024gtx}. 

\begin{figure}
    \centering
    \includegraphics[width=0.49\linewidth]{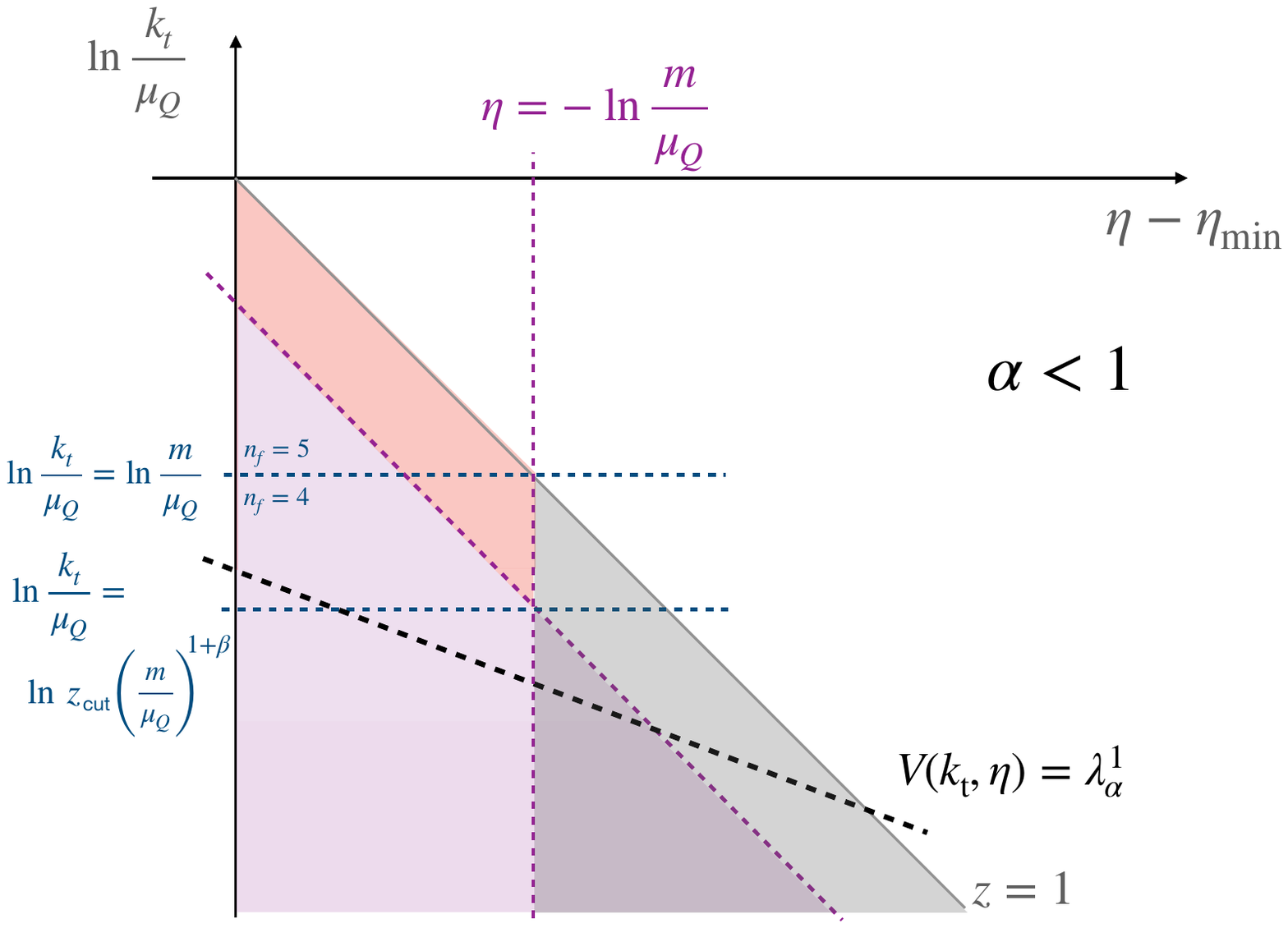} 
    \includegraphics[width=0.49\linewidth]{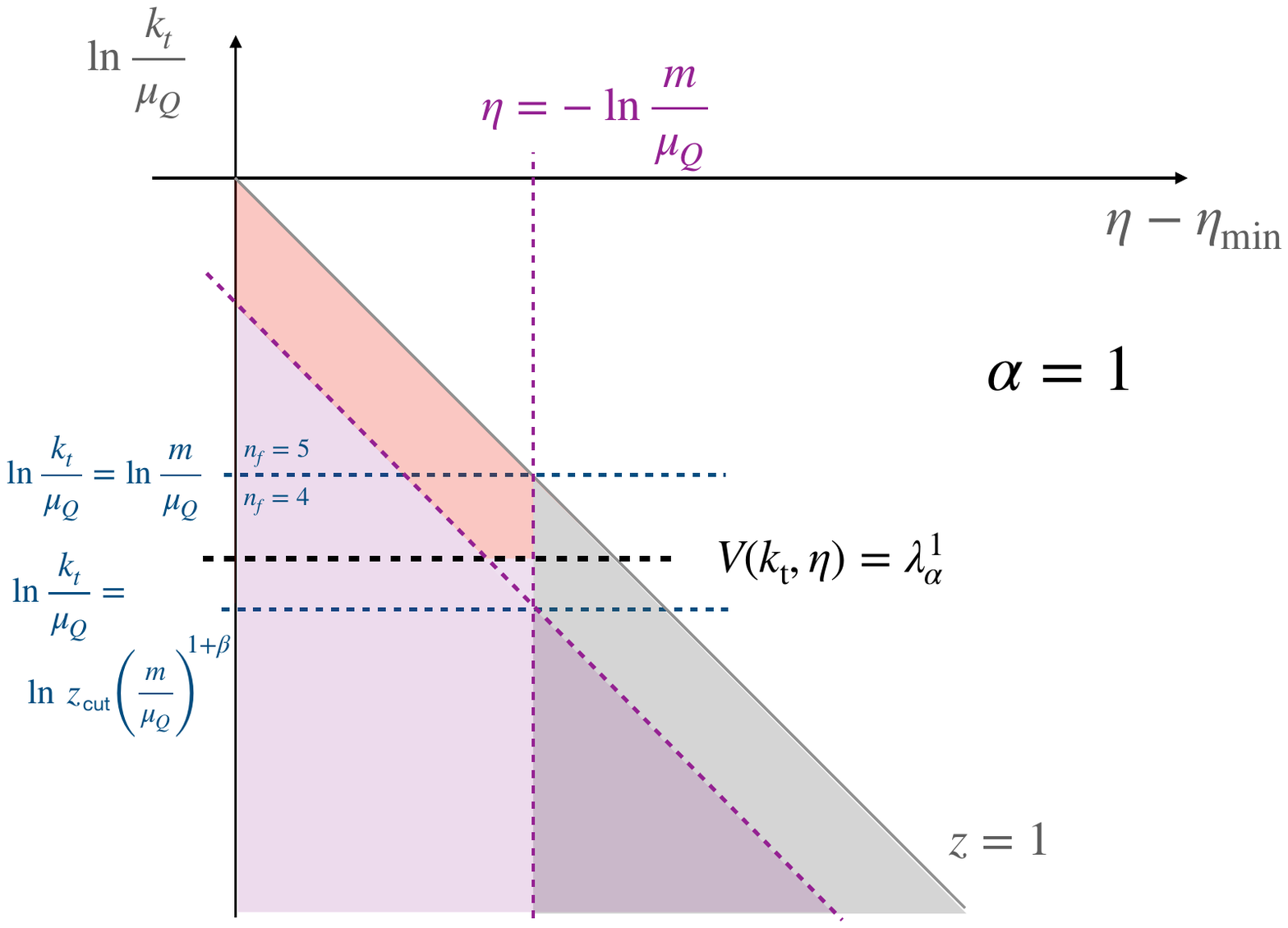}   
    \includegraphics[width=0.49\linewidth]{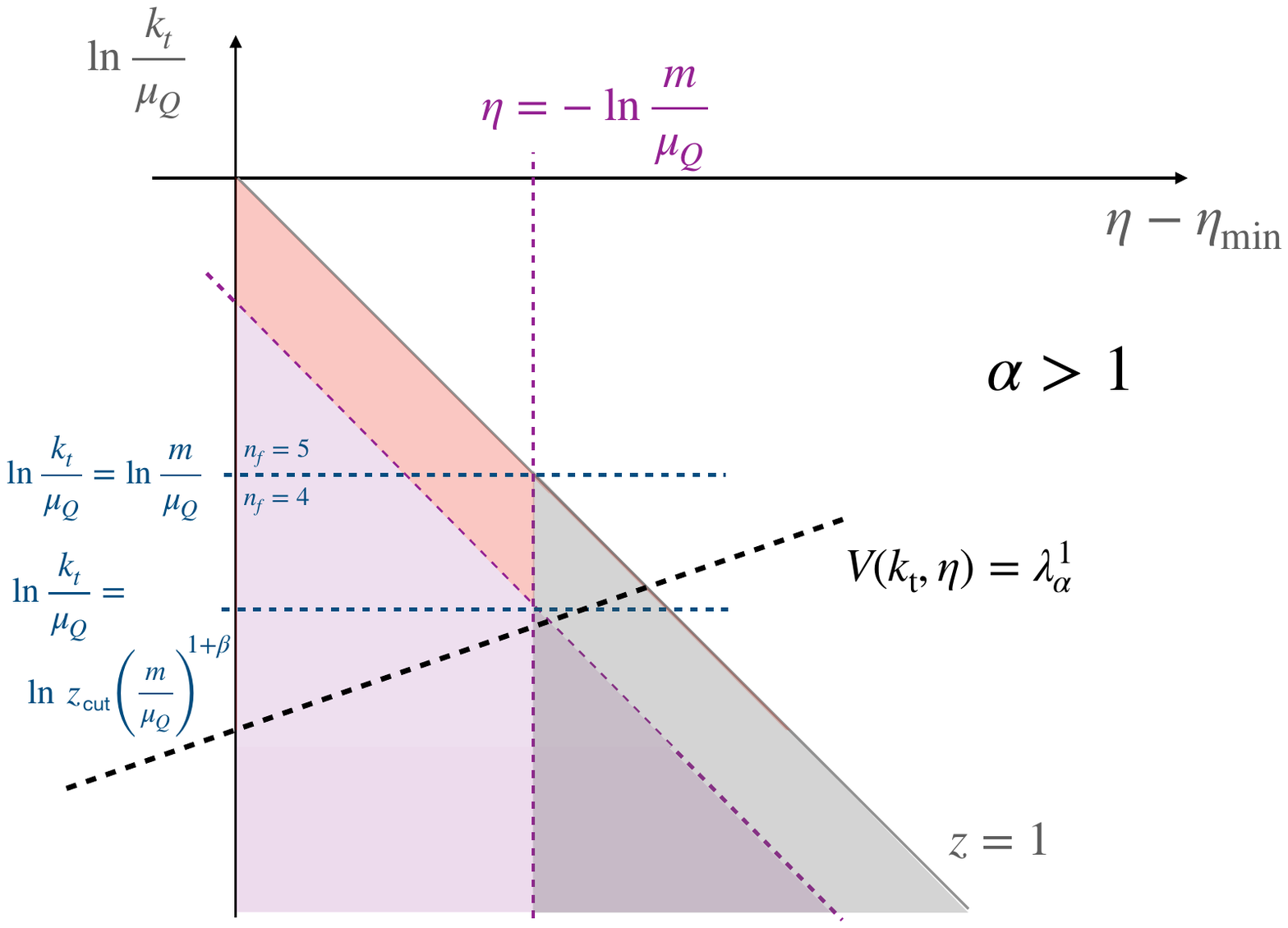}   
    \caption{Same as Fig.~\ref{fig:lund_ungroomed} for the case of a groomed jet. When \softdrop~is applied, the groomed-away region appears in light purple. 
    }
    \label{fig:lund_groomed}
\end{figure}

We note that for $L_\alpha > -\ln \zc - (\alpha + \beta)\ln \frac{m}{\mu_Q}$, the radiator no longer depends on the substructure observable.
To see this explicitly, we analyze the expression of the groomed radiator in the small observable limit. We have for $\beta=0$:
\begin{align}
\label{eq:exp_res_g}
    \bar{R}_b= \sum^{2}_{i,j=1} \bar{G}_{ij} \left(\frac{\as}{2\pi}\right)^iL_\alpha^j +\order{\as^3},
\end{align}
where the coefficients $\bar G_{ij}$ read
\begin{subequations}
\begin{align}
    \bar G_{10}=&\, 4\cf\left[-\ln\zc\left(H_1+\ln\left(\frac{\ptjet R_0}{\mu_Q}\right)\right)-\ln\left(\frac{m}{\mu_Q}\right)\left(B_1-\ln\zc\right)\right], \\
    \bar G_{11}=&\, \bar G_{12}=0\,, \\
    \bar G_{20}=&\, \frac{2}{3} C_F \Bigg\{
\ln^2\zc\Bigg[
4\pi (\beta^{(5)}_{0} - \beta^{(4)}_{0}) \ln \zc+3 \Big(
K^{(4)} - K^{(5)} +  4 H_1 \pi \beta^{(4)}_{0} \nonumber 
\\
&+ 4 \pi \beta^{(5)}_{0} \ln\left(\frac{\ptjet R_0}{\mu_Q}\right)
\Big)
\Bigg]+12 \pi \beta^{(5)}_{0} \left(B_1 - \ln\zc\right)\ln^2\!\left(\frac{m}{\mu_Q}\right)
\;\;\nonumber  \\
&\,-6 \ln \zc \ln\left(\frac{m}{\mu_Q}\right)\Big(
K^{(5)} + 4 H_1 \pi \beta^{(5)}_{0}
- 2 \pi \beta^{(5)}_{0} \ln \zc
\Big)
\; \Bigg\}\,,\\
\bar G_{21}=&\,\bar G_{22}=0.
\end{align}
\end{subequations}
This confirms that, in the small-observable limit, the radiator is indeed independent of $\lambda^1_\alpha$. This result does not hold for $\beta = 0$ only, but applies 
to all values of $\beta$.   
This is clearly visible on the Lund plane in Fig.~\ref{fig:lund_groomed}. The \softdrop\ line and the vertical line that indicates the dead cone cross in one point thus closing the phase space for (logarithmically enhanced) emissions. 
This was already noticed in~\cite{Caletti:2023spr}: while in the case of massless partons \softdrop\ with $\beta \ge 0$ behaves perturbatively as
a groomer, i.e. within resummed perturbation theory, it always returns a jet with $\lambda_\alpha^1>0$, the quark mass provides an
effective cutoff so that Sudakov suppression is frozen and there is a non-vanishing probability to find \softdrop\ jets with $\lambda_\alpha^1=0$. We expect non-perturbative effects to smear out this effect and will come back to this point when discussing them.

Also in this case, we examine the light-flavour distribution. The radiators read
  \begin{subequations}
\begin{align}
\label{eq:radiator_gromeed_massless}
    \bar R_q=\int^1_0 \de z\int^{\mu^2_Q}_{0}\frac{\de \kt^2}{\kt^2} &\;\frac{\as^{\text{CMW}}(\kt^2)}{2\pi} P_{gq}(z)  \Theta(\eta-\eta_{\text{min}}) \nonumber \\ 
    \times&\Theta\left(\frac{\kt e^{(1+\beta)\eta}}{2 \ptjet \cosh\yjet}-\zc'\right) \Theta\left( \frac{\kt}{\ptjet R_0} e^{-(\alpha-1)(\eta-\eta_{\text{min}})}-\lambda^1_\alpha\right), \\
    \bar R_g=\int^1_0 \de z\int^{\mu^2_Q}_{0}\frac{\de \kt^2}{\kt^2} &\;\frac{\as^{\text{CMW}}(\kt^2)}{2\pi} P_{xg}(z)  \Theta(\eta-\eta_{\text{min}}) \nonumber \\ 
    \times&\Theta\left(\frac{\kt e^{(1+\beta)\eta}}{2 \ptjet \cosh\yjet}-\zc'\right) \Theta\left( \frac{\kt}{\ptjet R_0} e^{-(\alpha-1)(\eta-\eta_{\text{min}})}-\lambda^1_\alpha\right).
\end{align}
\end{subequations}
Compared to the ungroomed case, here we have to deal with a significantly increased number of regions due to the presence of the new resolution scale $\zc$, both for heavy and light jets. Focusing on the regime where $\lambda_\alpha^1<\zc$, in the former case we obtain 5 regions for $\alpha=1$ and 13 regions for both $\alpha<1$ and $\alpha>1$, while in the latter case we have 3 regions for $\alpha=1$ and 6 regions for both $\alpha<1$ and $\alpha>1$. For the Lund-plane representation of jet angularities for \softdrop-groomed heavy-flavour jets see Fig.~\ref{fig:lund_groomed}.

\subsection{Validation against fixed order and matching} \label{sec:matching}

In this section, we combine our resummation expression derived in the previous section with exact fixed-order calculations for the process $pp\to Z+b$, employing the phase-space cuts and event-selection criteria described in Sec.~\ref{subsec: process and phase space cuts}. 
In analogy with eq.~\eqref{eq:caesar_master}, we introduce the fixed-order cumulative distribution
\begin{equation}
    \begin{aligned}
        \Sigma_{\rm fo} (v) &= \int_0^v \De V \frac{\De\sigma}{\De V} = \sigma-\int_v^1 \De V \frac{\De\sigma}{\De V} =: \sigma-\bar{\Sigma}_{\rm fo}(v)~, 
    \end{aligned}
\end{equation}
where in the last equality, we have introduced the subtractive $\bar{\Sigma}$ as the complement to the cumulative.
Both $\Sigma_{\rm fo}$ and $\Sigma_{\rm res}$ can be expanded in powers of $\as$ relative to the Born process:
\begin{equation}
    \Sigma= \Sigma^{(0)} + \Sigma^{(1)}+\Sigma^{(2)} + \ord{\as^4}, \qquad \Sigma^{(k)}\propto \alpha_{\text{EW}}\as^{1+k}~,
\end{equation}
and equivalently for $\bar{\Sigma}$. We use the shorthand
\begin{equation}
    \sigma_k = \Sigma^{(k)}(1)~,
\end{equation}
such that $\sigma_0$ is the Born inclusive cross section.
Explicitly, to NLO accuracy we have
\begin{equation}
  \begin{aligned}
    \Sigma_{\rm NLO} &= \sigma_0 + \Sigma_{\rm fo}^{(1)} + \Sigma_{\rm fo}^{(2)}~, \\
    \Sigma_{\rm exp.~NLO} &= \sigma_0 + \Sigma_{\rm res}^{(1)} + \Sigma_{\rm res}^{(2)}~.
  \end{aligned}
\end{equation}
The corresponding LO expressions, i.e. $\Sigma_{\rm LO}$ and $\Sigma_{\rm exp.~LO}$, are obtained by discarding the contributions $\Sigma^{(2)}$ in the above.
The cumulative $\Sigma_{\rm exp}$ is obtained by expanding eq.~\eqref{eq:caesar_master} in
$\as$. However, this is not the full information available in the matched
distribution, which will also contain a cross term with the constant prefactor
$C(\as)$. We therefore additionally include
\begin{equation}
  \Sigma_{\rm exp.~NLO+C} = \sigma_0 + \left(1+\frac{\Sigma_{\rm fo}^{(1)}-\Sigma_{\rm res}^{(1)}}{\sigma_0}\right)\Sigma_{\rm res}^{(1)} + \Sigma_{\rm res}^{(2)}~.
\end{equation}
The tree-level contributions to $\Sigma^{(1,2)}_{\text{fo}}$ are generated by \sherpa, using the toolchain for tree-level matrix elements, virtual corrections and subtraction of infrared divergencies described at the beginning of this section.
In this framework, we can obtain the Born cross section for our phase space, the real and virtual corrections to the $Z$+jet process that enter $\Sigma^{(1)}_{\rm fo}$,
and the 1-loop corrections to the $Z$+2 jets process that are needed for $\bar{\Sigma}^{(2)}_{\rm fo}$.
The latter are calculated down to small values of $v$, which are outside the plot range of any of the results we are going to present,
but remain finite so that only one parton at a time can be arbitrarily deep in the infrared--collinear region.
For the normalized distributions we are ultimately interested in, we do not need to compute $\sigma_2$, which would require genuine 2-loop contributions.
\begin{figure}[tb]
    \centering
    \includegraphics[width=0.45\linewidth]{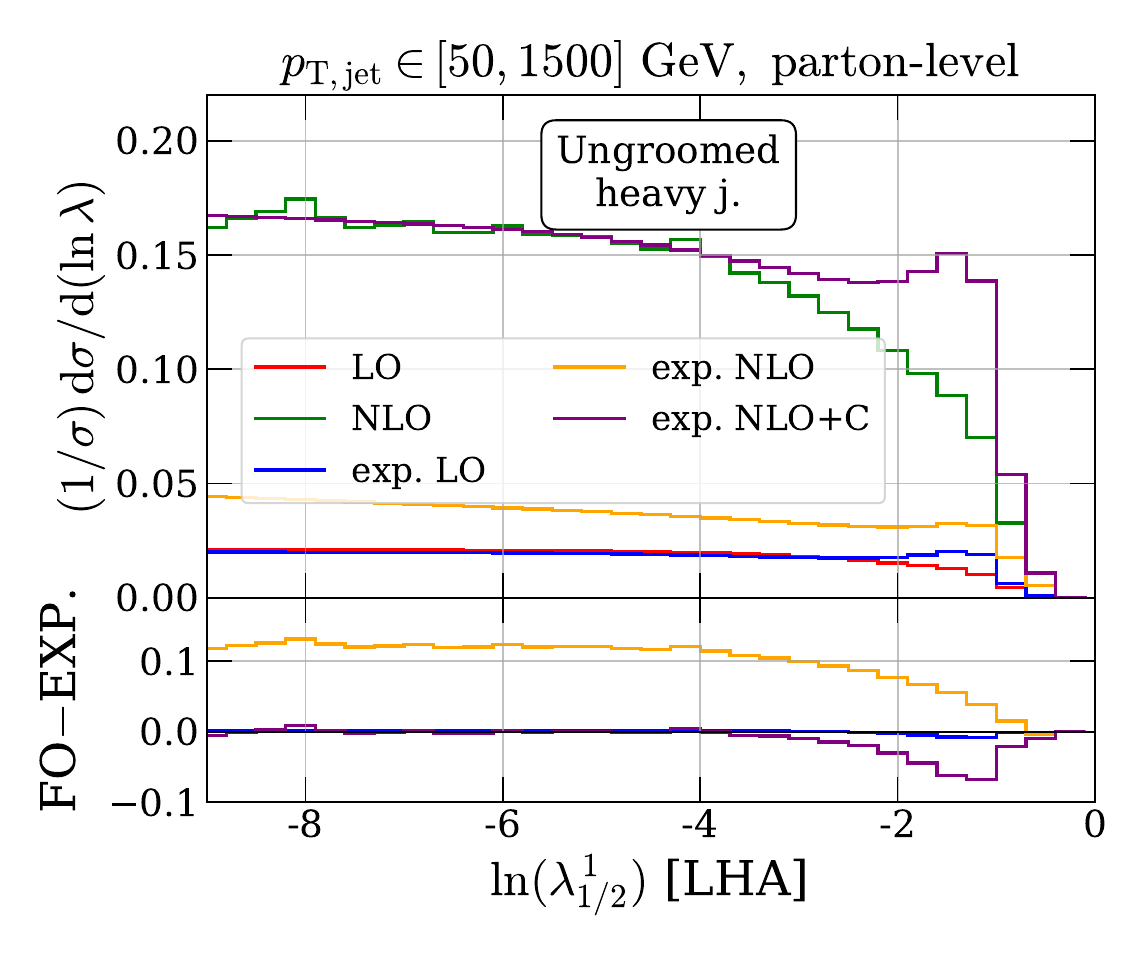}      
    \includegraphics[width=0.45\linewidth]{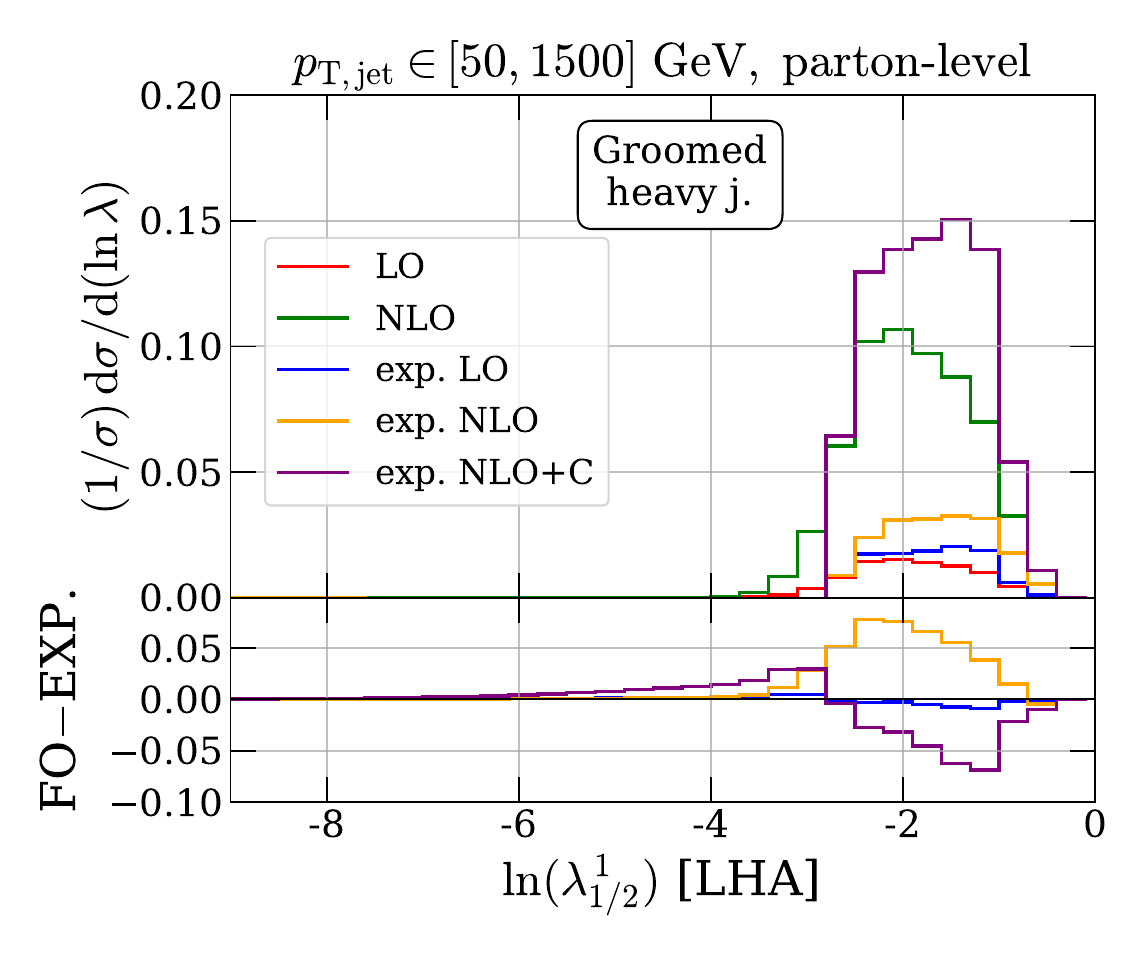}   \\ 
    \includegraphics[width=0.45\linewidth]{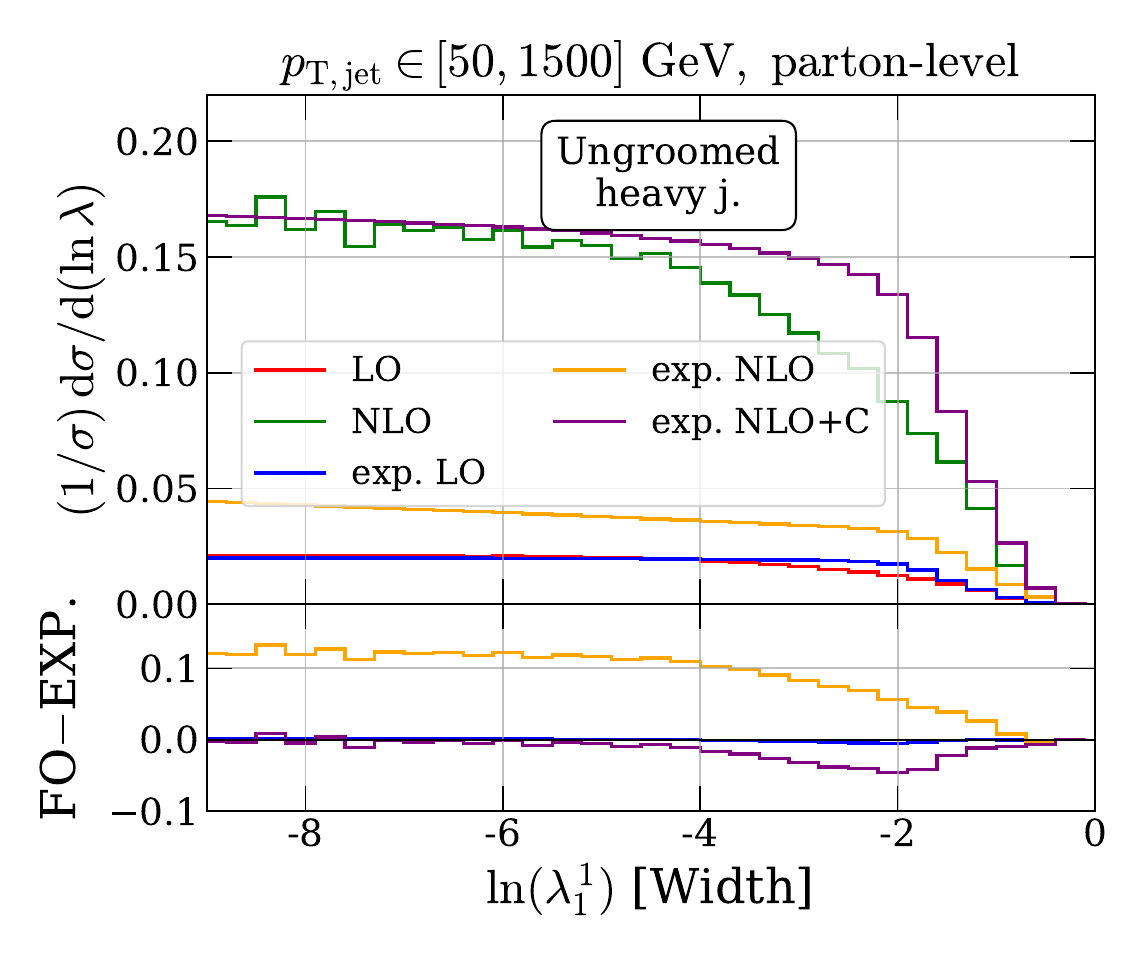}       
    \includegraphics[width=0.45\linewidth]{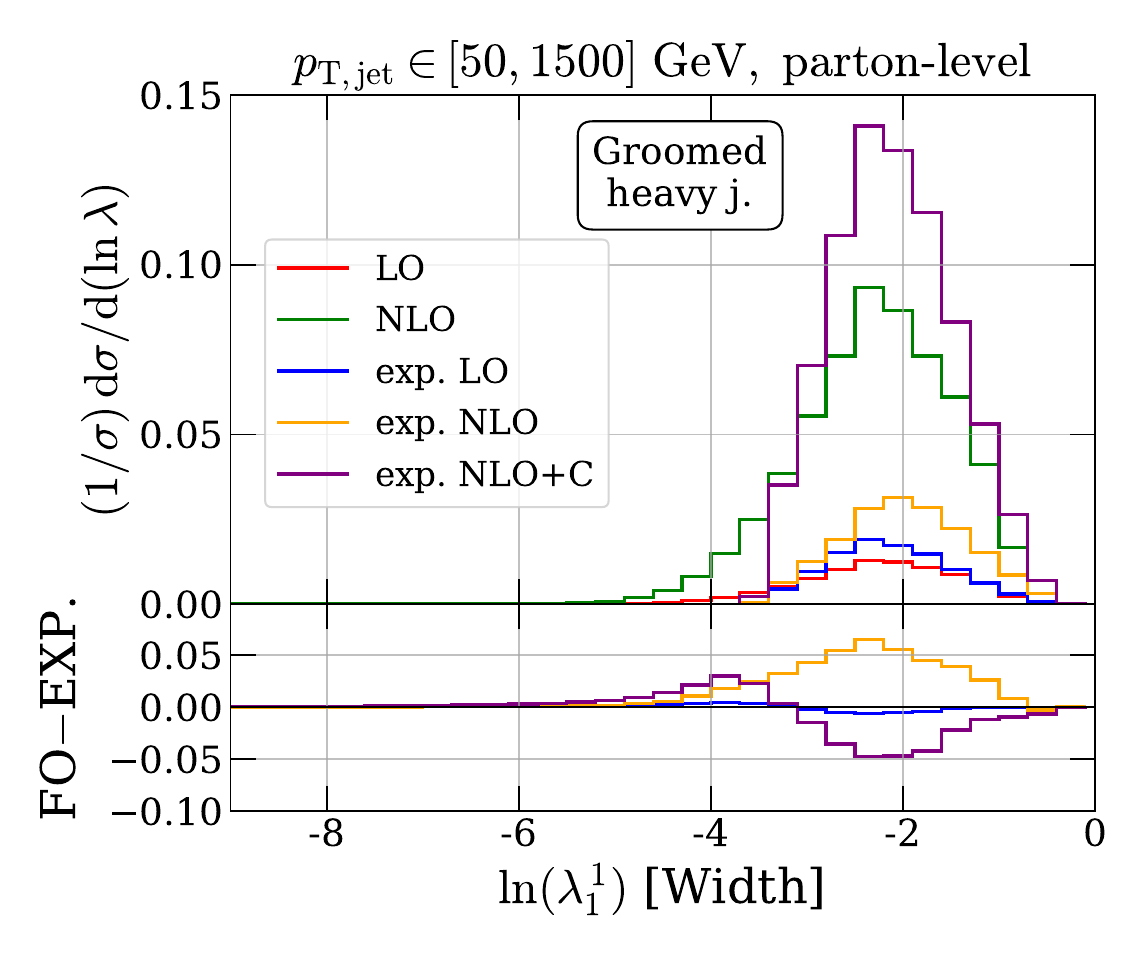}   \\
    \includegraphics[width=0.45\linewidth]{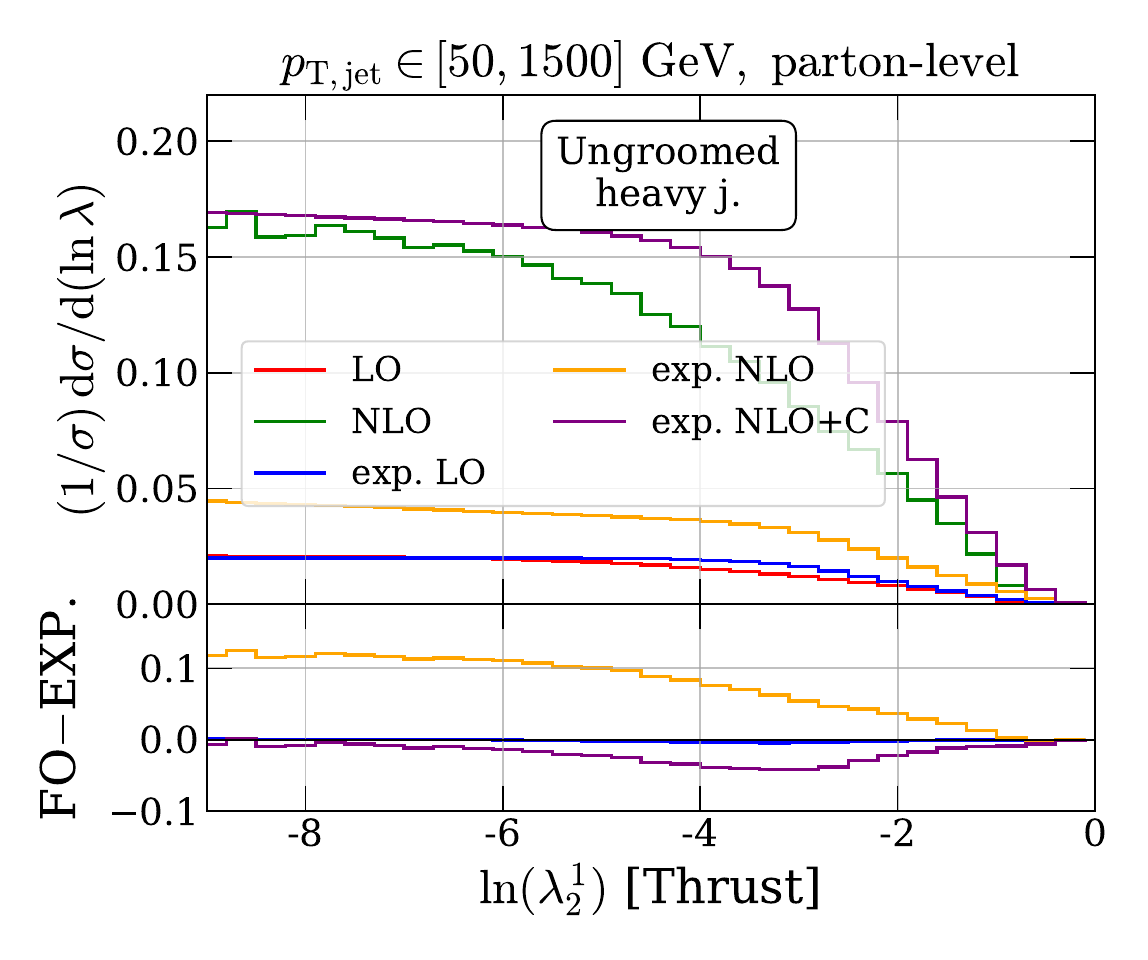} 
     \includegraphics[width=0.45\linewidth]{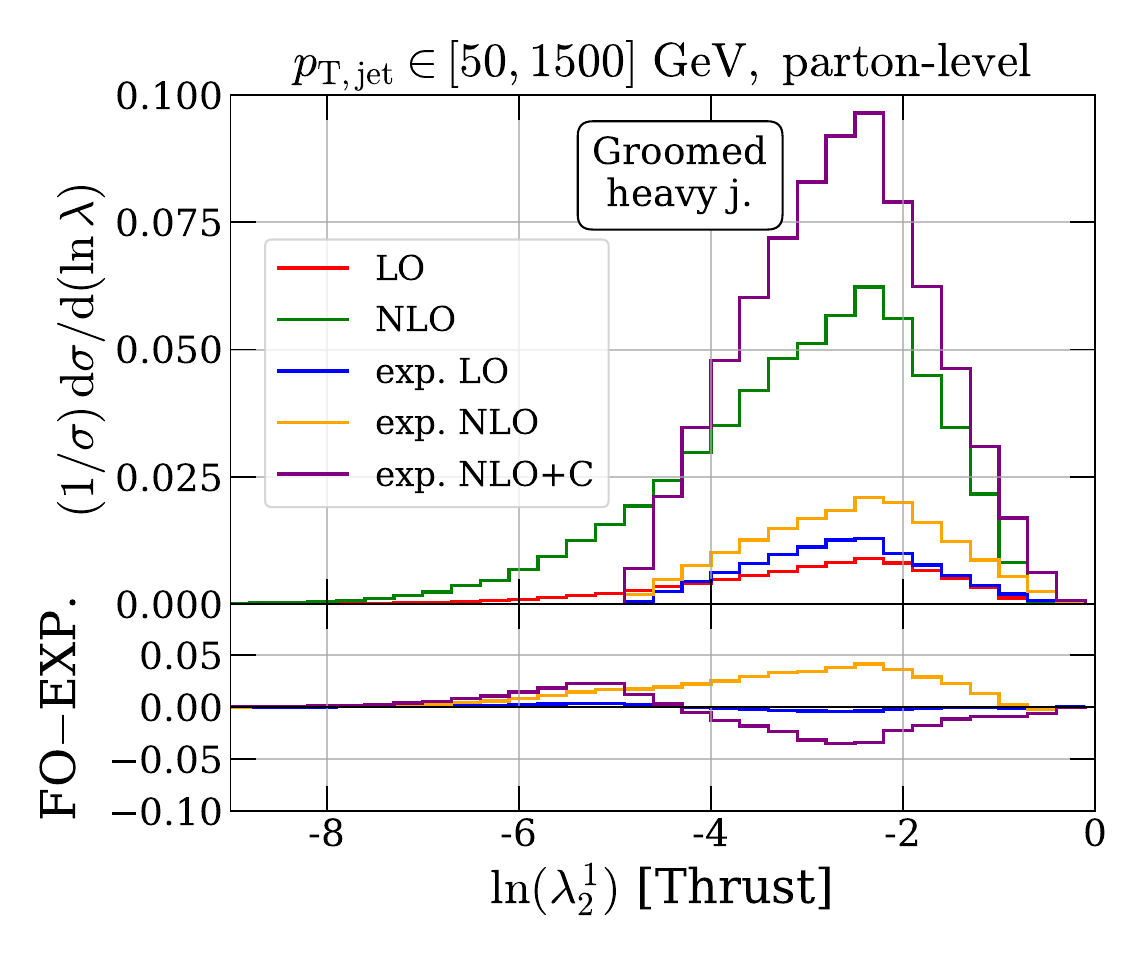}   
     \caption{ Fixed-order predictions for ungroomed (left column) and groomed (right column) angularities $\lambda^1_\alpha$, for $\alpha \in \{1/2,1,2\}$ (top to bottom),
       compared to the expansion of the resummation at the corresponding order of $\as$.
       The jet transverse momentum is constrained to $\ptjet \in [50, 1500]$ GeV.}
    \label{fig:lo_nlo_exp_test}
\end{figure}
In Fig. \ref{fig:lo_nlo_exp_test} we compare the cross sections introduced above, corresponding to fixed order or expanded resummation.
This serves as a consistency check, validating both the analytic expressions for the resummation and the setup for the fixed-order calculation.
We work with the fiducial phase space introduced in Sec.~\ref{sec:obs_def}, and present results for all three angularities without and with \softdrop grooming applied.
For each observable we include a panel showing the differences between the fixed-order result and the expansions of the resummation, up to order $\as$ and $\as^2$,  respectively.

We begin by considering the ungroomed distributions at leading order.
Compared to the massless case studied in \cite{Caletti:2021oor,Reichelt:2021svh}, the cumulative distribution is linear in the logarithm of the observable.
Its derivative thus approaches a constant for $L\to\infty$.
The fixed-order prediction follows this expectation from the resummation.
We observe that the difference between the fixed-order calculation (red curve) and the expansion of the resummed result (blue curve) approaches zero, as expected. We will examine the impact of power corrections in the jet radius and in the mass in App. \ref{app:dipoles-Zb}.
At NLO, the expanded resummed cumulative is quadratic in $L$ and hence the derivative plotted in Fig.~\ref{fig:lo_nlo_exp_test} is linear.
The fixed-order distribution also approaches a linear function with the slope as predicted by the resummation, however, with a significantly different offset.
This is most easily seen again by subtracting expansion and fixed order, and confirming that this approaches a constant difference.
We illustrate this with the yellow line in the respective lower panels of Fig.~\ref{fig:lo_nlo_exp_test}.
Up to power corrections in $\frac{m^2}{\ptjet^2R_0^2}$, which are discussed in App.~\ref{app:dipoles-Zb}, the offset is given by the cross term between the expansion of the soft function, $C$ and $G_{11}$. This yields a linear contribution to the $\mathcal{O}(\as^2)$ expansion of the cumulative and therefore produces an additional constant in the differential distribution.

We now turn our attention to the groomed case.
For the groomed angularities there are, in principle for $\beta =0$, deviations of $\ord{\zc}$.
However, they appear to be too small to be observed numerically here, as was also discussed in \cite{Baron:2020xoi} for event shapes. 
We observe that all distributions approach zero in the asymptotic limit.
This behavior arises because, in this region, the cumulative distributions become independent of the substructure observable, as can be observed from eq. \eqref{eq:exp_res_g}.
As in the ungroomed case, the inclusion of NLL$^\prime$ corrections has a significant impact on the resummed predictions. 
At NLO, the expansions of the resummation (shown by the yellow and purple curves) become negative over part of the spectrum.
This occurs because the derivative of the radiator turns negative in this region, an effect driven by the mass-dependent term of the splitting function in eq.~\eqref{eq:massive_splitting}.

We now want to match the resummed cross section with the exact result at NLO accuracy.
We first note that a change of the resummation scale $\mu_Q\to x_L\mu_Q$ can be accounted for 
by rescaling the argument of the logarithm to be resummed from $1/v$ to $x_L/v$, cf. eq.~\eqref{eq:AngParams}. We hence generalize our expressions by using
\begin{equation}
    L \to \ln \frac{x_L}{v}~, 
\end{equation}
and include the NLL term required by the scale independence of the overall cross sections up to that order. 
We vary this parameter together with the renormalization and factorization scales, as described above, to assess the perturbative uncertainties~\cite{Jones:2003yv}. 
As is common practice \cite{Catani:1992ua}, we modify the structure of the logarithms appearing in the resummed expressions to ensure they vanish in the fixed-order region.
We achieve this by further replacing
\begin{equation}
  \ln \frac{x_L}{v} \to \ln\left(\frac{x_L}{v}-\frac{x_L}{v_{\rm max}}+1\right)~,
\end{equation}
where we determine the endpoint $v_{\rm max}$ numerically from the NLO calculation.

We perform a multiplicative matching, for each partonic Born channel separately.
To achieve this, we use the procedure described in \cite{Caletti:2021oor}.
We iteratively apply the BSZ flavour algorithm \cite{Banfi:2006hf} to determine the flavour of a pre-determined final-state jet.
This way, we can assign each phase-space point, especially of the multi-parton final states entering $\Sigma^{(1,2)}_{\rm fo}$, to a unique flavour channel $\delta$.
We distinguish light-quark jets $\delta=q$ (including also anti-quarks), gluon jets $\delta=g$, and massive (anti-)quark jets $\delta=b$, as channels.
Any of the cumulative or integrated cross sections above can hence be split into contributions $\Sigma^\delta$ and $\sigma^{\delta}$ from these channels.

The final matched expression for the cumulative distribution reads:
\begin{align}\label{eq:matched}
    \Sigma^\delta_{\text{matched}} &= \Sigma^\delta_{\text{res}}\, \left(1 + \frac{\Sigma^{\delta(1)}_{\text{fo}} - \Sigma^{\delta(1)}_{\text{res}} }{\sigma_0^{\delta}}  
    -\frac{\bar \Sigma^{\delta(2)}_{\text{fo}} +\Sigma^{\delta(2)}_{\text{res}} }{\sigma_0^{\delta}} 
    -
    \frac{\Sigma^{\delta(1)}_{\text{res}} }{\sigma_0^{\delta}}
    \frac{\Sigma^{\delta(1)}_{\text{fo}} - \Sigma^{\delta(1)}_{\text{res}} }{\sigma_0^{\delta}}  \right)\Bigg|_{n_f=5},
\end{align}
where the resummed contribution $ \Sigma^\delta_{\text{res}}$ is computed with flavour thresholds, as detailed above, while the fixed order and the expansion, i.e.\ the terms in the brackets, are computed with fixed flavour number $n_f=5$.
We then sum over all channels
\begin{equation}
    \Sigma_{\rm matched} = \sum_{\delta = q,g,b} \Sigma^\delta_{\rm matched}~,
\end{equation}
to arrive at our final predictions.
Given the agreement between fixed order and expanded resummation at each order shown above, eq.~\eqref{eq:matched} will reduce to the resummed distribution in channels $\delta$ in the small observable limit.
In the other extreme, the expansion up to order $\as$ or $\as^2$ reproduces the leading- or next-to-leading-order distributions, up to terms that are induced by the difference between $\Sigma$ calculated with fixed $n_f=5$ flavours, and calculated with flavour thresholds. 
At large scales, above the $b$-quark mass, we hence reproduce exactly the fixed-order calculation up to its order in $\as$. 
At lower scales, we instead use the resummed expression including the threshold in the running of $\as$.
This changes formally the structure of the fixed order $\as$ and $\as^2$ contributions.
However, we argue that this only happens in a range where the fixed-order description is not adequate, and the resummed expression is more appropriate anyway.

\section{Particle-level predictions} \label{sec:HLpredictions}

To complement our \NLOpNLLp accurate perturbative predictions with non-perturbative corrections, 
we perform full particle-level simulations with the \sherpa\ event-generator framework. While these
can, in principle, be directly compared with experimental data, here we furthermore use them to 
extract parton-to-hadron-level transfer matrices, differential in the angularity observables and 
the  jet kinematics. Upon folding these with our resummed results, we can account for effects of
hadronization and the underlying event, which facilitates a more direct comparison of our resummed predictions with experimental measurements. 

\subsection{Particle-level Monte Carlo simulations}

As our resummation calculations, all particle-level simulations are performed with version 3.0.3 of the \sherpa\ event generator~\cite{Sherpa:2024mfk}. The event selections, calculational tools, and input-parameter settings are kept identical to the resummed calculations, cf. Secs.~\ref{subsec: process and phase space cuts} and \ref{sec:caesar_review}. We generate two dedicated event samples: 

\begin{itemize}
\item[(i)] \textbf{heavy-flavour production:} we consider the partonic process $gg\to Zb\bar{b}$ at NLO QCD, treating the final-state $b$-quarks as massive, combined with the parton shower 
using the MC@NLO approach~\cite{Hoeche:2011fd}\footnote{We use a four-flavour scheme as at present \sherpa\ does not support shower simulations off massive initial-state partons at MC@NLO accuracy. Note that the hard-process calculation is fully regularized by the finite $b$-quark mass, such that the second $b$-jet can fully be integrated over.},
\item[(ii)] \textbf{inclusive production:} we merge the NLO QCD matrix elements for $jj\to Zj$ and $jj\to Zjj$, assuming four massless quark flavours, based on the truncated-shower formalism~\cite{Hoeche:2009rj,Hoeche:2012yf}. The merging scale is thereby set to $Q_{\text{cut}}=20\,\text{GeV}$.
\end{itemize}
All hard-process events, i.e. prior to parton showering, are required to feature 
at least one $R_0=0.4$ anti-$k_t$ jet with $\ptjet\geq 35\;\text{GeV}$, and $|y_\text{jet}|<3$. As factorization and 
renormalization scale for the reconstructed $2\to 2$ core processes we use 
\begin{equation}
    \mu_\text{F}=\mu_\text{R} = p_{T,Z}\,.
\end{equation}
For the potential second hard-process jet the strong coupling is evaluated at the 
reconstructed shower-emission scale. For parton showering we use \sherpa's dipole shower~\cite{Schumann:2007mg}, its starting scale is given by
\begin{equation}
    \mu_\text{Q}=\ptjet R_0\,.
\end{equation}
To model the hadronization of partons into primary hadrons we use \sherpa's cluster fragmentation model~\cite{Chahal:2022rid}. The decays of unstable 
particles are also accomplished by \sherpa. Note, that for the \textit{heavy-flavour 
production} we treat all $B$-hadrons as stable particles, which experimentally
corresponds to fully reconstructing them from their decay products. All final-state 
$B$-hadrons are required to satisfy the pseudorapidity condition $|\eta_B| < 5$. 
We define $b$-jets as jets that have exactly one $B$-hadron as a constituent~\footnote{
Because jets featuring three (or more) $B$ hadrons are rare, this labeling is essentially equivalent to the so-called \emph{mod2 flavour} scheme advocated in~\cite{Behring:2025ilo}.}.
In contrast, for the  \textit{inclusive production} sample, we let $B$-hadrons decay, 
allowing us to compare to the CMS results for not flavour-tagged jets~\cite{CMS:2021iwu}. 
To estimate theoretical uncertainties, we consider $7$-point scale variations of 
$\mu_\text{R}$ and $\mu_\text{F}$ by factors $1/2$ and $2$ in both the hard process 
and the parton shower, evaluated on the fly~\cite{Bothmann:2016nao}.

For the analysis of particle-level events and for the extraction of the 
parton-to-hadron transition matrices, we use the \textsc{Rivet} package~\cite{Bierlich:2024vqo}.

\subsection{Non-perturbative corrections via transfer matrices}\label{sec:transfer_matrices}

To incorporate non-perturbative corrections into our \NLOpNLLp results we employ the 
transfer-matrix approach first presented in \cite{Reichelt:2021svh} and later
used in~\cite{Knobbe:2023ehi,H1:2024aze,H1:2024pvu,Chien:2024uax}. We thereby attempt 
to map the event kinematics, subject to event-selection cuts, and the value of the 
observable of interest from parton to hadron level. This is achieved by analyzing 
individual simulation events twice, first right after the parton-shower evolution of the 
hard-process configuration, and a second time after the underlying-event simulation 
and the hadronization process are accomplished. To this end, we use the truth-level 
information provided by the \textsc{HepMC} event record~\cite{Buckley:2019xhk}. 
This allows us to build a multi-differential map of a parton-level configuration 
$\mathcal{P}$ onto a hadron-level configuration $\mathcal{H(P)}$. This map can 
then be applied to a differential parton-level prediction for a given observable, 
such as our \NLOpNLLp calculation of the jet angularities $\lambda^1_\alpha$, to account 
for non-perturbative corrections.

We assume a generic set of variables $\vec{V}=(V_1,\dots,V_m)$ that can be measured on 
$\mathcal{P}$ and $\mathcal{H(P)}$\footnote{For concreteness, consider the jet-kinematics
variables and the jet-angularity observable, 
i.e.\ $\vec{V}=\left(p_{T,\rm jet},\,y_{\rm jet},\,\lambda^1_\alpha\right)$.}. 
The conditional probability of observing at the hadron level the observable values $\vec{v}^h$ given $\vec{v}^p$ at the parton level can then be given by 
\begin{align}\label{eq:transferOperator}
\mathcal{T}\left(\vec{v}^h|\vec{v}^p\right) = \frac{\int \de\mathcal{P}\frac{\de\sigma^p}{\de\mathcal{P}}\delta^{(m)}\left(\vec{v}^p-\vec{V}(\mathcal{P})\right)\delta^{(m)}\left(\vec{v}^h-\vec{V}(\mathcal{H(P)})\right)}{\int \de\mathcal{P}\frac{\de\sigma^p}{\de\mathcal{P}}\delta^{(m)}\left(\vec{v}^p-\vec{V}(\mathcal{P})\right)}\,.
\end{align}
%
The hadron-level cross section fully differential in $\vec{v}^h$ then reads
\begin{align}
    \frac{\de^m\sigma^h}{\de v^h_1\dots \de v^h_m} = \int \de^m\vec{v}^p\;\mathcal{T}\left( \vec{v}^h|\vec{v}^p\right)\frac{\de ^m\sigma^p}{\de v^p_1\dots \de v^p_m}\,.
\end{align}
In practice, we consider binned distributions of the kinematic variables and observables, resulting in the transfer operator of eq.~\eqref{eq:transferOperator} being given by a $m\times m$ matrix of 1D histograms. Note that in order to 
properly describe the hadron-level variables, the parton-level input needs to be sufficiently inclusive. For example, hadronization corrections tend to reduce the transverse momentum of jets, the underlying event, however, will typically increase it~\cite{Salam:2010nqg}. To account for this cross-section migration, we introduce underflow and overflow bins in the jet-kinematic variables; see also Ref.~\cite{Reichelt:2021svh}. 

\begin{figure}[h!]
    \centering
    \includegraphics[width=0.48\linewidth]{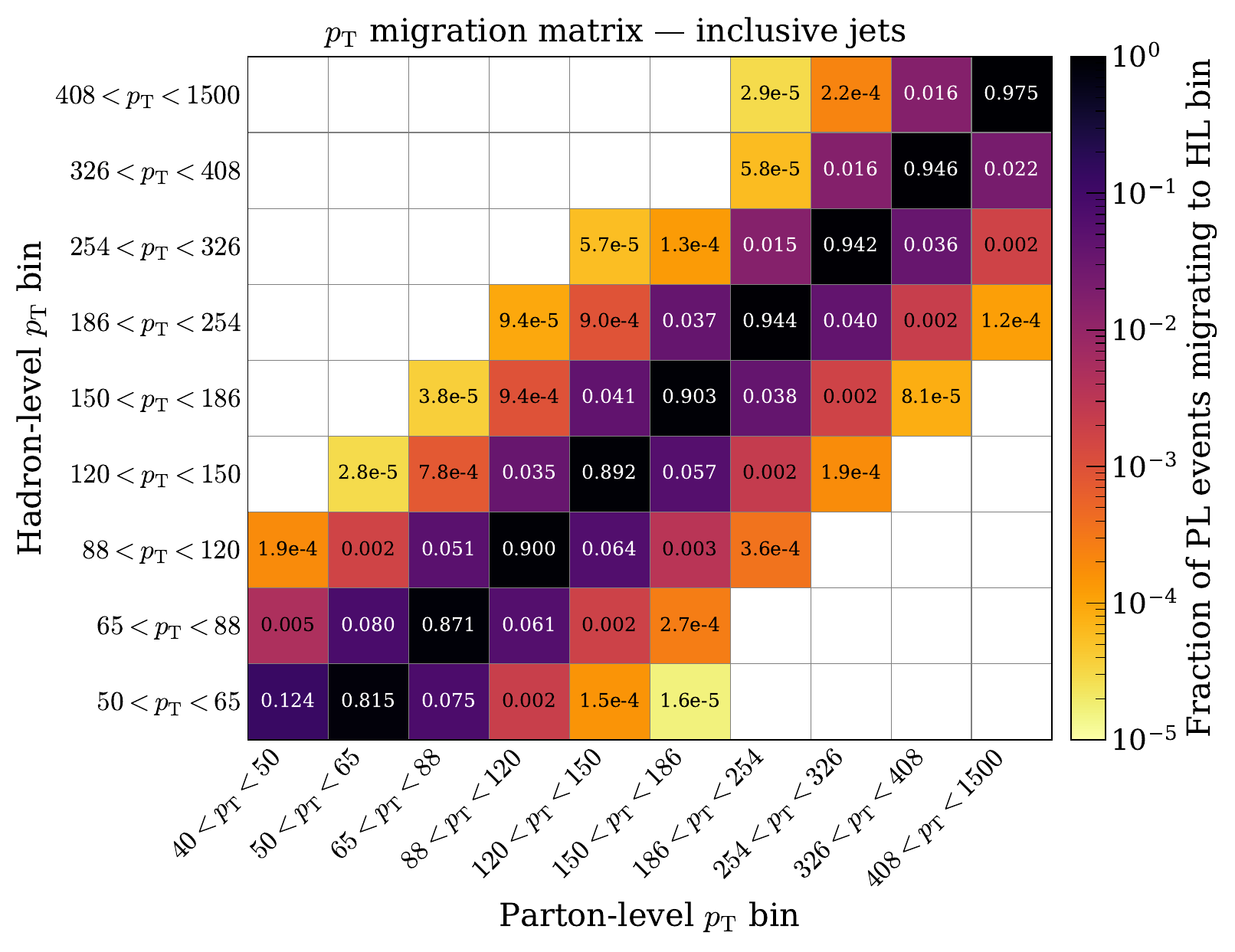}\hfill
    \includegraphics[width=0.48\linewidth]{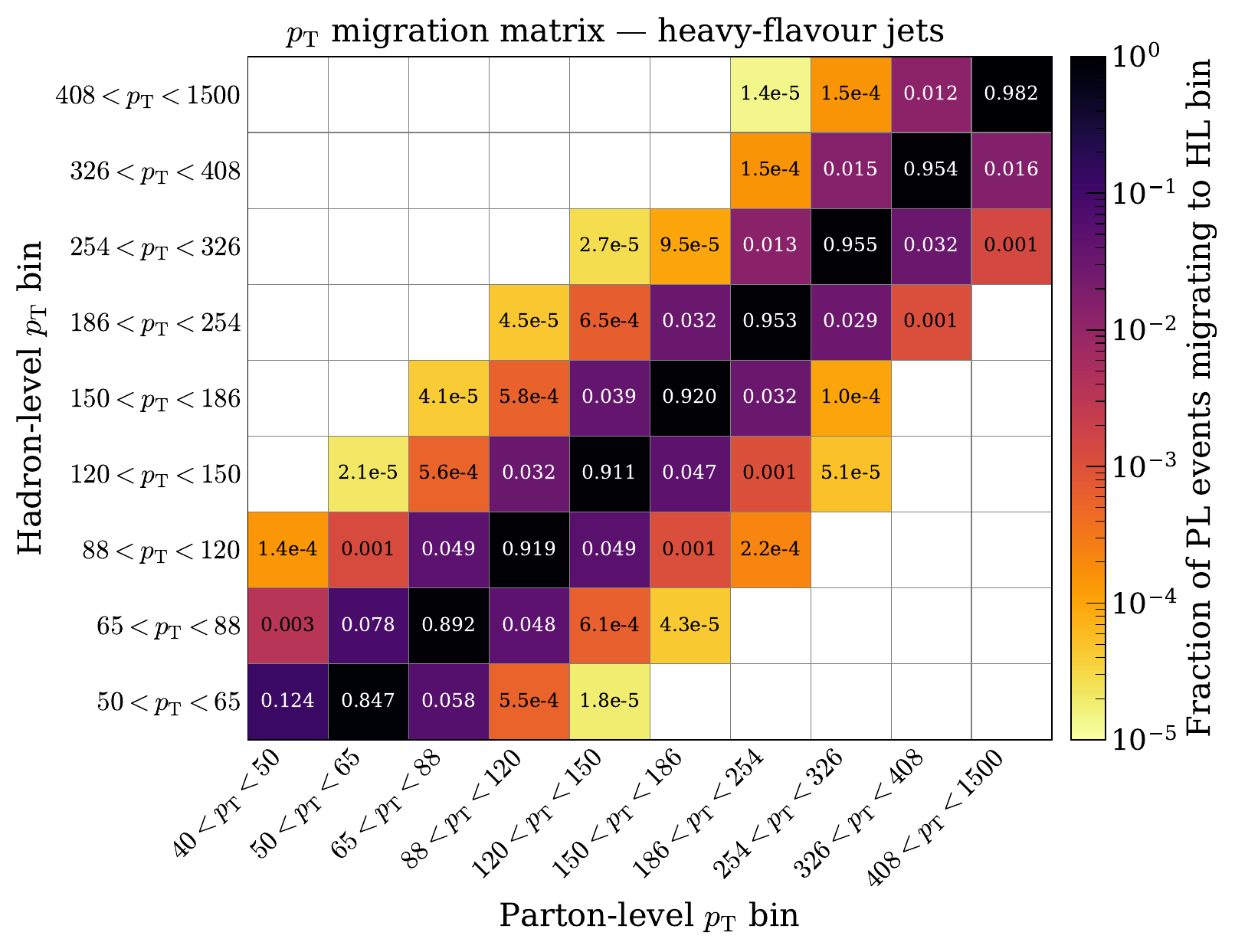}\\
    \caption{The effect of jet-transverse-momentum migration from parton to hadron level for the case of
    inclusive jets (left) and heavy-flavour jets (right).}
    \label{fig:tm_pT}
\end{figure}

To illustrate the change in jet-$p_T$ upon inclusion of non-perturbative corrections, we present in Fig.~\ref{fig:tm_pT} the corresponding transfer
matrices. For the considered bins in jet transverse momentum, these show the distribution of parton-level events of given $\ptjet^{\text{PL}}$ onto the observed hadron-level jets with $\ptjet^{\text{HL}}$. We here present results for the case of inclusive jets (left) and heavy-flavour jets (right). Both migration matrices are rather diagonal, but a non-negligible population of the neighbouring jet-$p_T$ slices is observed. The effect is slightly more pronounced for inclusive jets and in particular for lower jet transverse momenta. We do not expect the underlying-event activity to differ significantly for the two scenarios; however, in the massive case, we treat $B$-hadrons as stable, resulting in fewer fragmentation products, which reduces migration effects. We furthermore note that indeed jets from the parton-level underflow bin, i.e. with $\ptjet^{\text{PL}}<50\,\text{GeV}$, contribute to hadron-level jets with significantly higher transverse momentum. Although the largest fraction of these events exhibits $\ptjet^{\text{HL}}<50\,\text{GeV}$, and are correspondingly discarded by our event selection and therefore not considered in the transfer matrices, about $13\%$ of the parton-level cross section is
spilled into the higher $\ptjet$ bins.

\begin{figure}[h!]
    \centering
    \includegraphics[width=0.48\linewidth]{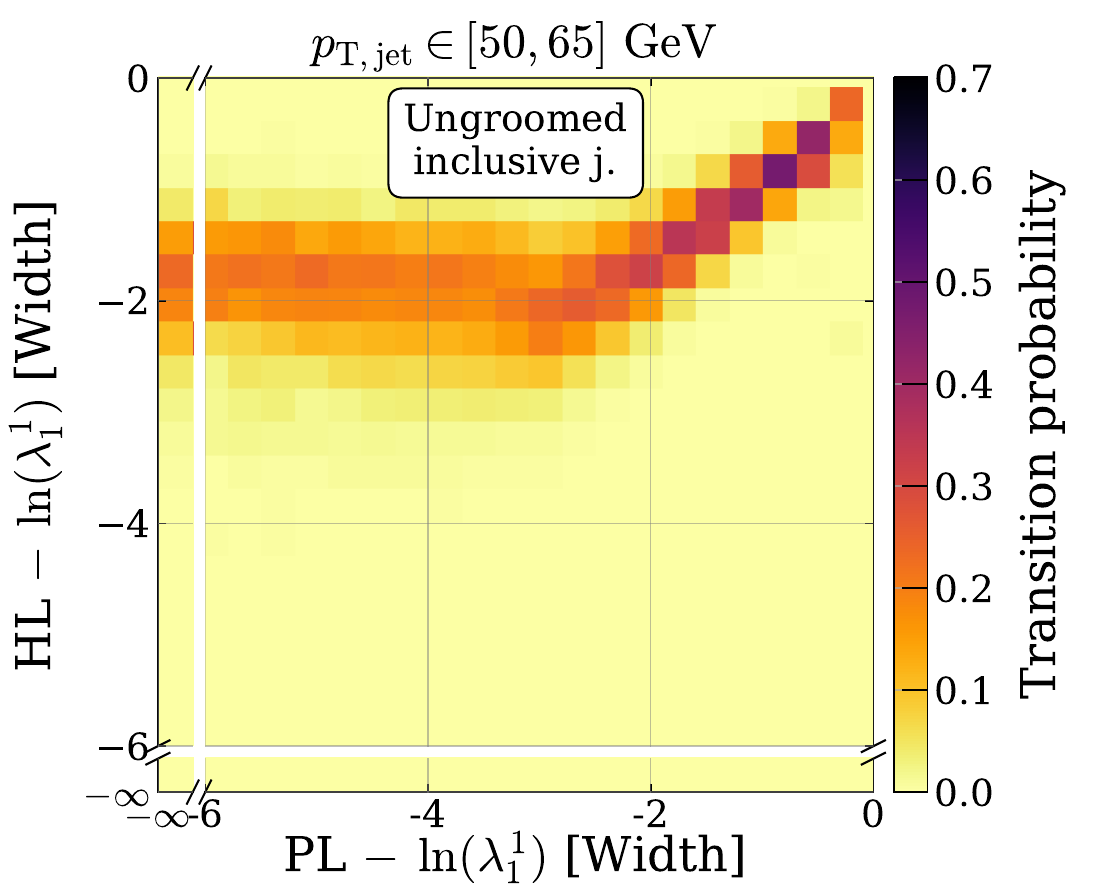}\hfill
    \includegraphics[width=0.48\linewidth]{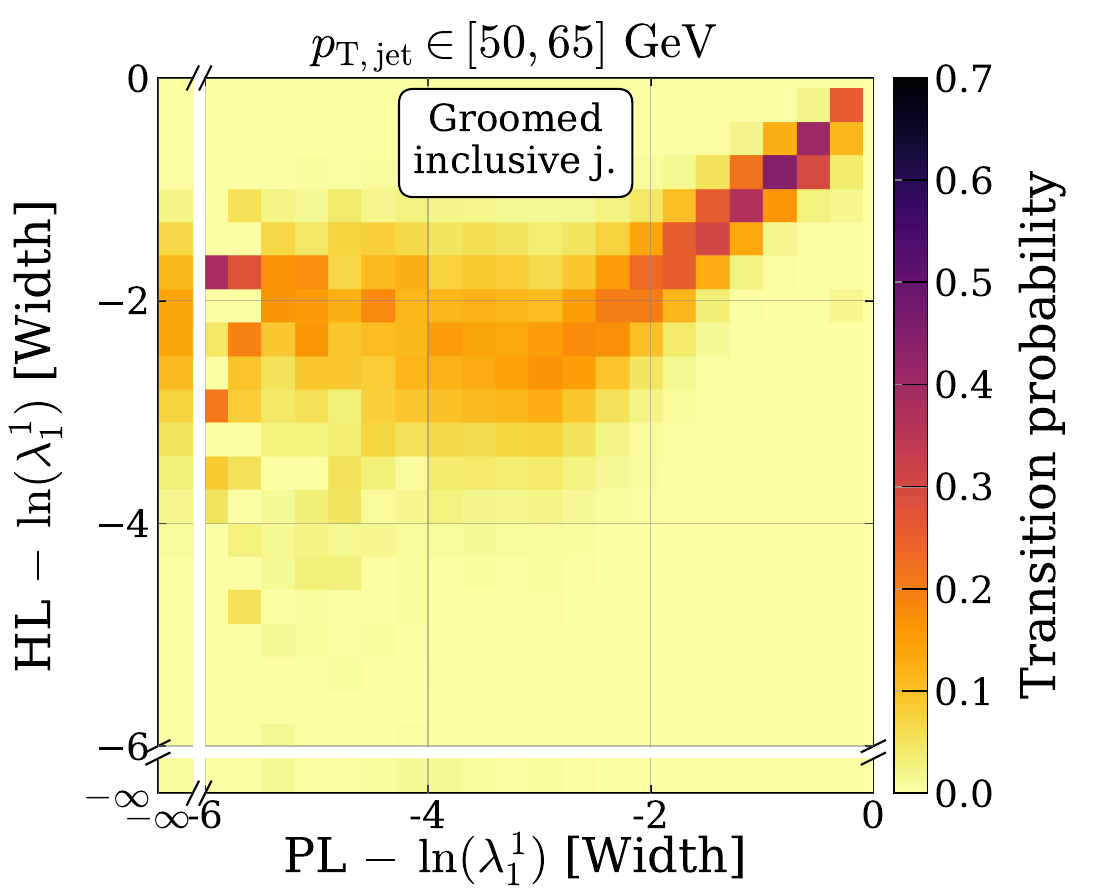}\\ 
    \includegraphics[width=0.48\linewidth]{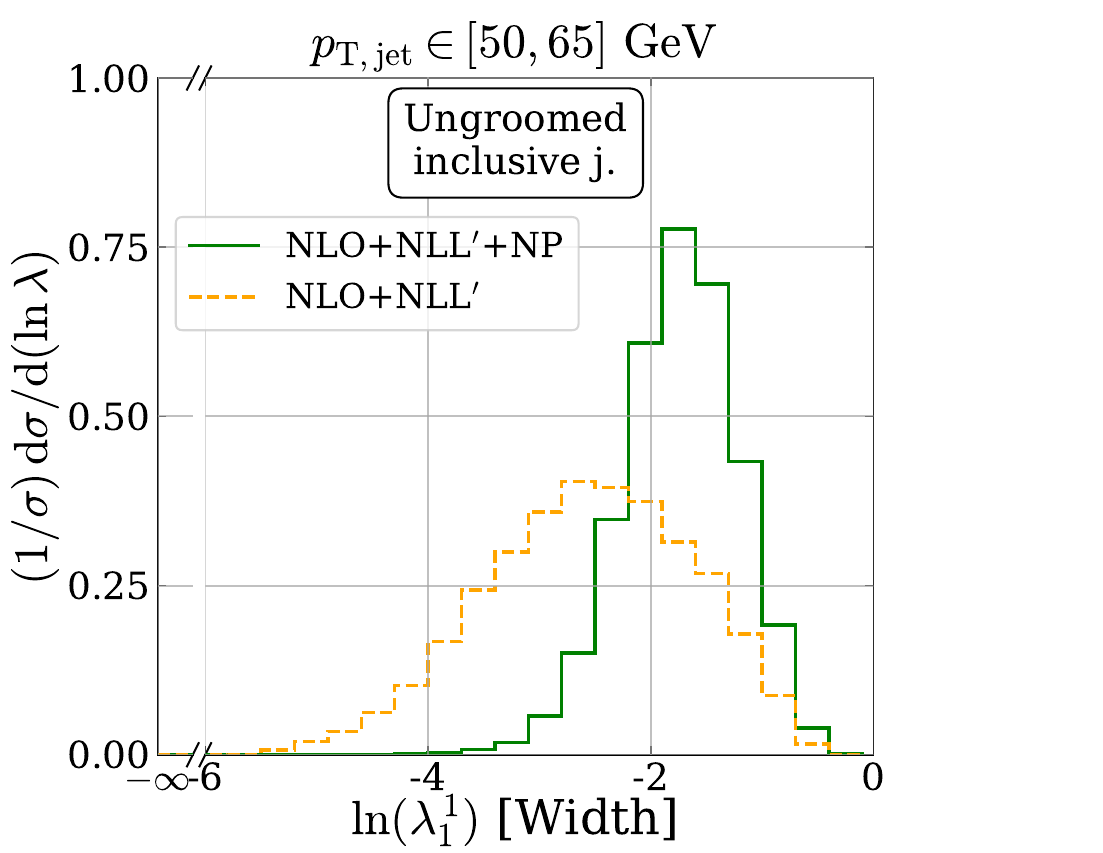} \hfill
    \includegraphics[width=0.48\linewidth]{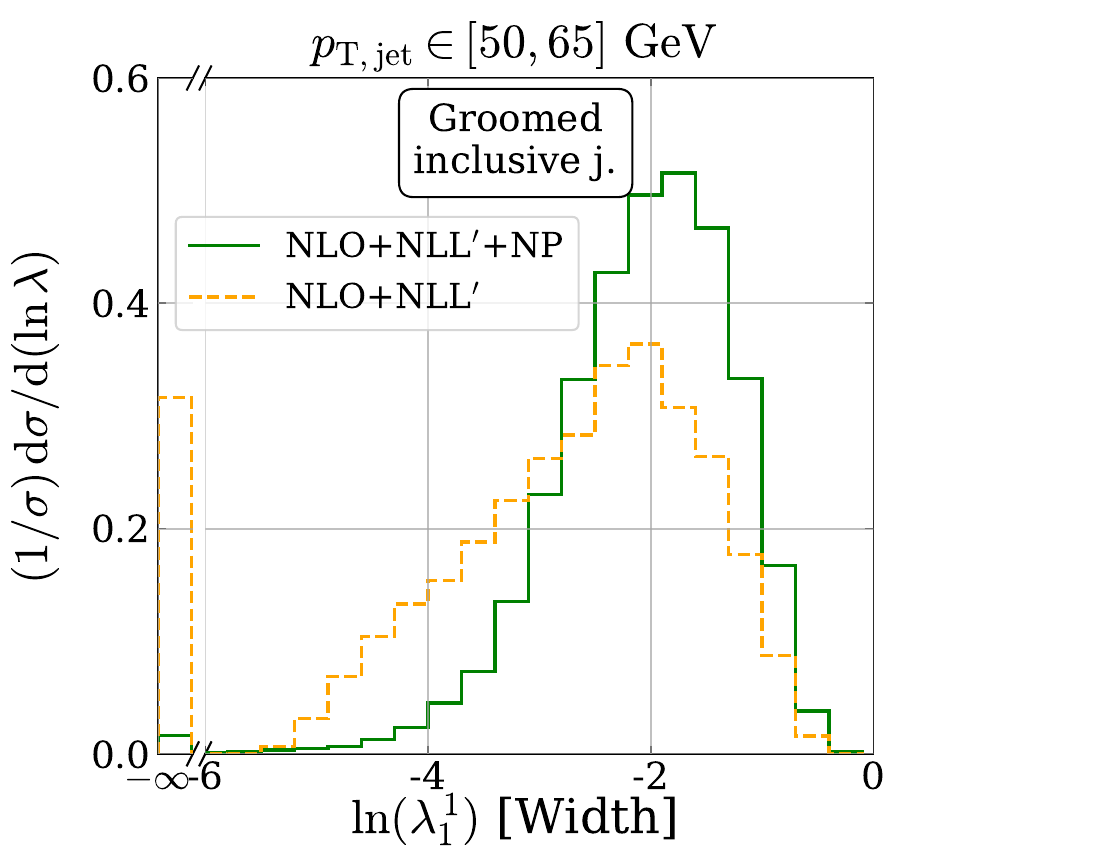}\\ 
    \caption{Parton-to-hadron-level transfer matrices for the Width angularity ($\lambda^1_1$) for untagged ungroomed (left) and groomed (right) jets with $\ptjet\in [50,65]\,\text{GeV}$ at parton and hadron level. The lower panels show the corresponding \NLOpNLLp predictions at parton level and at hadron level, the latter accounting also for migration from lower and higher $\ptjet^{\text{PL}}$ bins.}
    \label{fig:tm_inclusiveB}
\end{figure}

\begin{figure}[h!]
    \centering
    \includegraphics[width=0.48\linewidth]{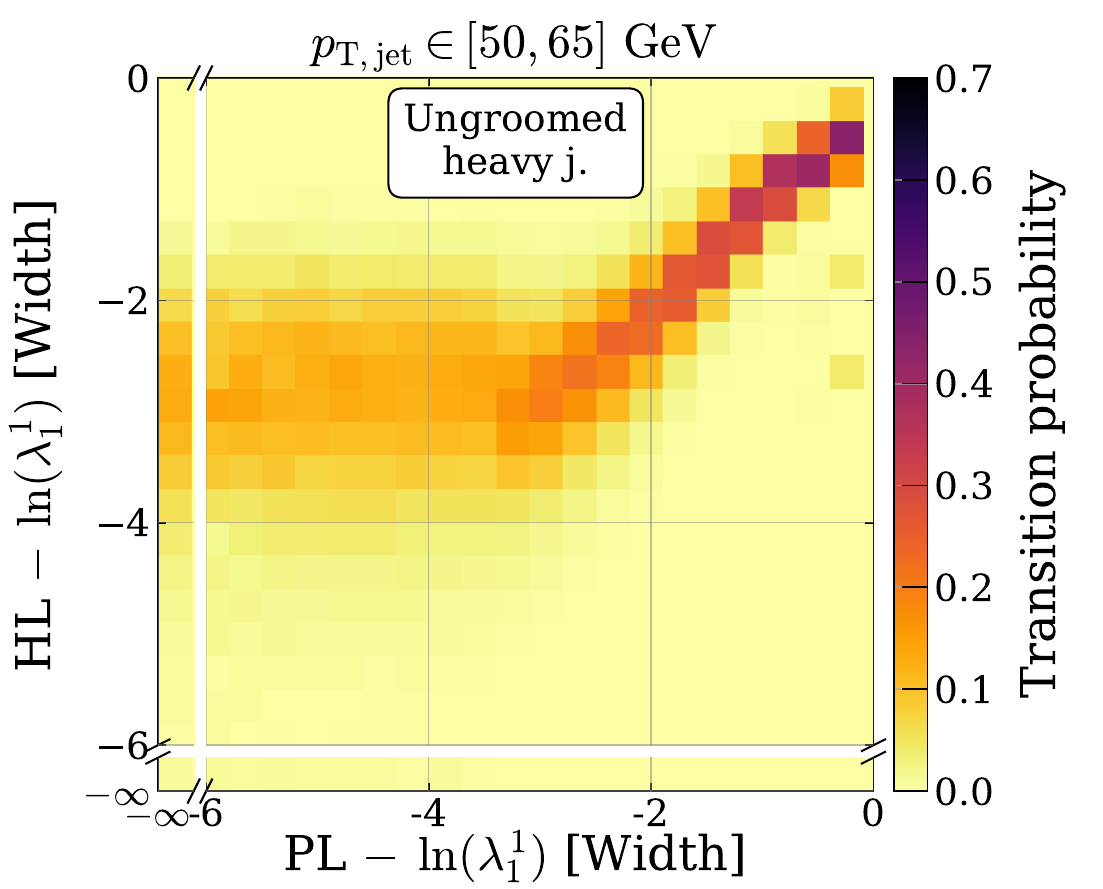}
    \includegraphics[width=0.48\linewidth]{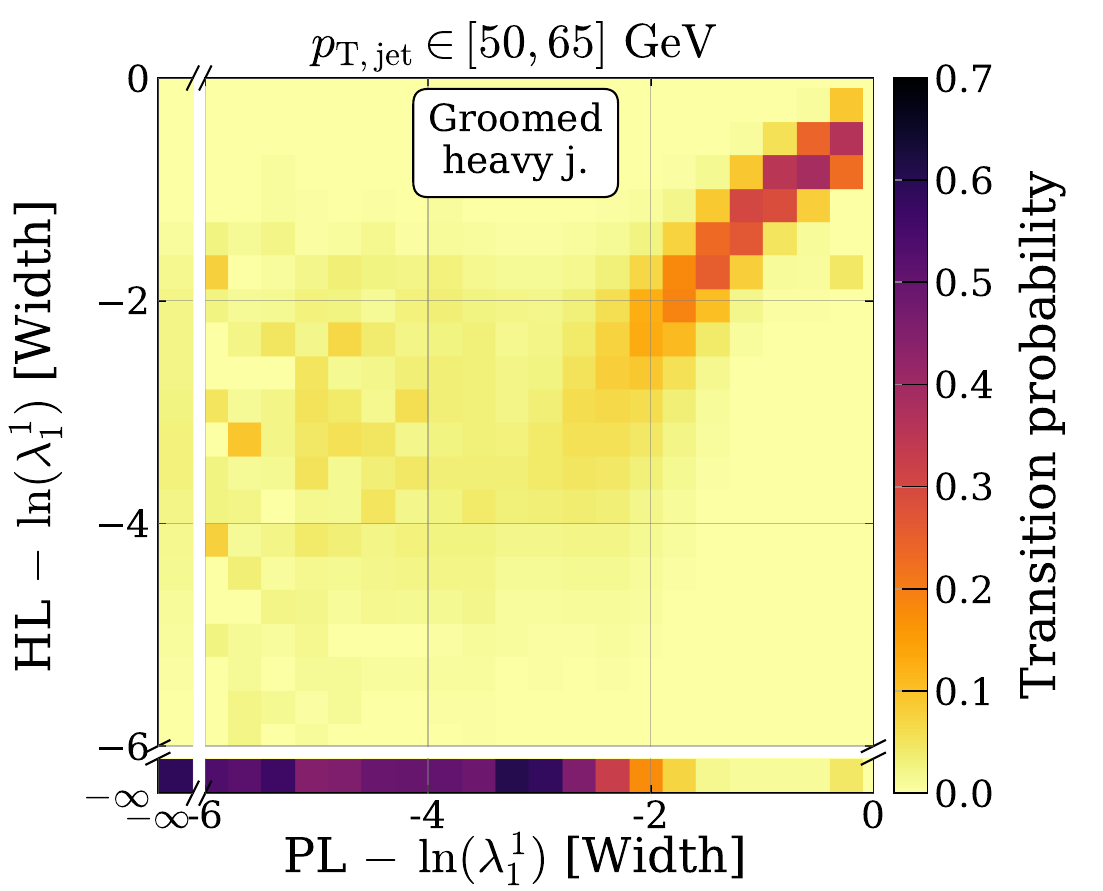}\\ 
    \includegraphics[width=0.48\linewidth]{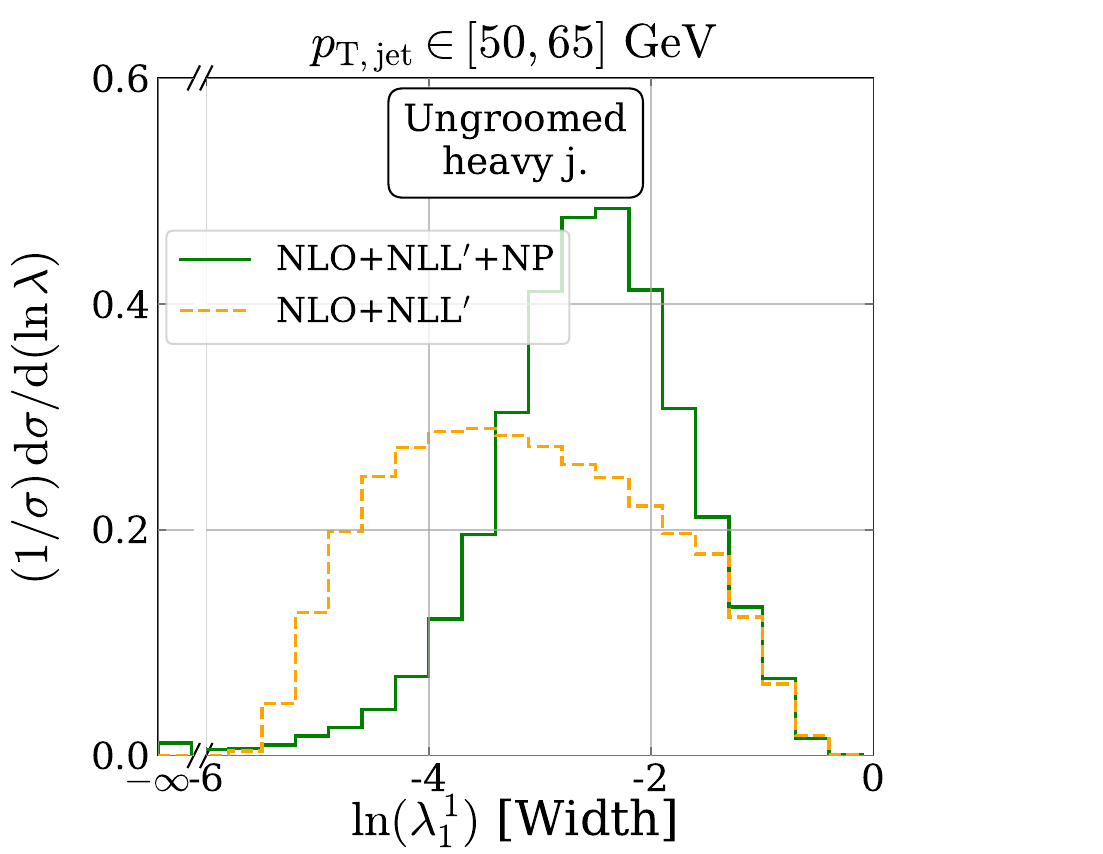} 
    \includegraphics[width=0.48\linewidth]{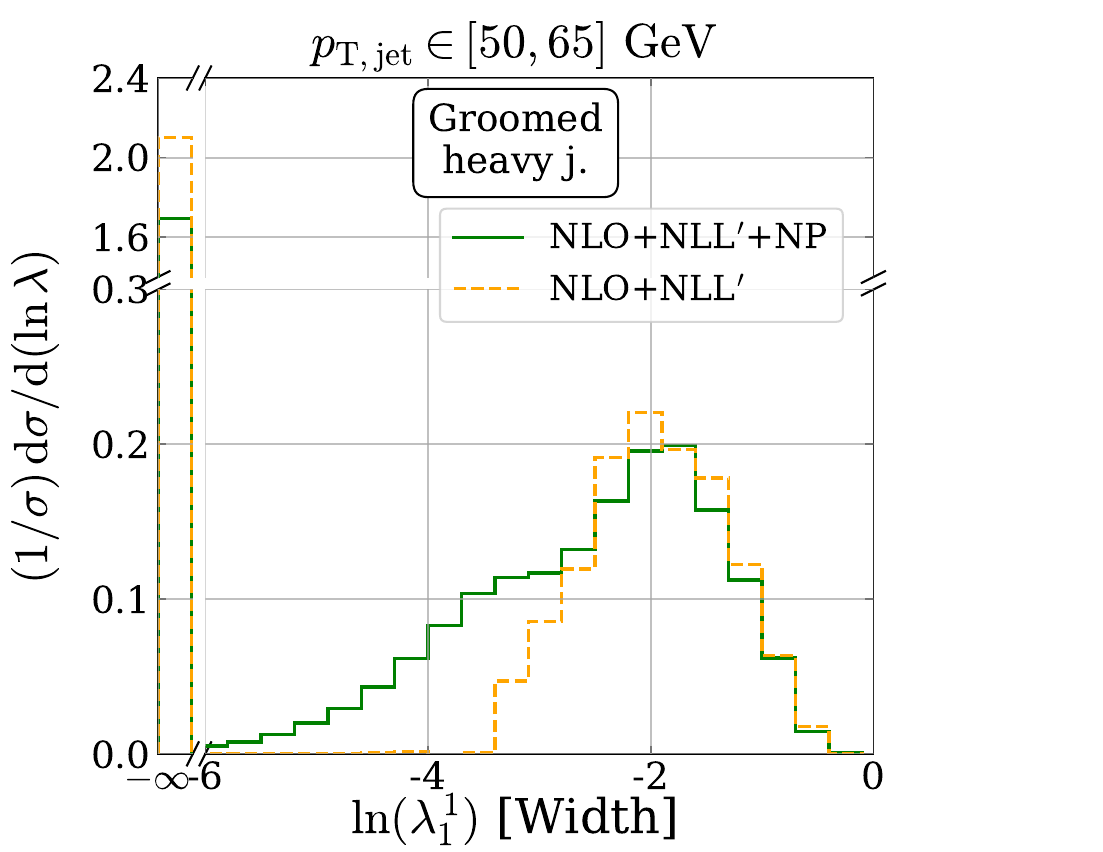}\\ 
    \caption{Parton-to-hadron-level transfer matrices for the Width angularity ($\lambda^1_1$) for heavy-flavour ungroomed (left) and groomed (right) jets with $\ptjet\in [50,65]\,\text{GeV}$ at parton and hadron level. The lower panels show the corresponding \NLOpNLLp predictions at parton level and at hadron level, the latter accounting also for migration from lower and higher $\ptjet^{\text{PL}}$ bins.}
    \label{fig:tm_massiveB}
\end{figure}

\softdrop\ grooming does not affect the  transverse momentum of the candidate jets used to evaluate the angularity observables. However, obviously the actual observable values get affected.  Accordingly, the parton-to-hadron-level migration in the observables differs for ungroomed and groomed jets. In Figs.~\ref{fig:tm_inclusiveB} and \ref{fig:tm_massiveB} we show examples of transfer matrices for ungroomed and groomed jets for the $\lambda^1_1$ observable for inclusive and heavy-flavour jets, respectively. While the upper panels contain the observable-migration matrices for the lowest jet-$p_T$ bin, i.e.\ for jets with $\ptjet\in[50,65]\,\text{GeV}$ both at parton and hadron level, the lower panels show the corresponding observable distributions prior to including non-perturbative effects, labelled \NLOpNLLp (yellow dashed), and upon their inclusion, labelled \NLOpNLLp+NP (green solid). Note that the hadron-level predictions thereby include the effect of migration from lower and higher $\ptjet^\text{PL}$ bins.

We observe that both for inclusive and heavy-flavour jets the transfer matrices are largely diagonal, down to observable values of about $\ln(\lambda^1_1)\approx -2 \dots -3$. Smaller observable values at the parton level get pushed towards these values at the hadron level. There is a notable difference for \softdrop\ groomed jets. Both at parton and hadron level events appear, which have a vanishing angularity value, that populate the underflow bin in the histogram. For the case of inclusive jets this effect is largely reduced when going from parton to hadron level. However, as mentioned already in Sec.~\ref{sec: NLL SD}, for heavy-flavour jets this effect is more pronounced and is prominent also at the hadron level. In fact, most parton-level events with $\ln(\lambda^1_1)<-2.5$ result in $\lambda^1_1=0$ at hadron level. 

When looking at the corresponding observable distributions, the effect of the non-per\-turb\-ative corrections is largely a shift towards higher observable values and a narrowing of the distribution. For groomed heavy-flavour jets, however, the tail towards small $\lambda^1_1$ values is more pronounced at hadron level and there remains a significant fraction of events with $\lambda^1_1=0$. 

\begin{figure}[h!]
    \centering
      \includegraphics[width=0.48\linewidth]{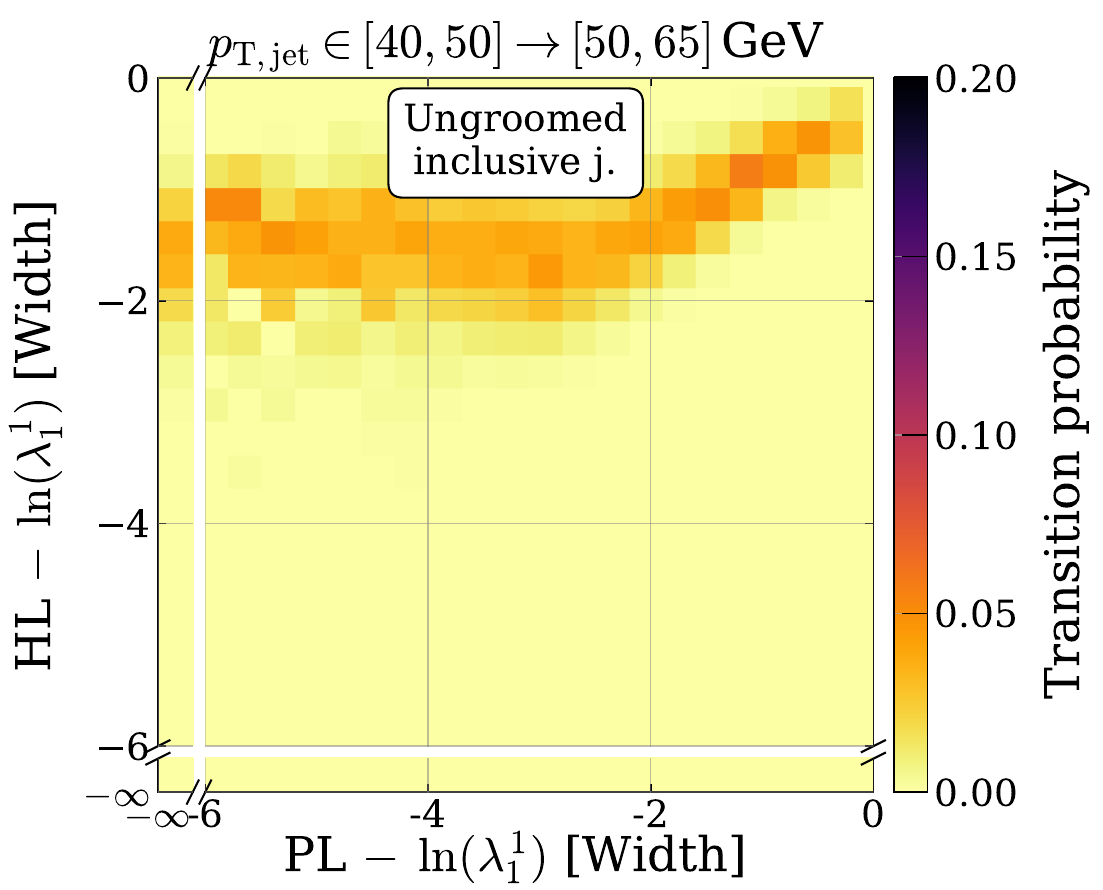}
    \includegraphics[width=0.48\linewidth]{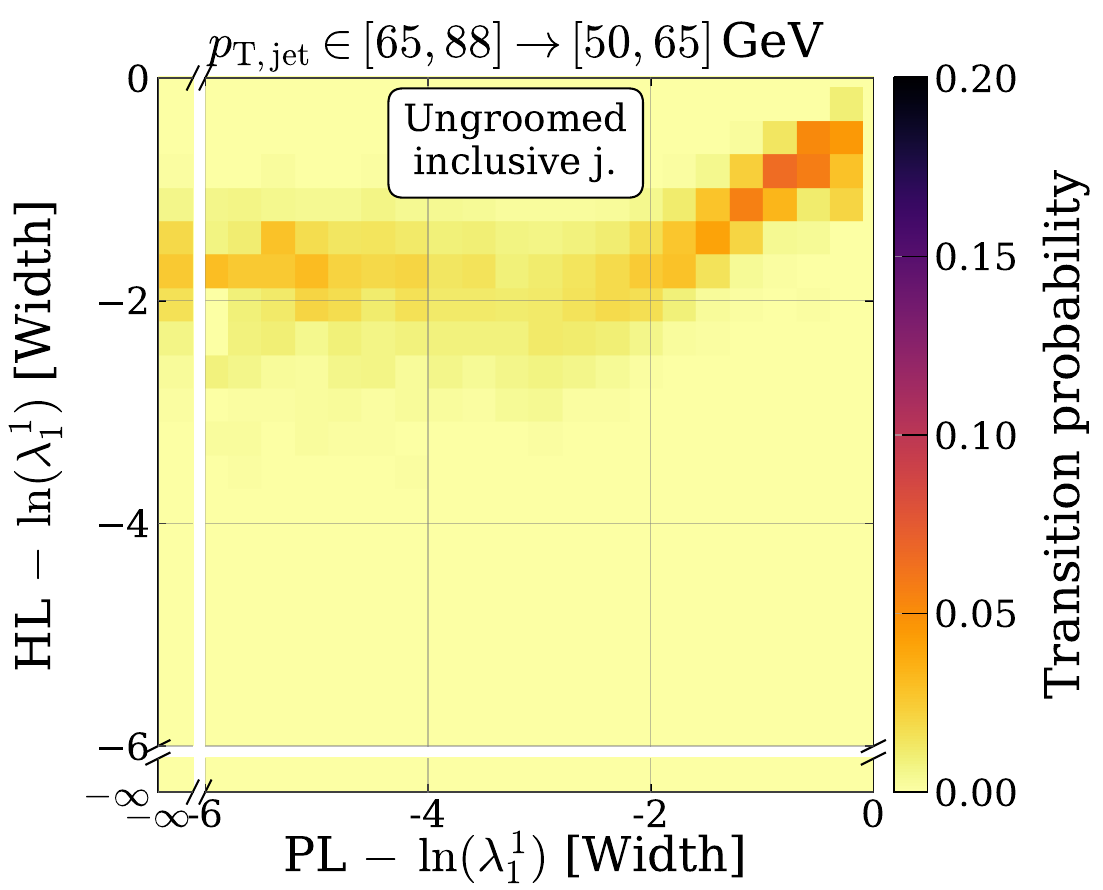}\\ 
    \includegraphics[width=0.48\linewidth]{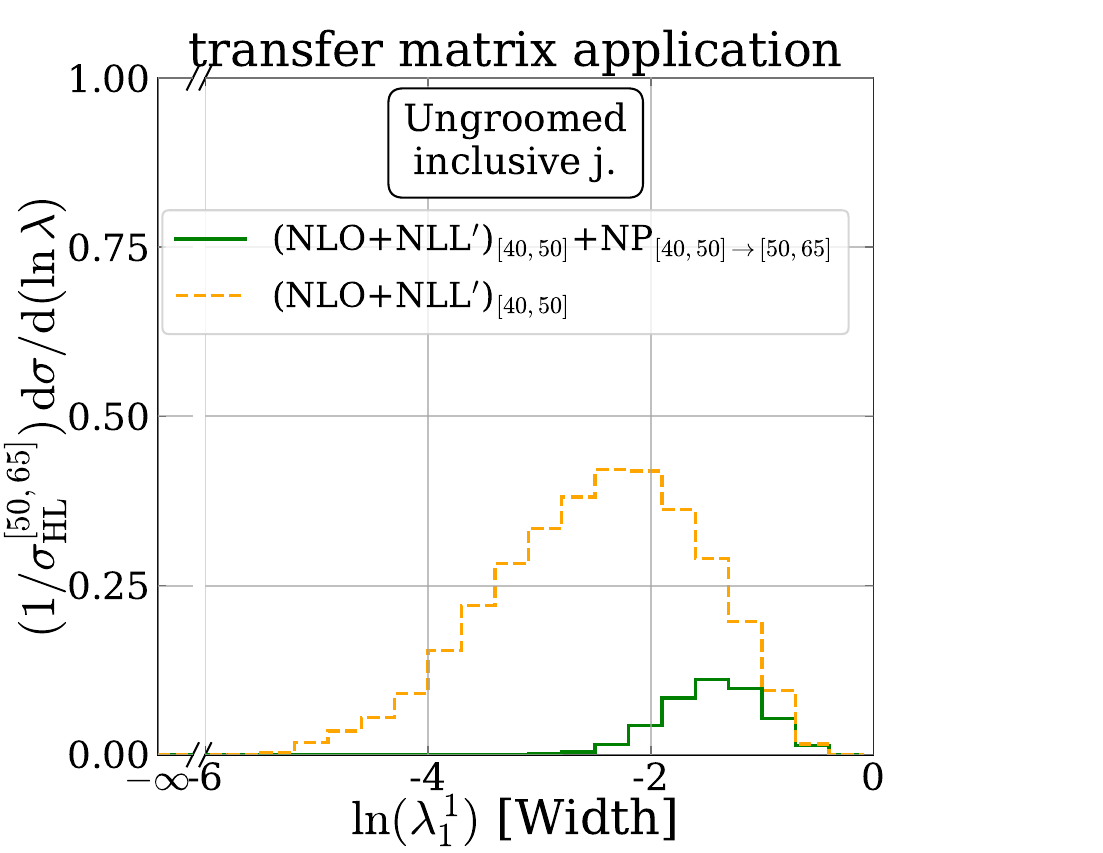} 
    \includegraphics[width=0.48\linewidth]{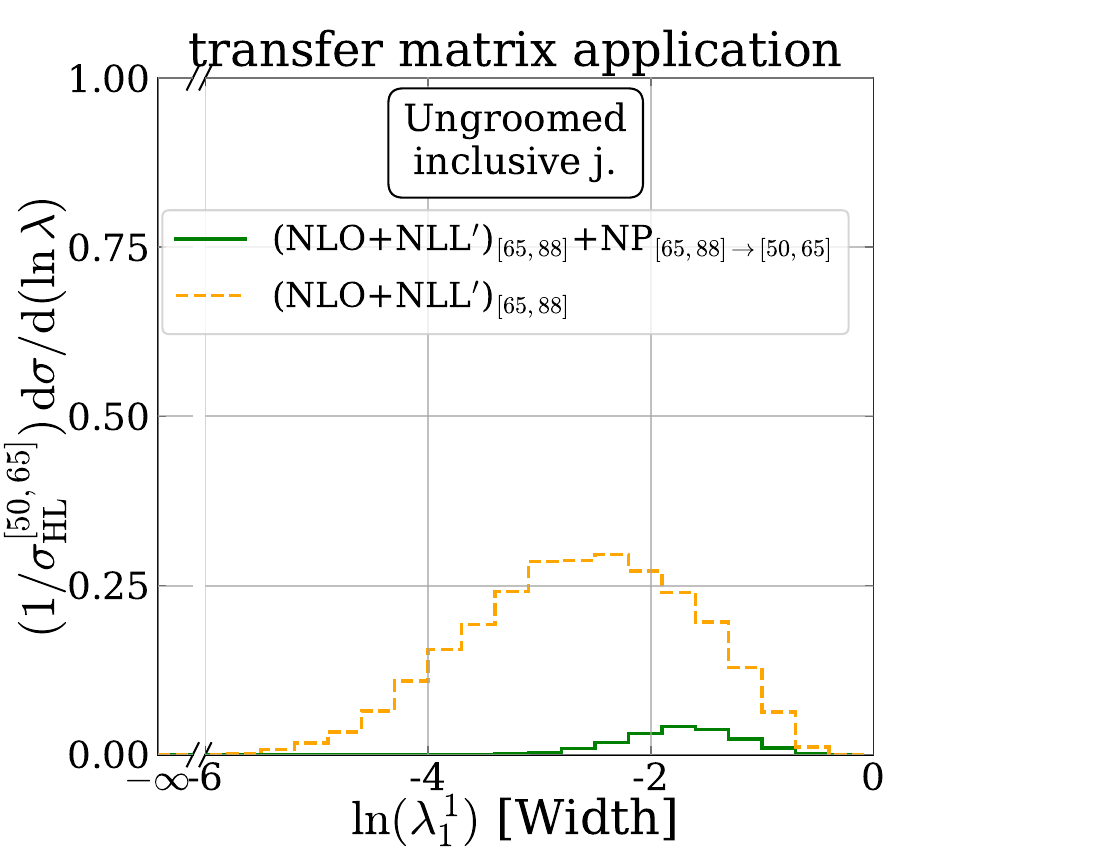}\\ 
    \caption{Parton-to-hadron-level transfer matrices for the Width angularity ($\lambda^1_1$) for inclusive jets 
    migrating from $\ptjet^{\text{PL}}\in [40,50]\,\text{GeV}$ (left) and from $\ptjet^{\text{PL}}\in[65,88]\,\text{GeV}$ (right) to $\ptjet^{\text{HL}}\in[50,65]\,\text{GeV}$, respectively. The lower panels show the corresponding \NLOpNLLp parton-level distribution and the resulting contribution to the hadron-level distribution.}
    \label{fig:tm_incl_offdiagonal}
\end{figure}

\begin{figure}[h!]
    \centering
    \includegraphics[width=0.48\linewidth]{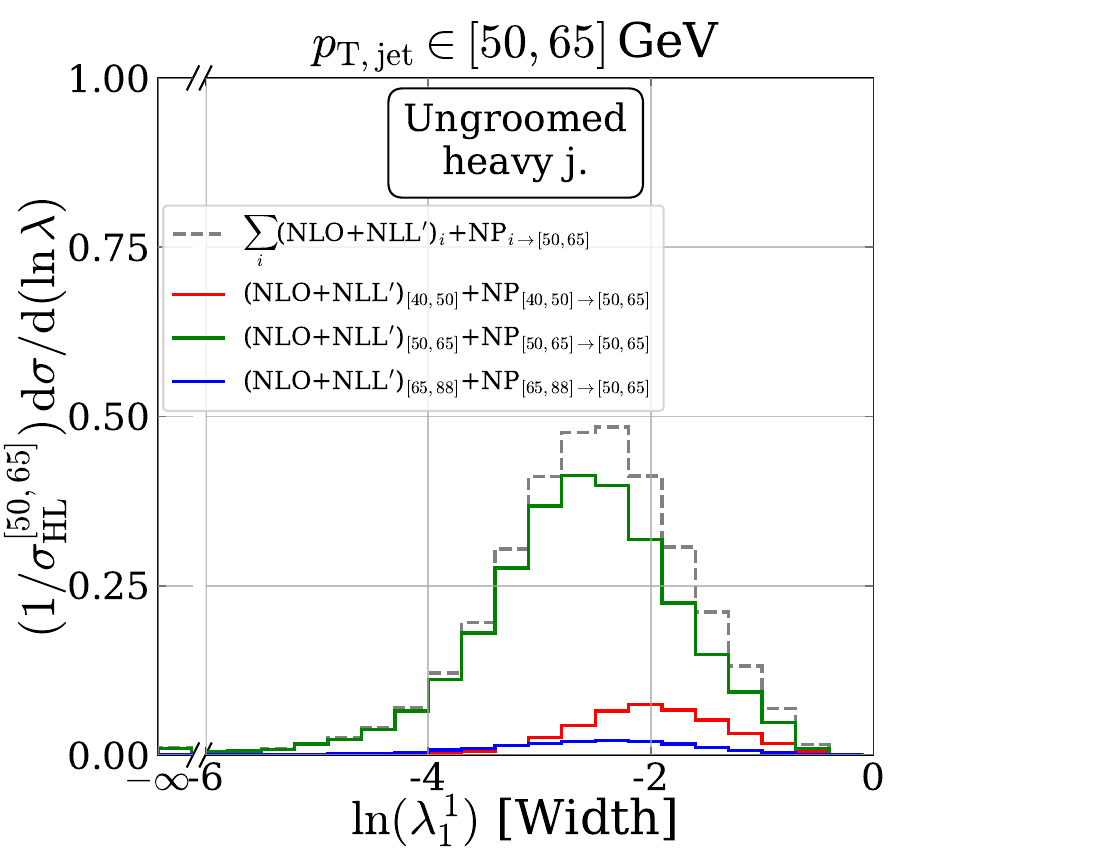}
    \includegraphics[width=0.48\linewidth]{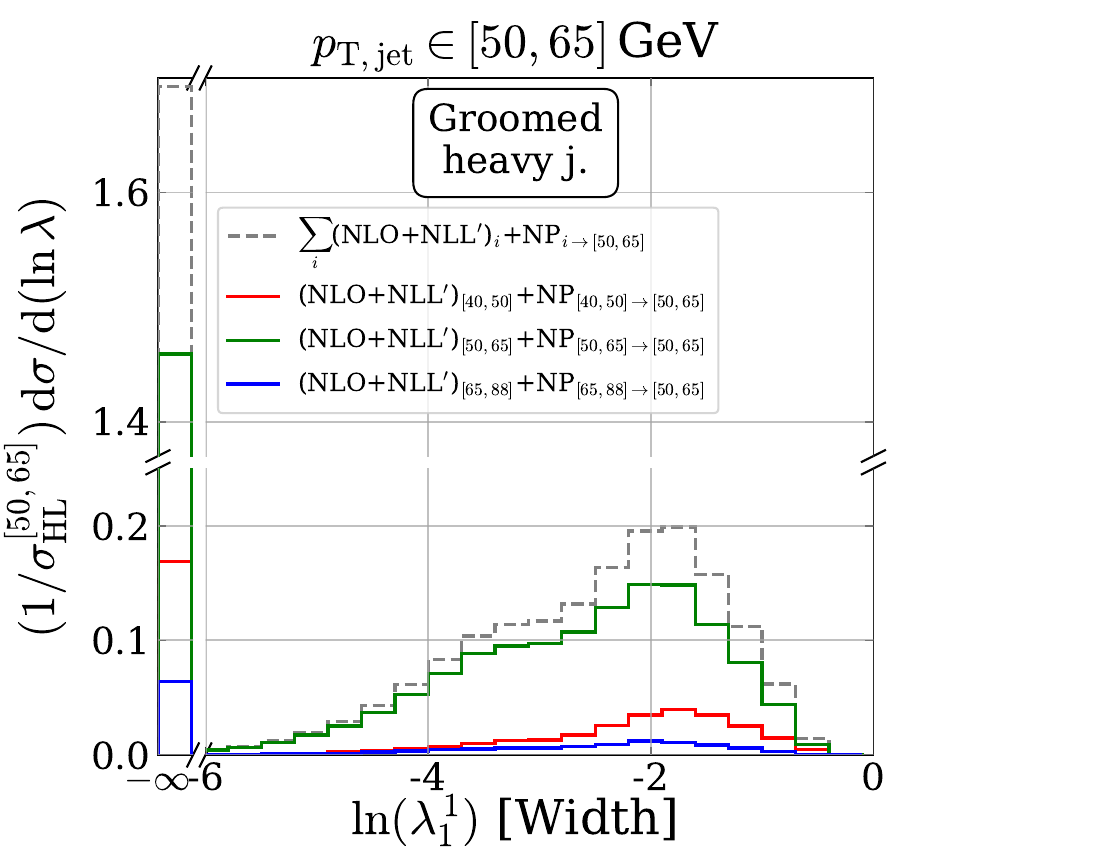}
    \caption{Contributions from different parton level jet transverse momentum bins to the Width angularity ($\lambda^1_1$) in the hadron level $\ptjet\in [50,65]\,\text{GeV}$ bin. The gray dashed line corresponds to the green line in Fig.~\ref{fig:tm_massiveB}.}
    \label{fig:tm_neighborPT}
\end{figure}

To further illustrate the efficacy of our approach to incorporate non-perturbative corrections, we show in Fig.~\ref{fig:tm_incl_offdiagonal} the migration matrices and the resulting observable distributions across $\ptjet$ bins. For concreteness, we here consider the case of ungroomed inclusive jets and show results for the $\lambda^1_1$ angularity. For events with $\ptjet\in [50,65]\,\text{GeV}$ at hadron level we 
show the migration of events from neighbouring parton-level bins, i.e.\ with $\ptjet^{\text{PL}}\in[40,50]\,\text{GeV}$ (left) and $\ptjet^{\text{PL}}\in [65,88]\,\text{GeV}$ (right). For both cases the transfer matrices have rather large off-diagonal elements, populating significantly larger observable values at hadron than at parton level. As can be inferred from Fig.~\ref{fig:tm_pT}, the probability of parton-level events with $\ptjet^{\text{PL}}\in [40,50]\,\text{GeV}$ being pushed to $\ptjet^{\text{HL}}\in [50,65]\,\text{GeV}$ is about $12.4\%$. In contrast, for events with $\ptjet^{\text{PL}}\in [65,88]\,\text{GeV}$ this amounts to $7.5\%$ only. However, even if the sheer size of the migrated cross section seems not too worrying, there can be significant changes to the shape of the observable distribution, due to the correlation between the shift in jet transverse momentum and the angularity measured on the jet. We illustrate this in Fig.~\ref{fig:tm_neighborPT}, where we separately plot the contributions to the hadron level with $\ptjet\in [50,65]\,\text{GeV}$ from jets that at parton level had $\ptjet^{\text{PL}}\in [40,50], [50,65], [65,88]\,\text{GeV}$, i.e.\ the diagonal and the first and largest off-diagonal elements of the transfer matrices in Fig.~\ref{fig:tm_pT}. In both the ungroomed (left plot) and groomed (right plot) cases, we find that the jets migrating towards higher transverse momentum also exhibit a larger mean angularity value than jets that have similar $p_T$ at hadron and parton level. The downward migrated jets are distributed more similar to the unmigrated jets, with a distribution peaking at roughly the same value. Due to these non-trivial migration effects a simple ratio-based approach for the extraction of non-perturbative corrections as was initially used in Ref.~\cite{Caletti:2021oor} turns out to be quite limited. Similar effects we observe when applying \softdrop-grooming as well as for heavy-flavour jets. 

\FloatBarrier

\subsection{Final Results: Resummed predictions with non-perturbative corrections}
\begin{figure}
\centering
\includegraphics[width=0.32\linewidth]{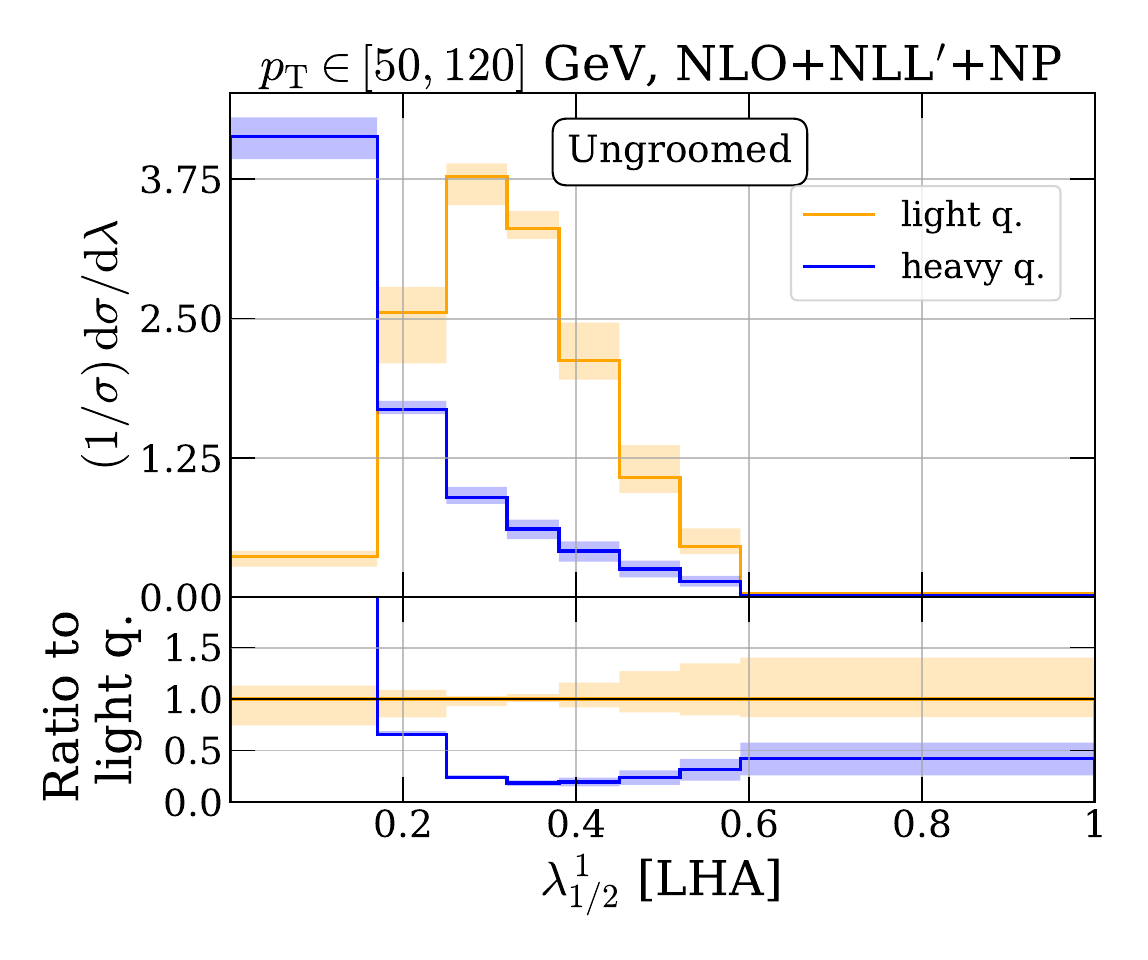}
\includegraphics[width=0.32\linewidth]{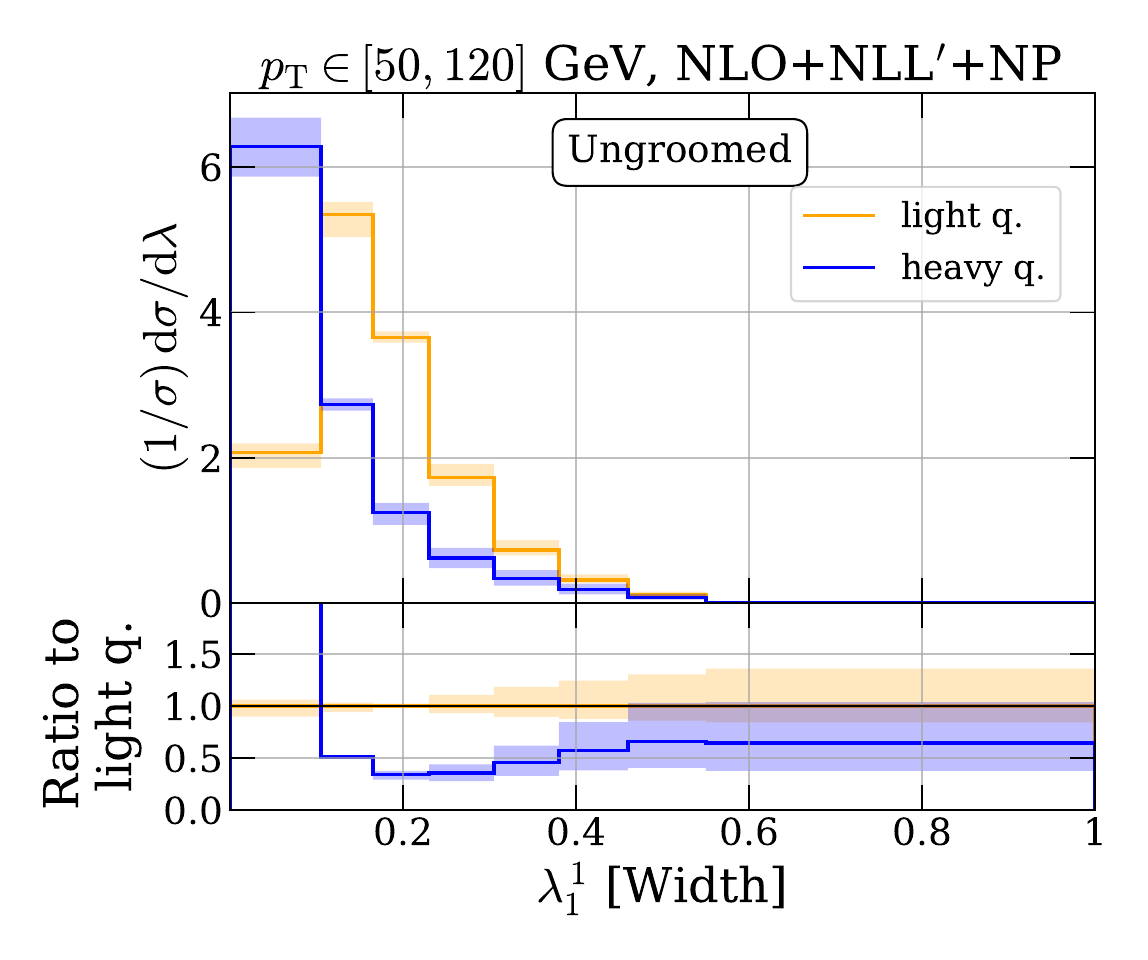}
\includegraphics[width=0.32\linewidth]{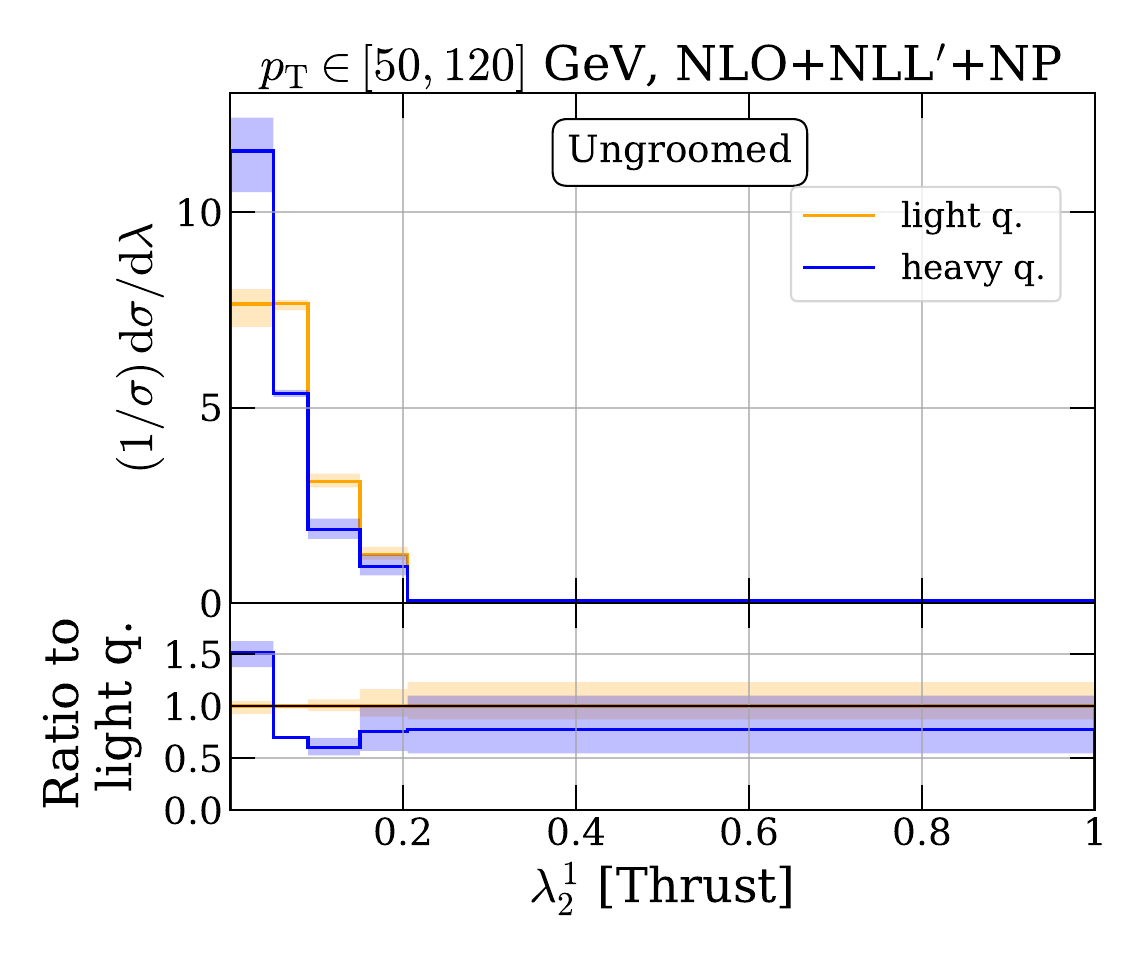}
\includegraphics[width=0.32\linewidth]{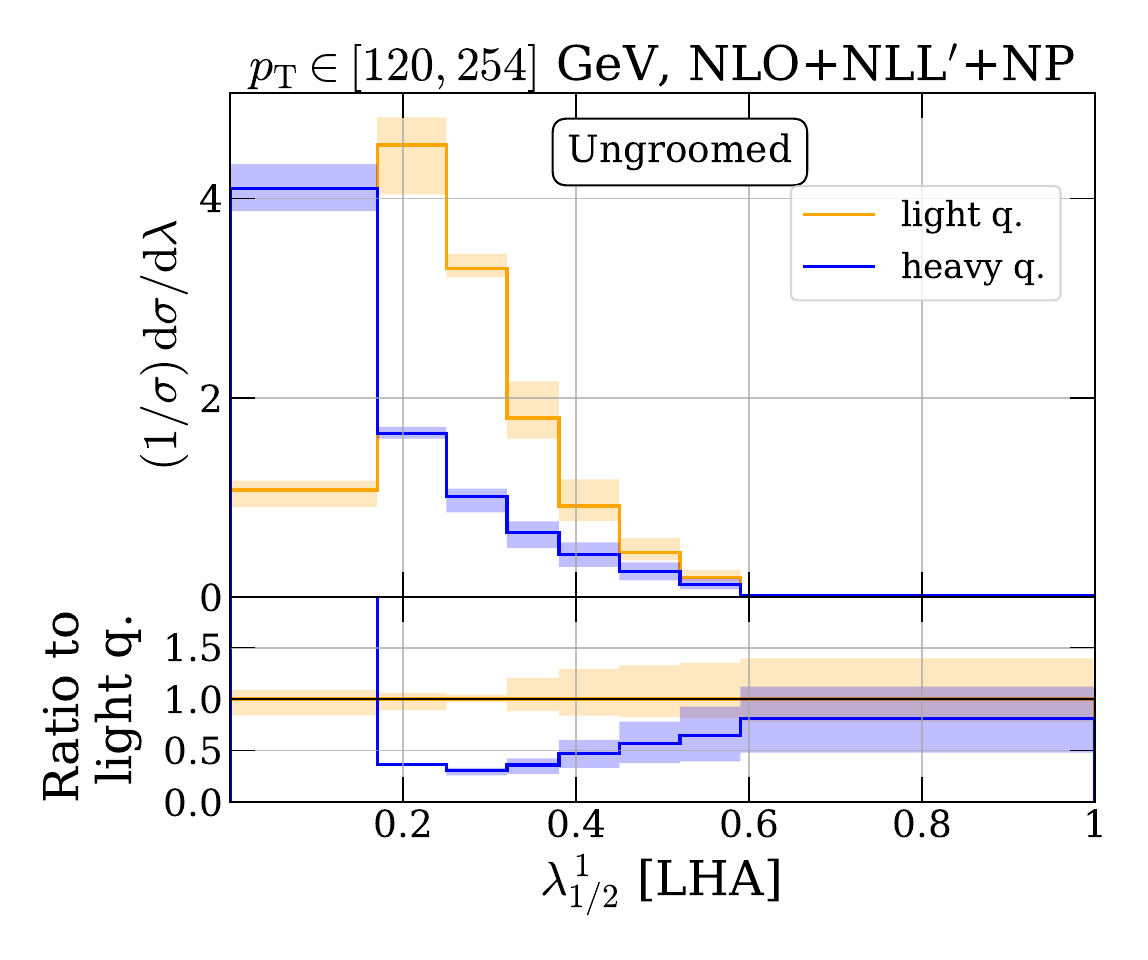}
\includegraphics[width=0.32\linewidth]{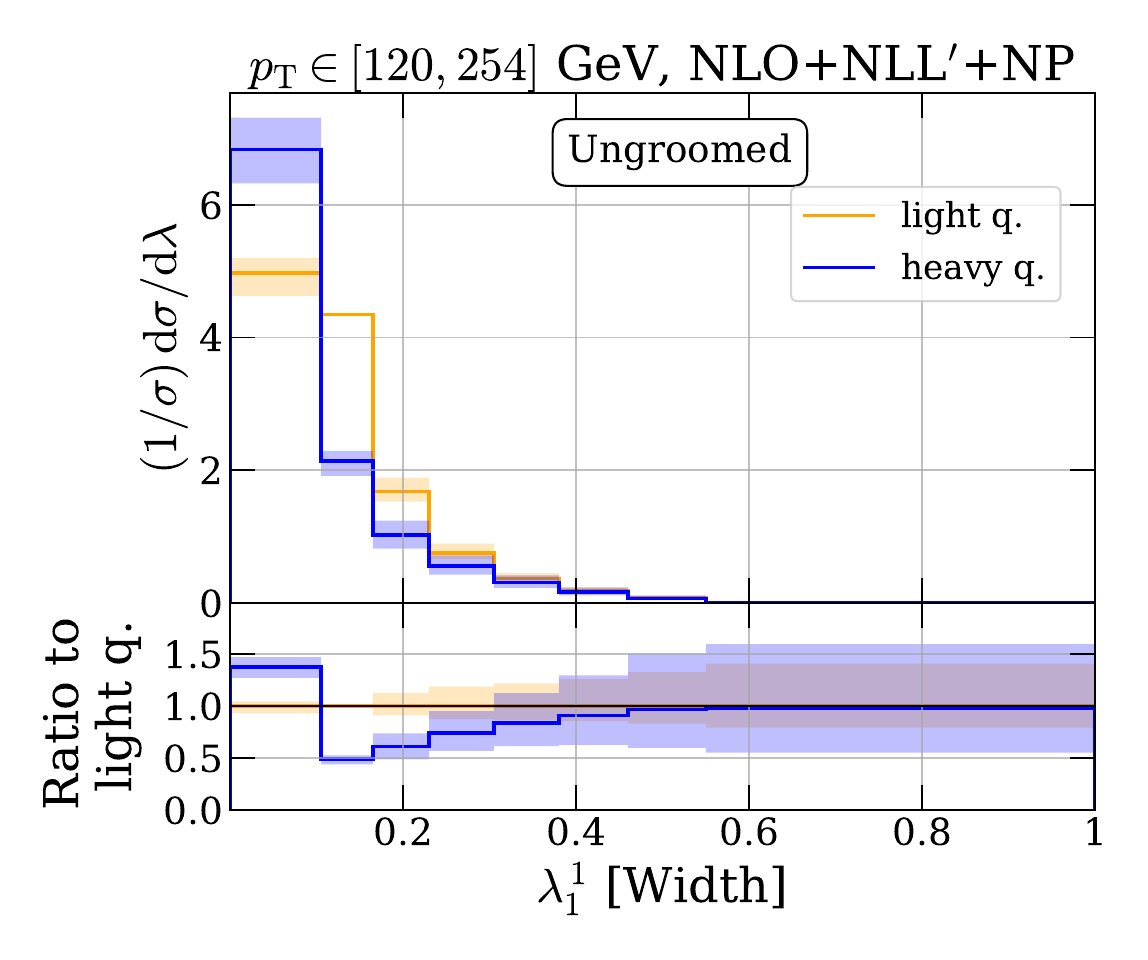}
\includegraphics[width=0.32\linewidth]{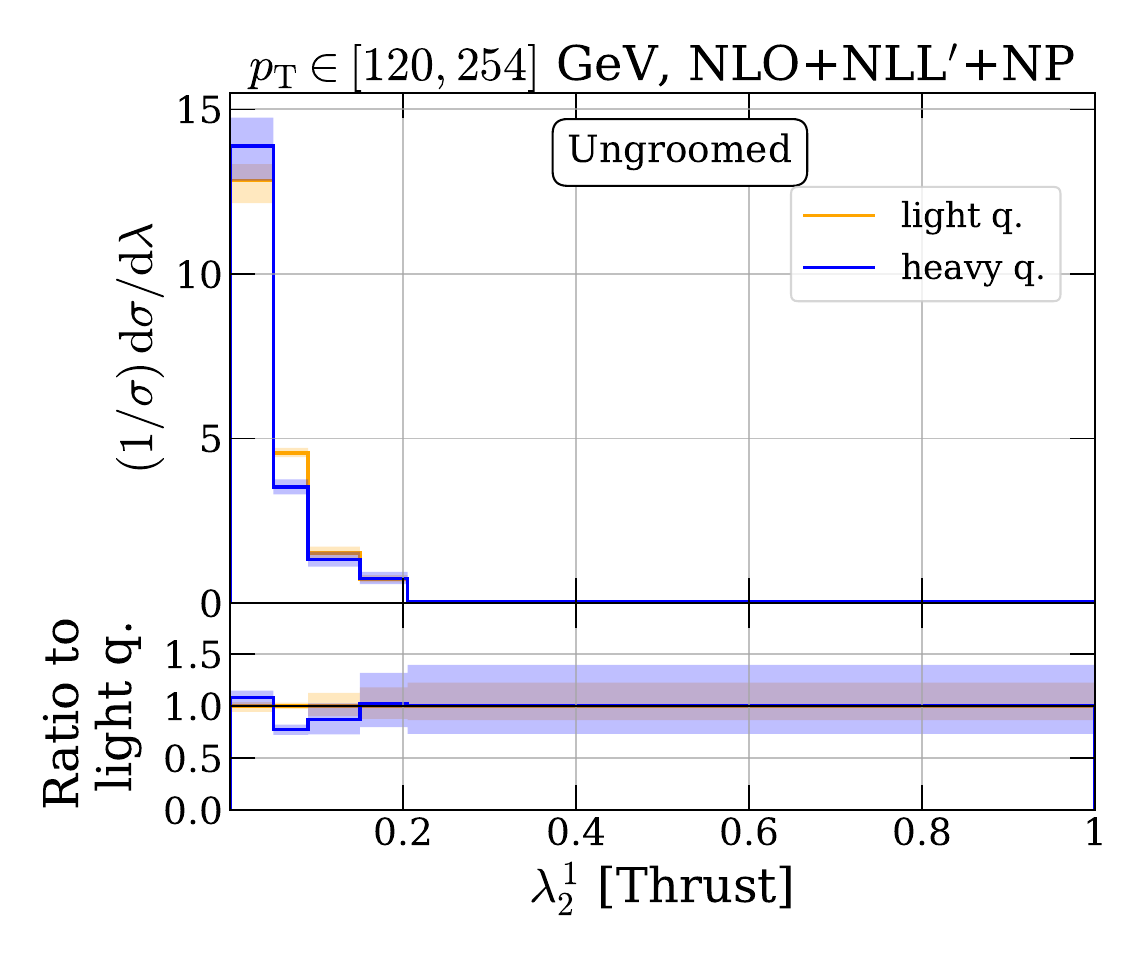}
\includegraphics[width=0.32\linewidth]{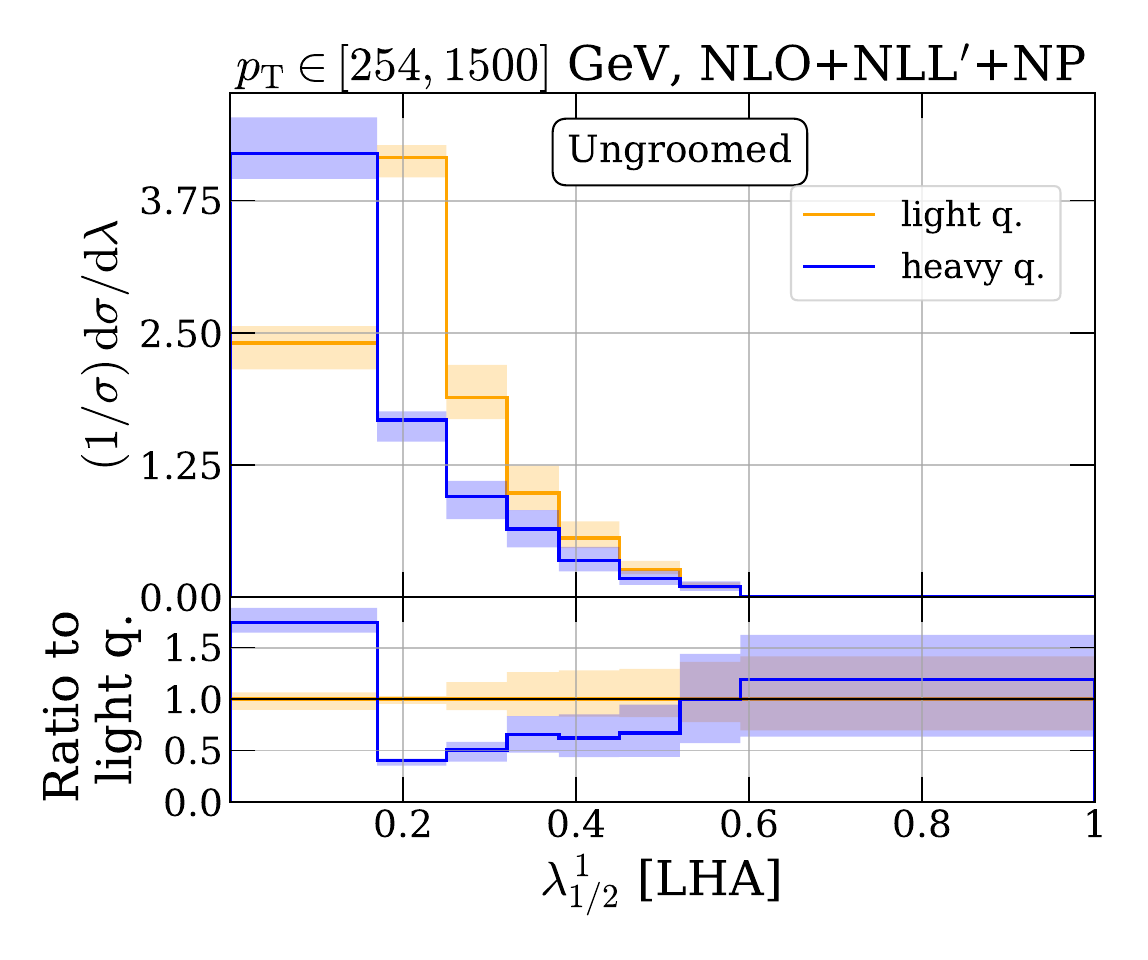}
\includegraphics[width=0.32\linewidth]{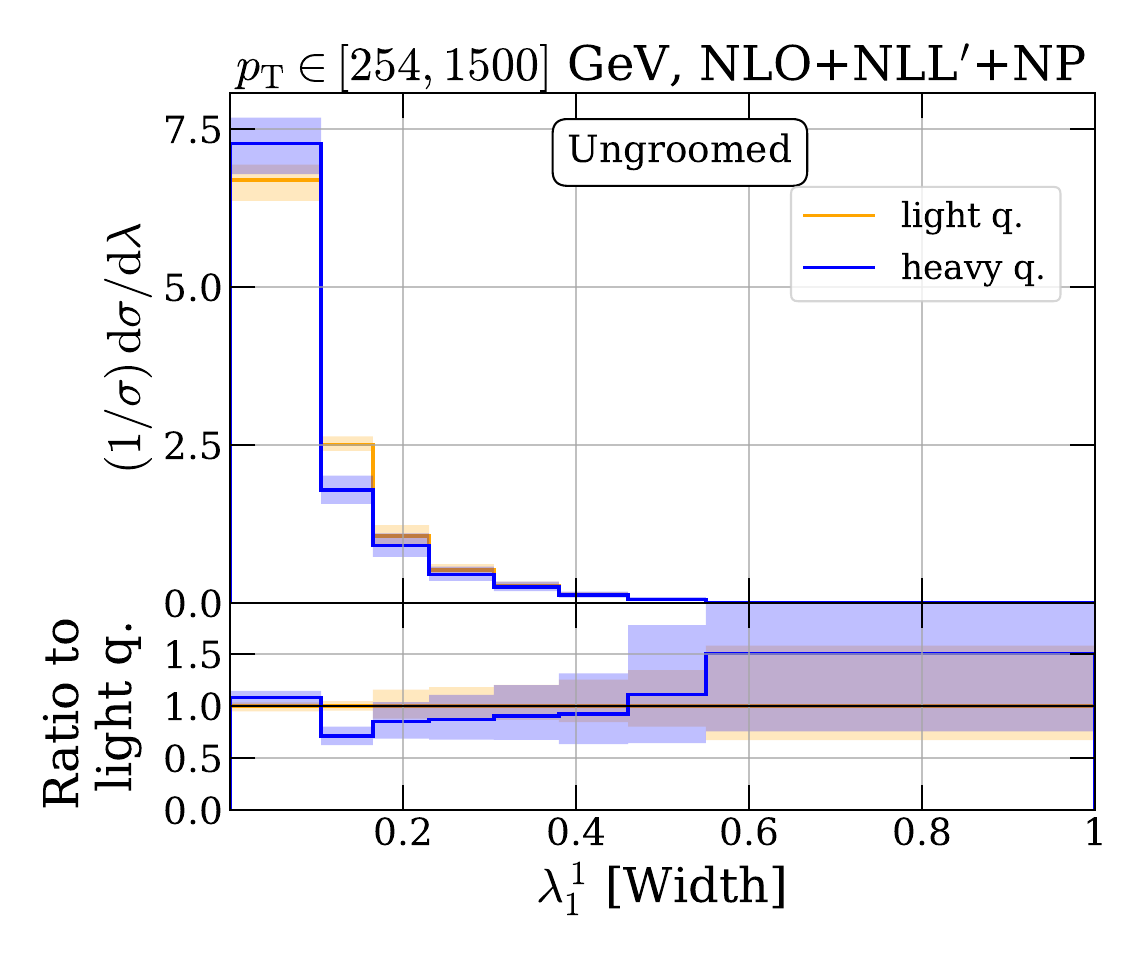}
\includegraphics[width=0.32\linewidth]{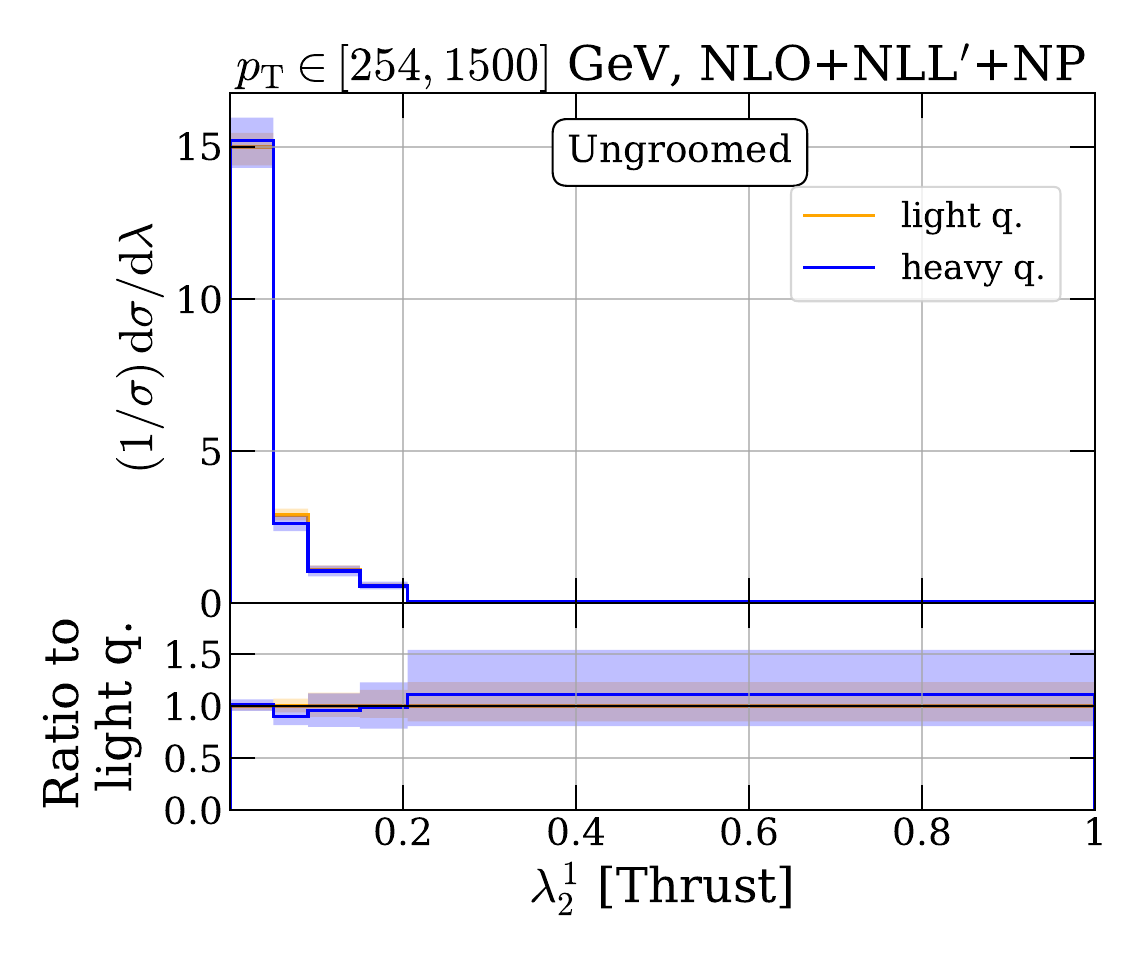}
    \caption{Comparison between jet angularities measured on $b$-jets (with undecayed $B$-hadrons) and on a light-quark-jet sample. From top to bottom, we show the distribution of the same angularity for the low, medium and high transverse-momentum slice, respectively. The three different jet angularities, LHA, Width and Thrust, are shown from left to right.}
    \label{fig:heavy_vs_light_coarse}
\end{figure}

\begin{figure}
\centering
\includegraphics[width=0.32\linewidth]{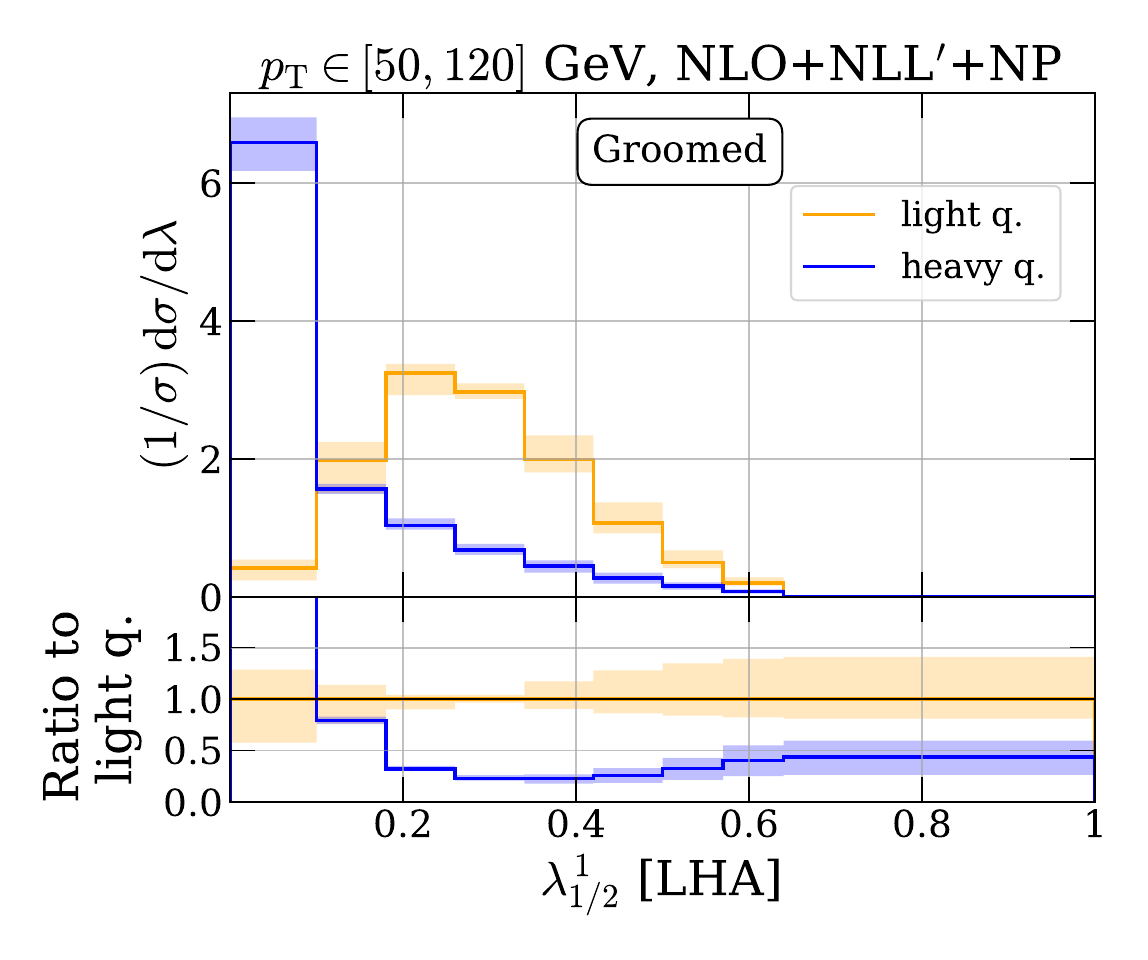}
\includegraphics[width=0.32\linewidth]{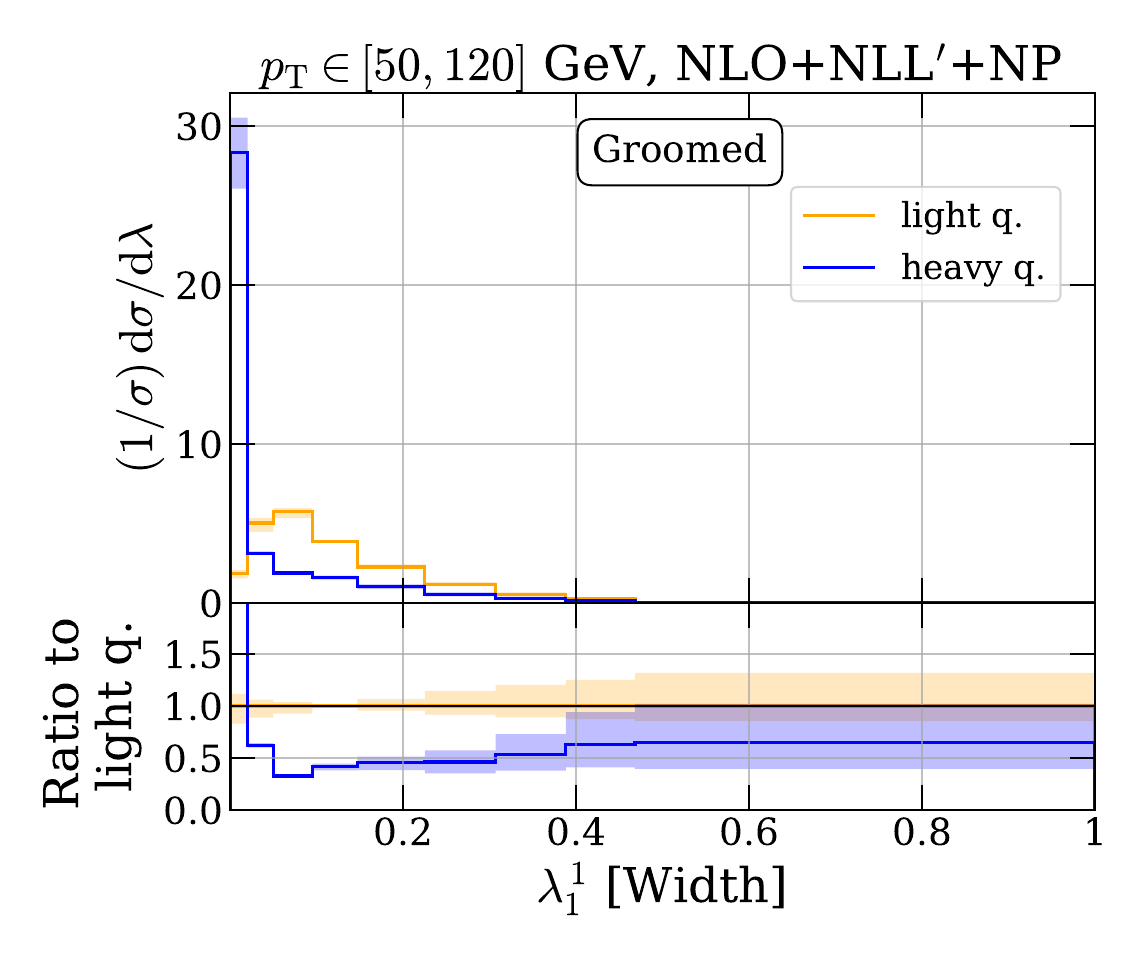}
\includegraphics[width=0.32\linewidth]{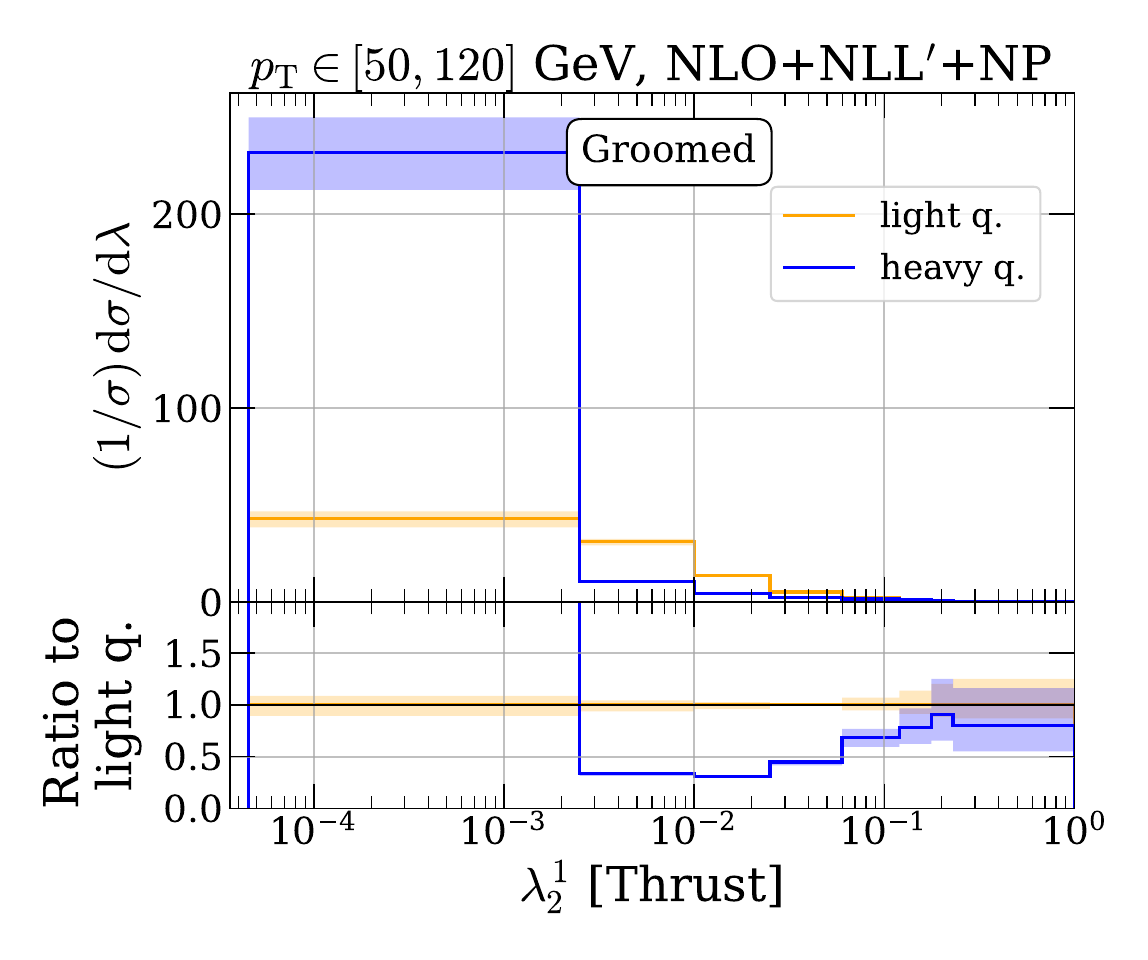}
\includegraphics[width=0.32\linewidth]{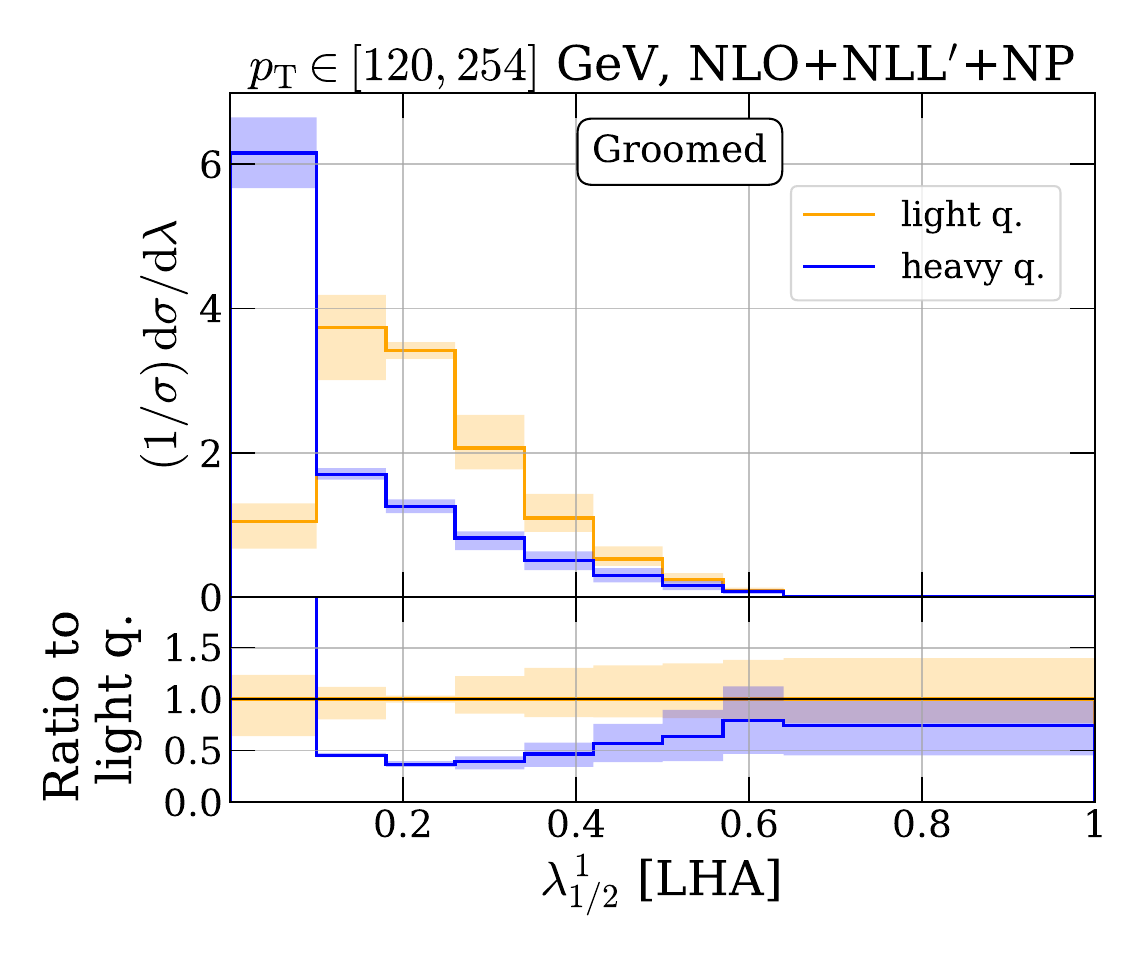}
\includegraphics[width=0.32\linewidth]{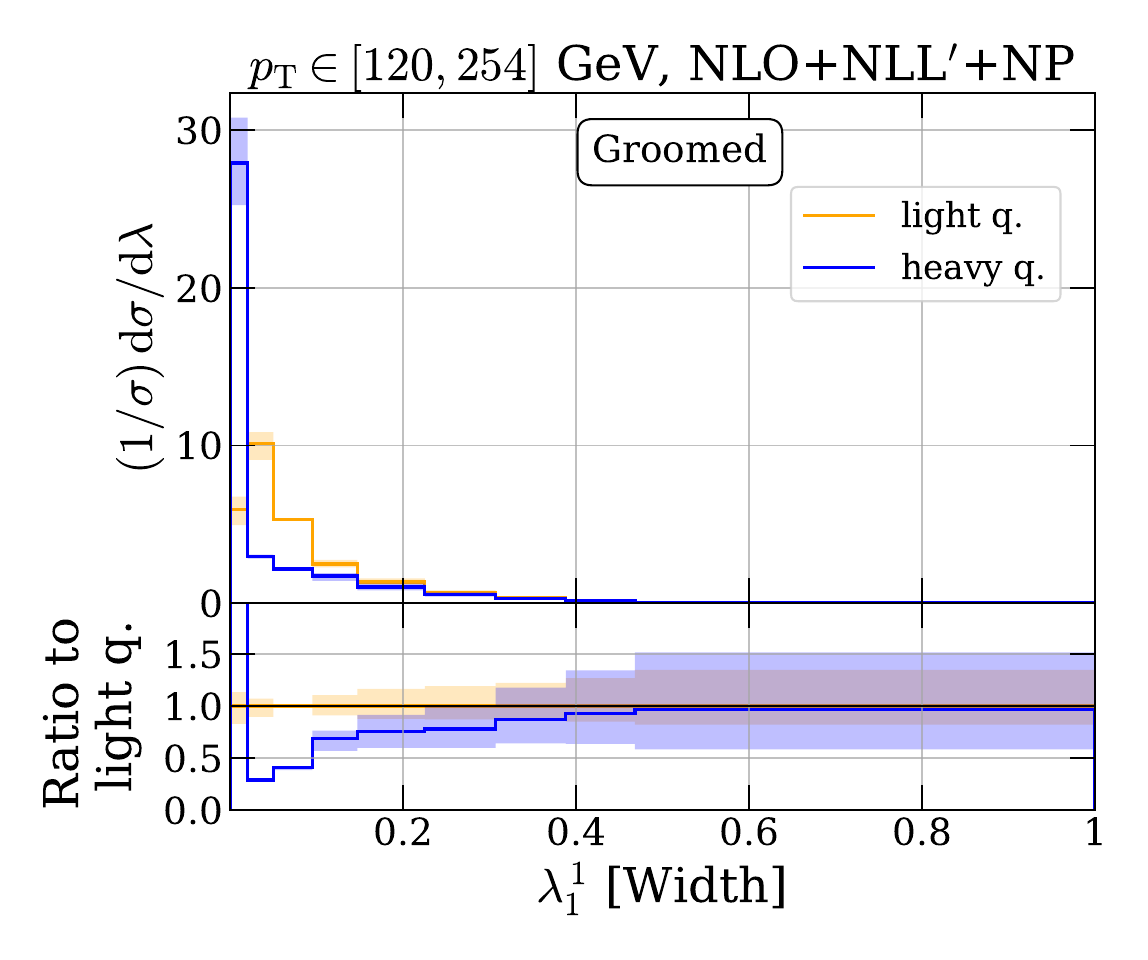}
\includegraphics[width=0.32\linewidth]{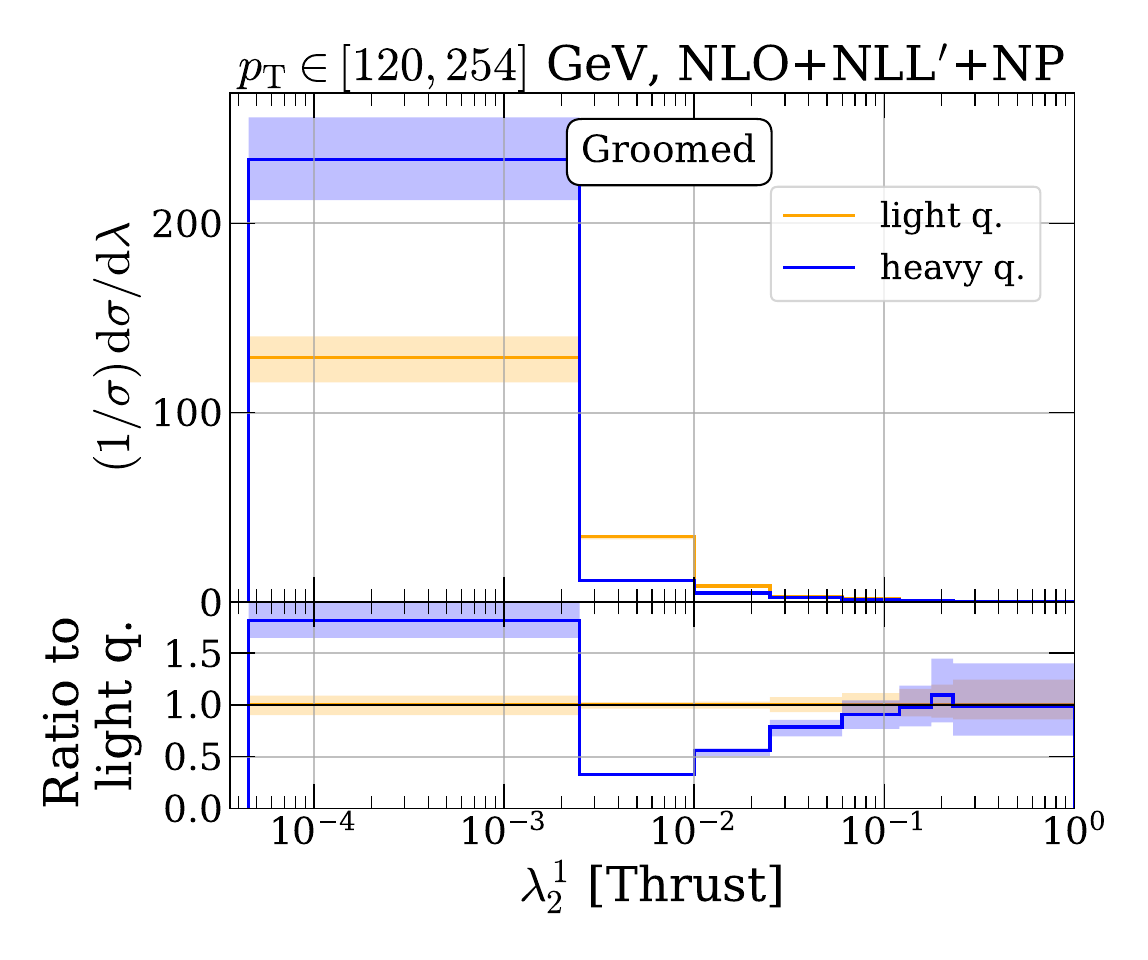}
\includegraphics[width=0.32\linewidth]{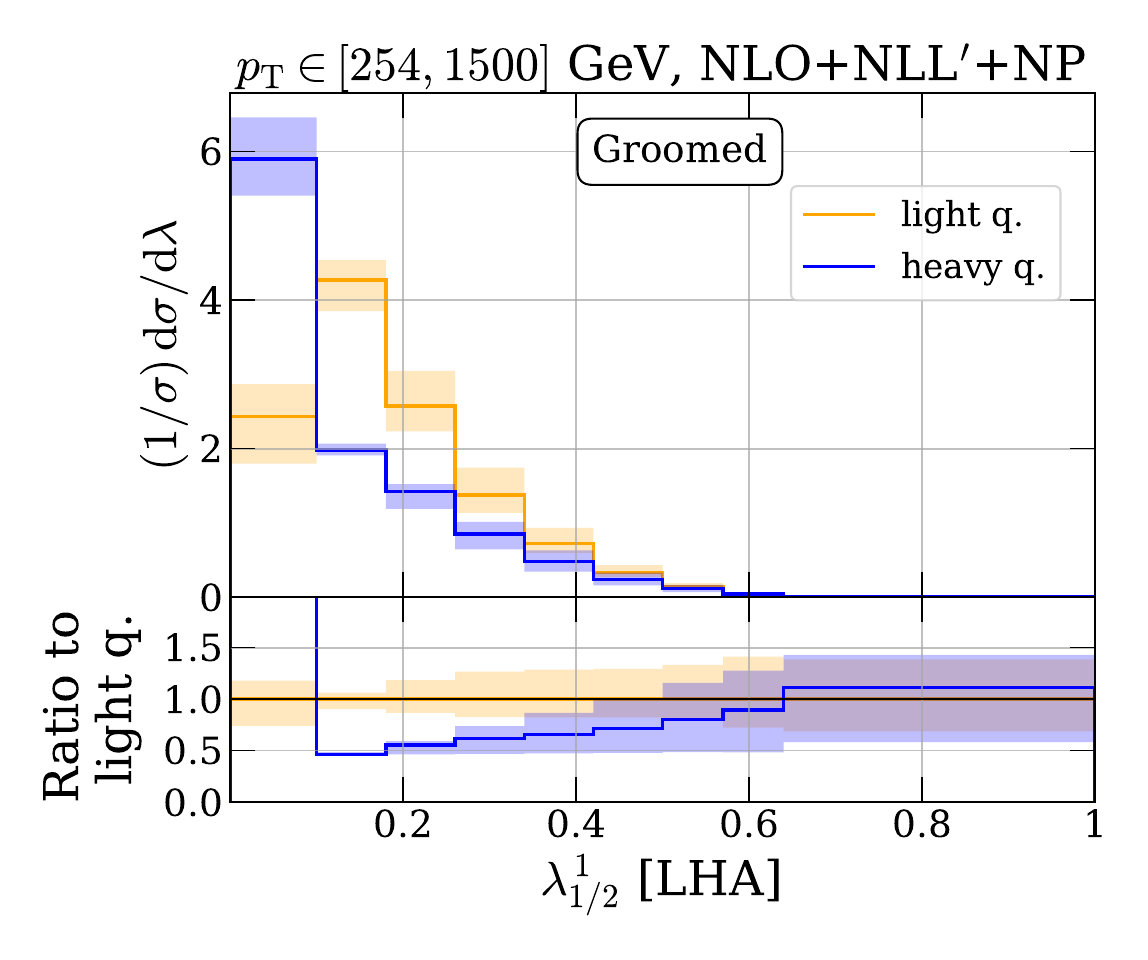}
\includegraphics[width=0.32\linewidth]{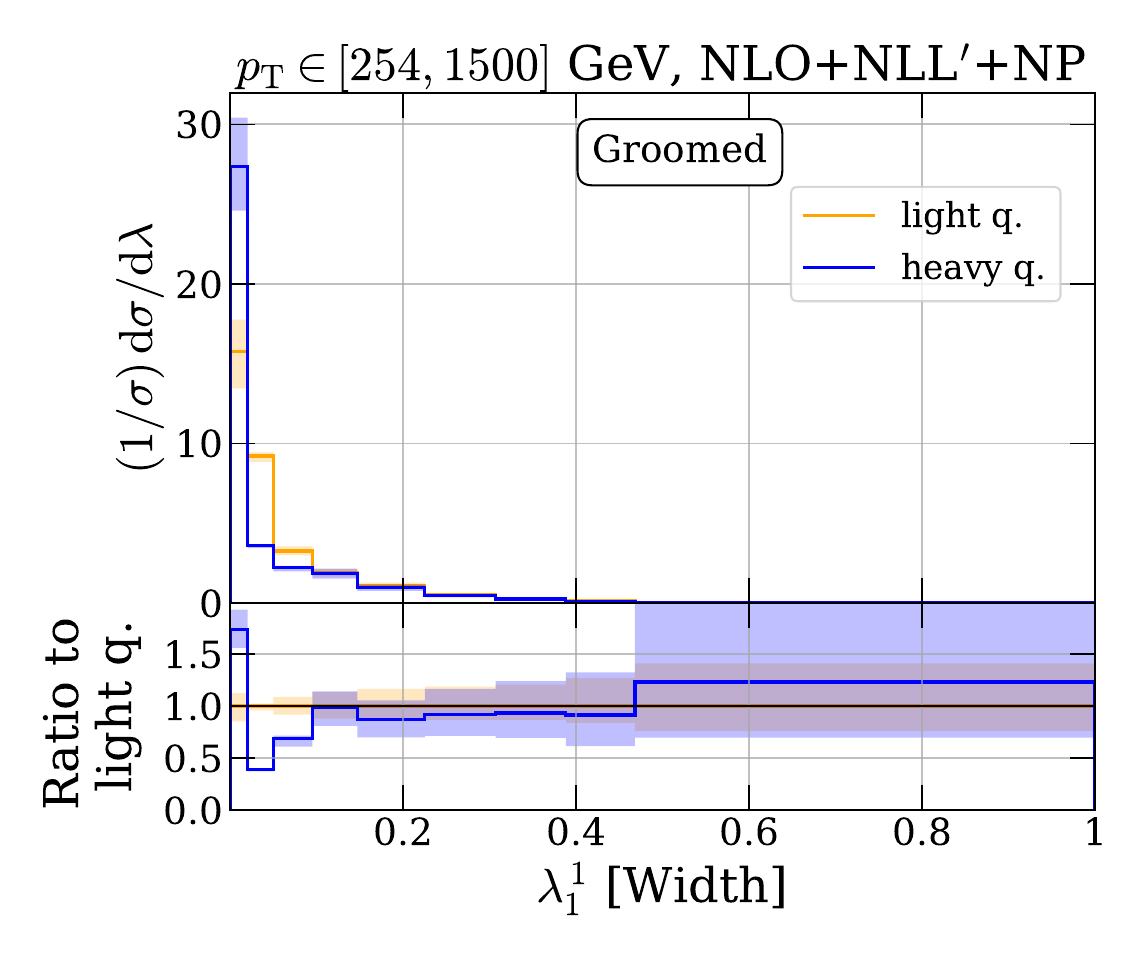}
\includegraphics[width=0.32\linewidth]{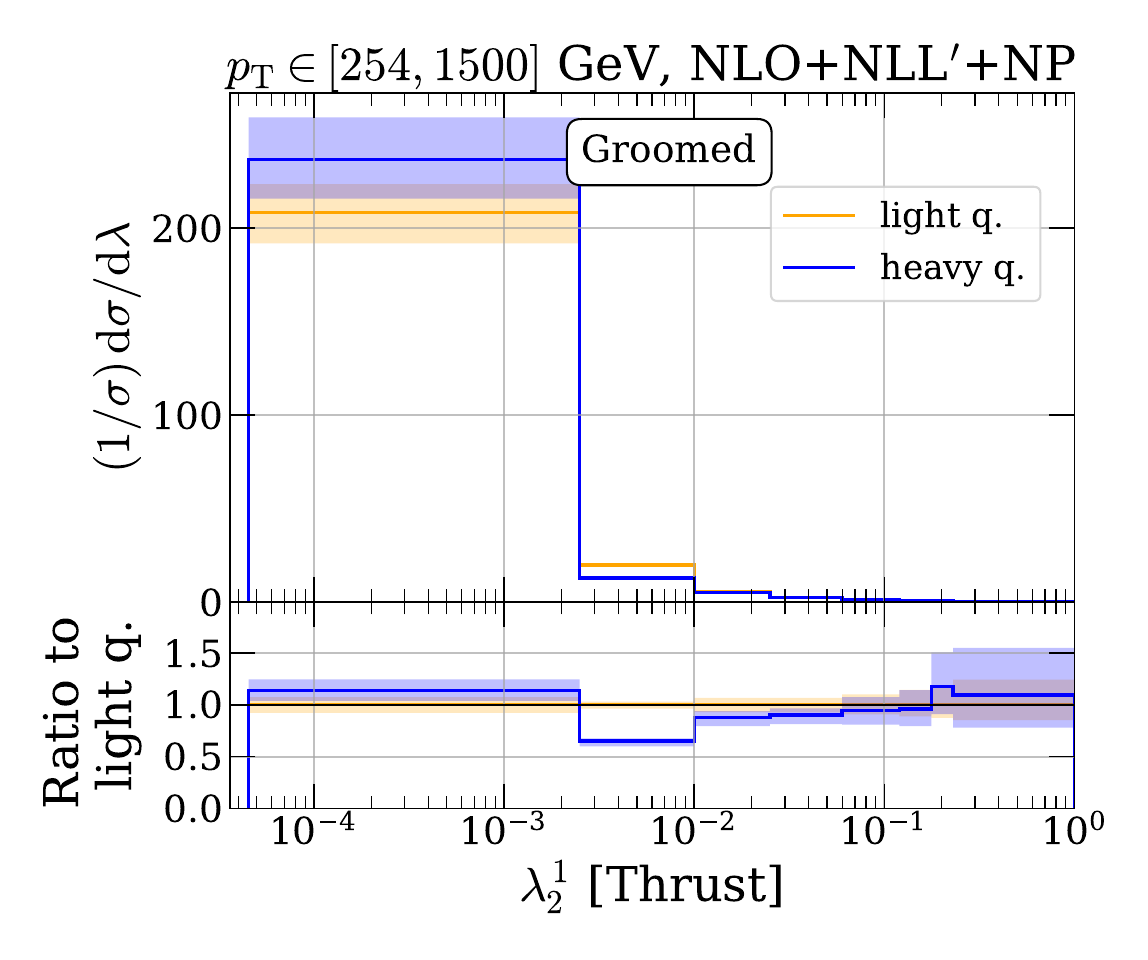}
    \caption{Same as Fig.~\ref{fig:heavy_vs_light_coarse}, but for \softdrop-groomed jets.}
    \label{fig:heavy_vs_light_coarse_SD}
\end{figure}

We are now ready to present and discuss our final predictions at \NLOpNLLp accuracy, including non-perturbative (NP) corrections. We show the angularity distributions $\lambda^1_\alpha$ for different values of the angular parameter $\alpha=\frac{1}{2},1,2$, namely LHA, jet Width, and jet Thrust, for both ungroomed jets and jets groomed with \softdrop, using parameters $\beta=0$ and $\zc=0.1$.
We consider three transverse-momentum slices: low $\ptjet \in [50,120]$~GeV, medium $\ptjet \in [120,254]$~GeV, and high $\ptjet \in [254,1500]$~GeV. 
%
To consider a realistic resolution of the observable values, we adopt the same binning as in the CMS measurement of jet angularities in dijet and $Z$+jet events~\cite{CMS:2021iwu}.
We begin by comparing the hadron-level resummed and matched predictions for $b$-jets with those for jets initiated by light quarks.
The results for standard (ungroomed) jets are shown in Fig.~\ref{fig:heavy_vs_light_coarse}. From top to bottom, we display the same angularity distribution for the low, medium, and high transverse-momentum slices, respectively. From left to right, we show the three angularities: LHA, jet Width, and jet Thrust.
Most distributions exhibit a clear separation between $b$-quark and light-quark jets. This separation becomes more pronounced as the angular exponent $\alpha$ decreases. This behaviour is expected, since mass effects primarily influence the collinear region, which is strongly suppressed for larger values of $\alpha$, such as in jet Thrust ($\alpha=2$).
Indeed, it is known~\cite{Dhani:2024gtx} that the spectrum becomes sensitive to mass effects below
\begin{equation}\label{eq:transpoint}
\lambda^1_\alpha \sim \left(\frac{m}{\ptjet R_0}\right)^\alpha.
\end{equation}
As $\ptjet$ increases, this transition shifts to smaller values, and the $b$-jet and light-jet spectra become increasingly similar. In the high-$\ptjet$ region, for $\alpha=1$ and $\alpha=2$, the two distributions essentially coincide. Although sensitivity to dead-cone effects is lost in this regime, $b$-tagged jets remain useful as they provide a relatively pure sample of quark-initiated jets.
Similar considerations apply, at least qualitatively, to groomed jets, whose results are shown in Fig.~\ref{fig:heavy_vs_light_coarse_SD}. In this case, the distinction between $b$-quark and light-quark jets remains visible even for jet Thrust. This is because \softdrop removes soft radiation, enhancing the relative importance of the collinear region within the jet.

\begin{figure}
\centering
\includegraphics[width=0.32\linewidth]{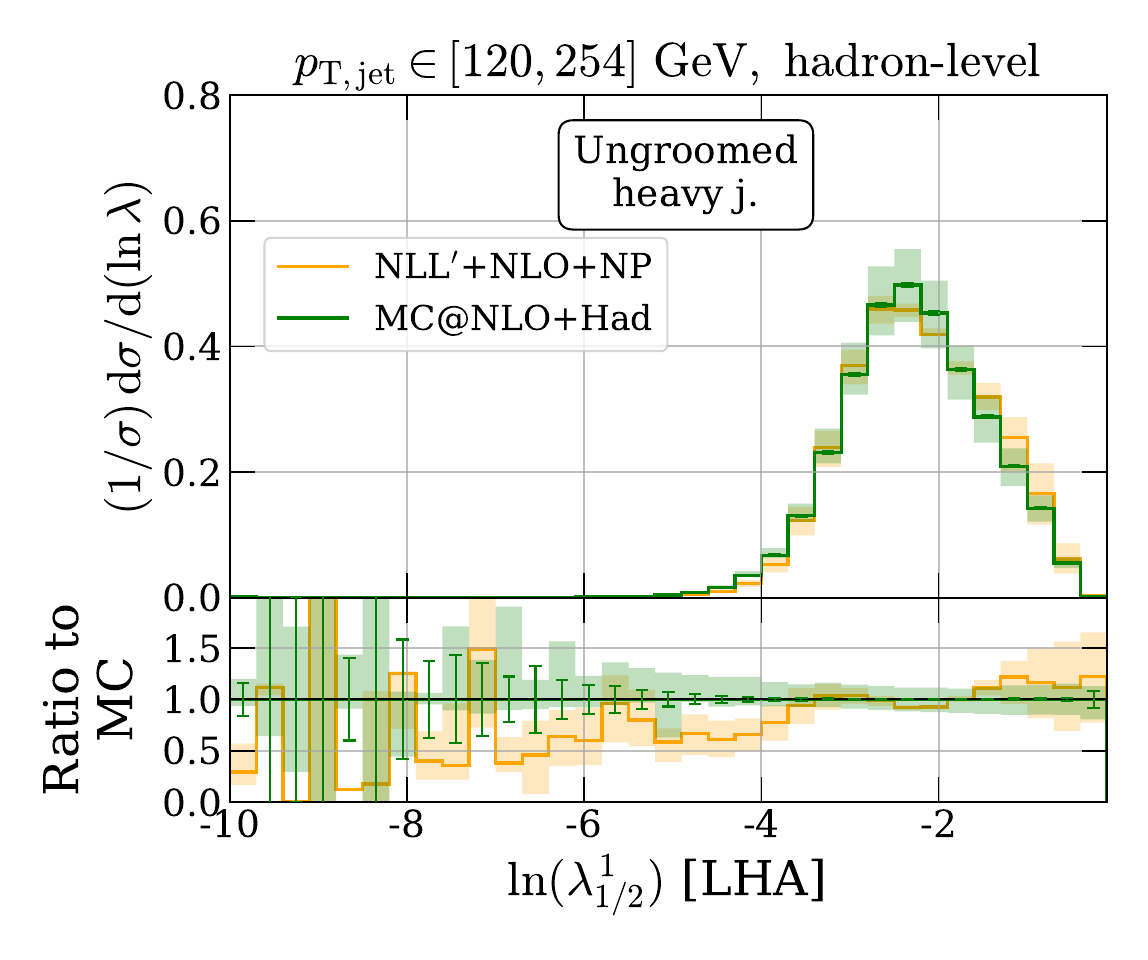}
\includegraphics[width=0.32\linewidth]{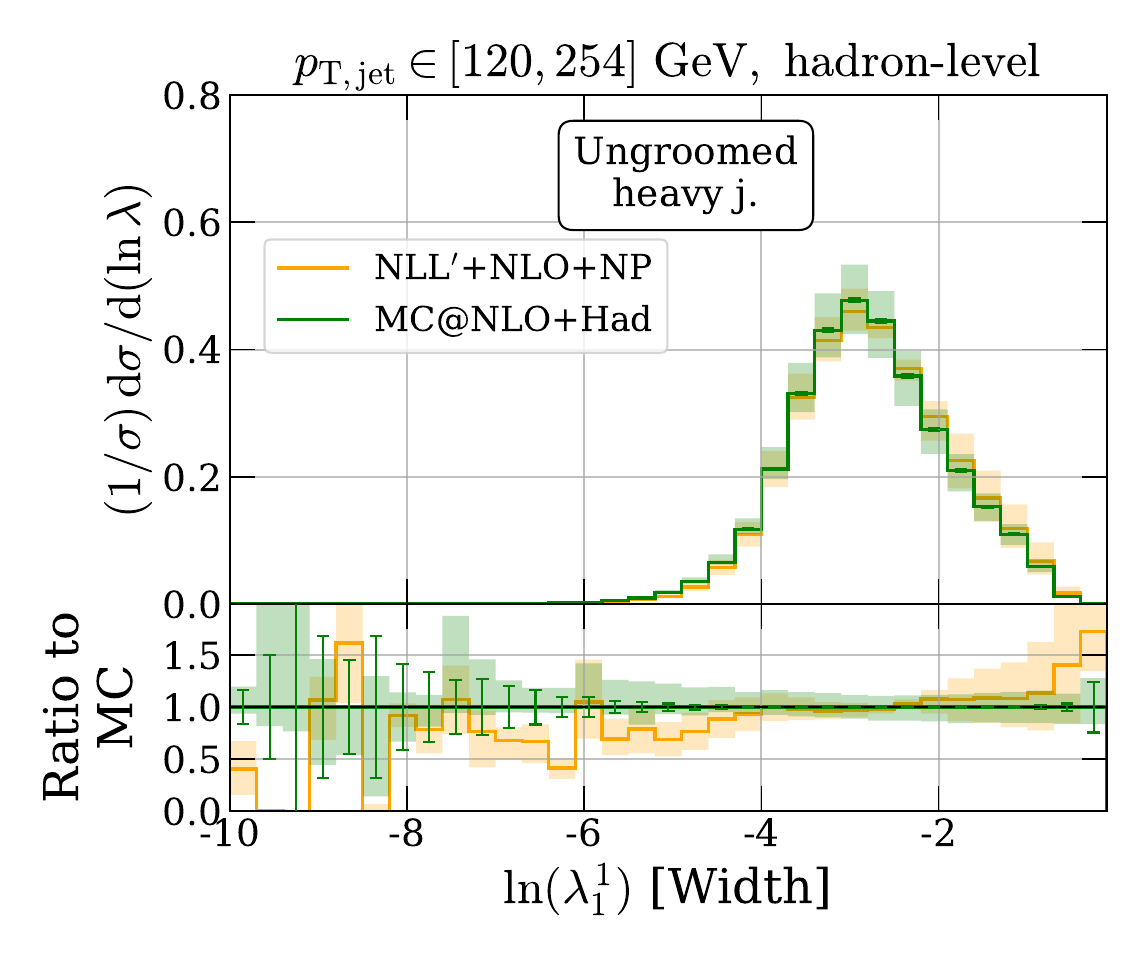}
\includegraphics[width=0.32\linewidth]{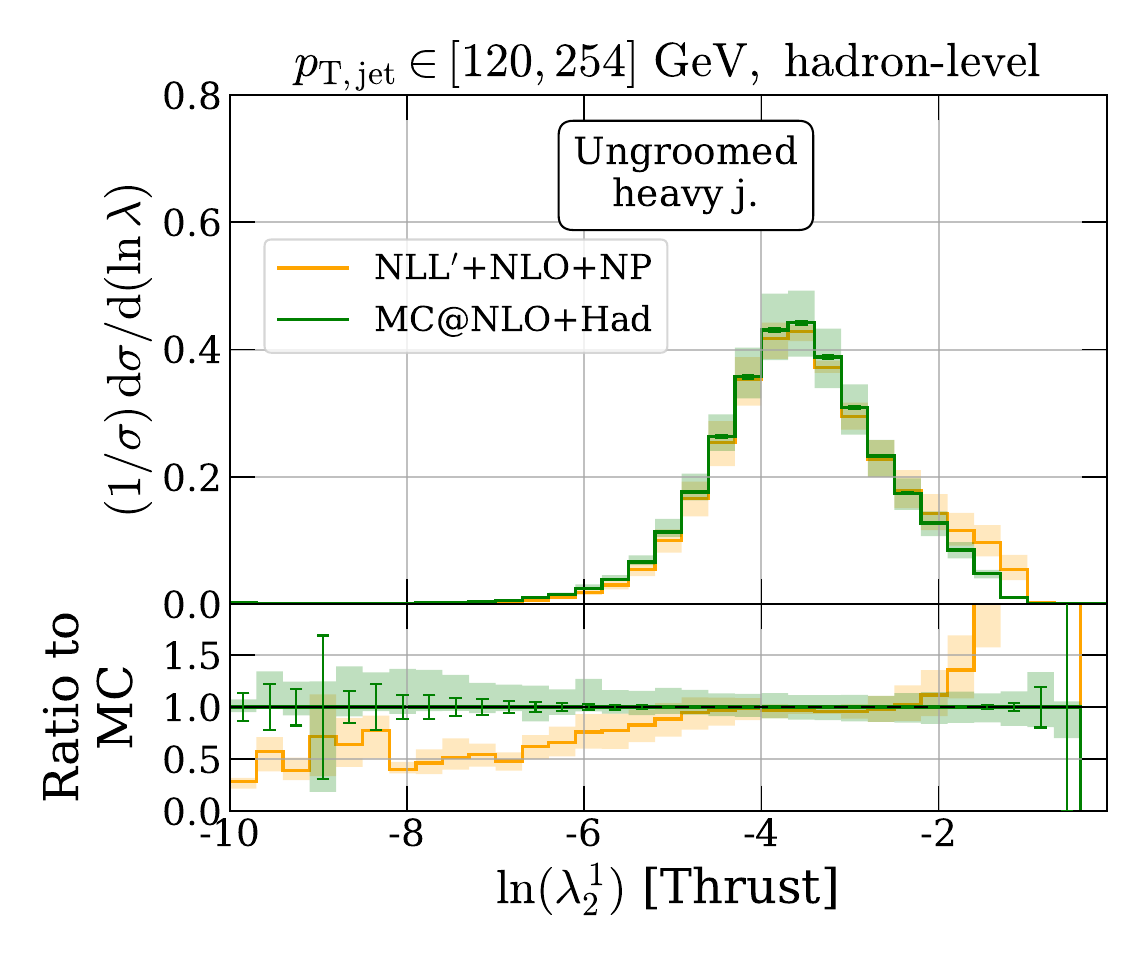}
\includegraphics[width=0.32\linewidth]{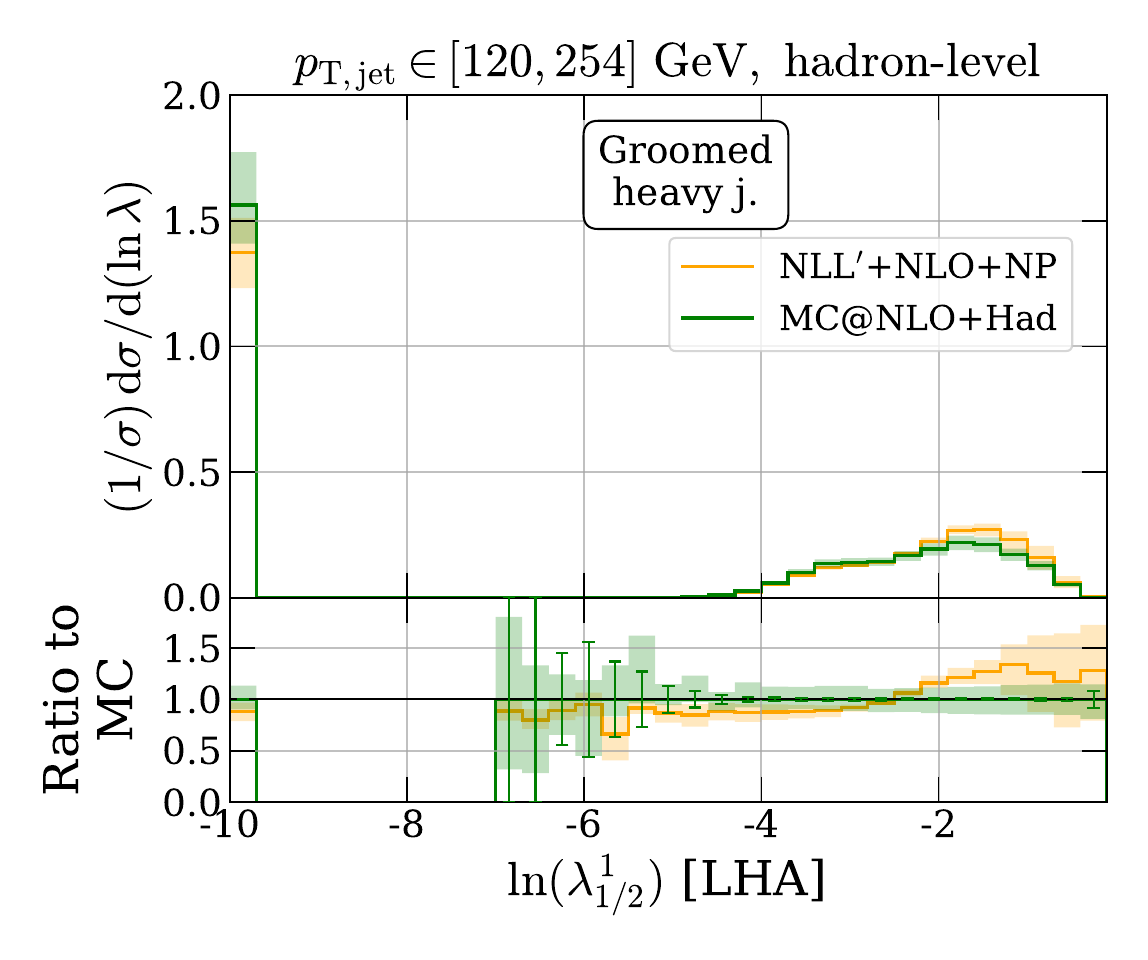}
\includegraphics[width=0.32\linewidth]{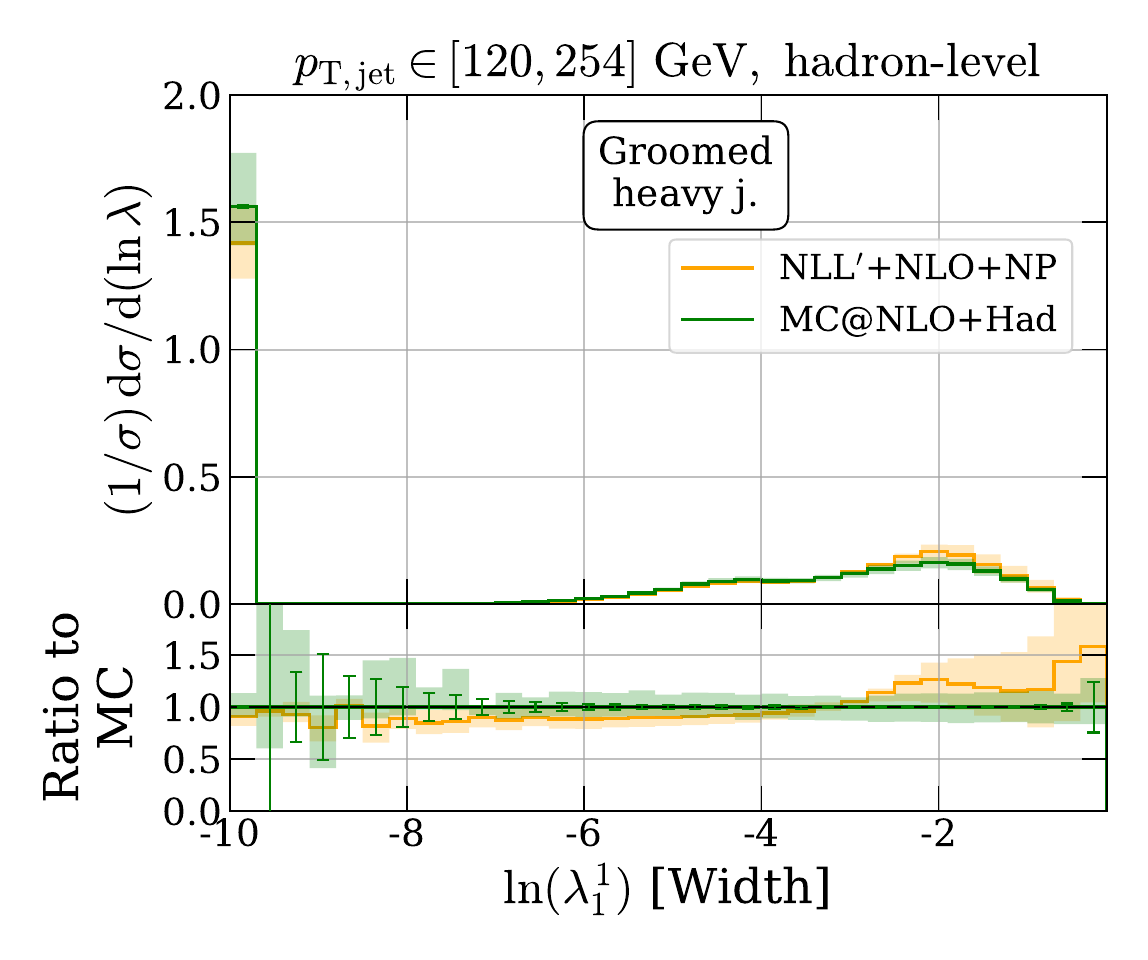}
\includegraphics[width=0.32\linewidth]{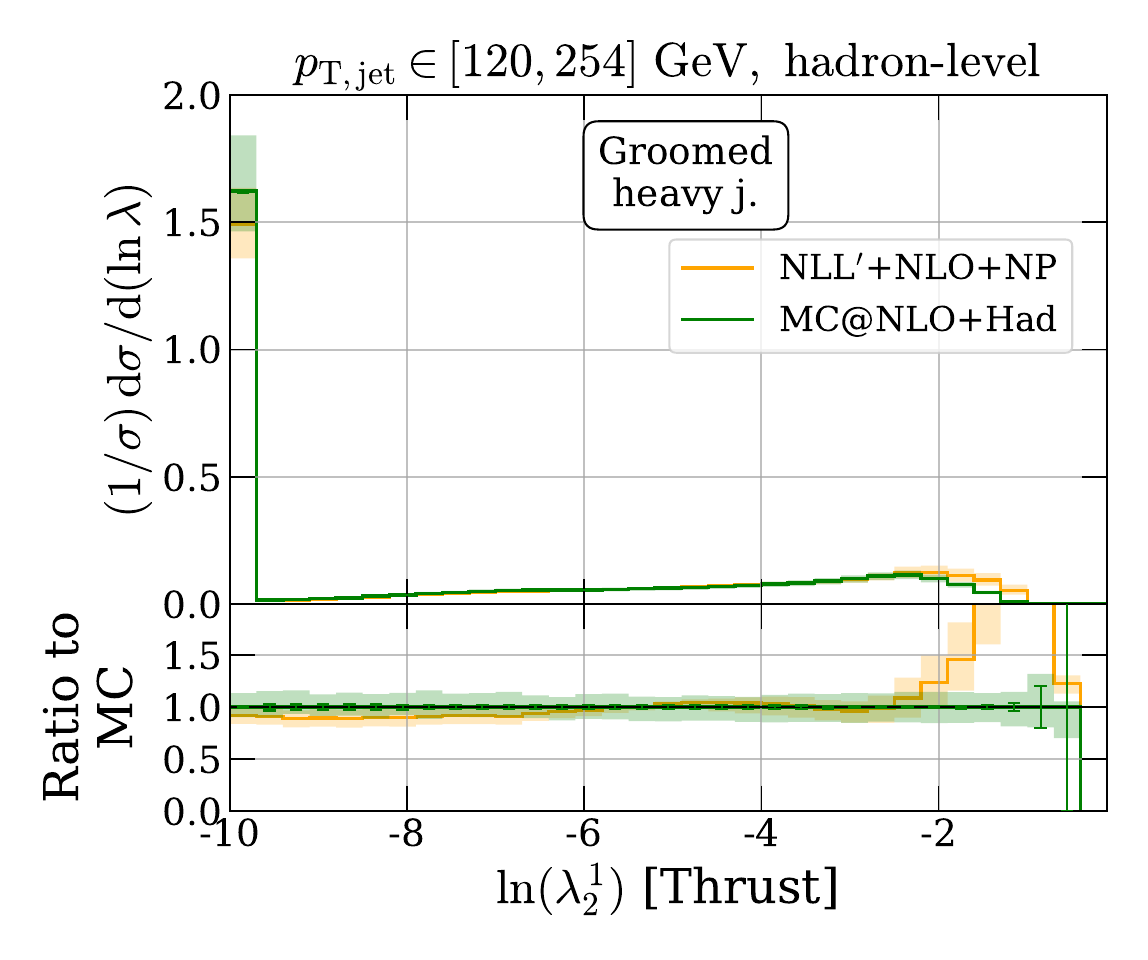}
    \caption{Comparison of our \NLOpNLLp+NP results for $b$-jets with predictions from the Monte Carlo event generator \sherpa\ at MC@NLO accuracy. The plots at the top are for standard jets, while the ones at the bottom for \softdrop-groomed jets. From left to right, we show the LHA, Width and Thrust angularities.}
    \label{fig:comparison_MC}
\end{figure}

The most striking difference between the $b$-jet and light-jet spectra is the presence, in the former, of a pronounced peak in the first bin. This is particularly evident for groomed jets.
In the following, we aim to better understand the origin of this feature.
We start by comparing our resummed and matched predictions for $b$-jets with those from a general-purpose Monte Carlo event generator, namely \sherpa\ at MC@NLO accuracy. Figure~\ref{fig:comparison_MC} shows this comparison for a representative $\ptjet$ slice. The upper panels correspond to standard jets, while the lower ones show \softdrop jets. As before, the three angularities $\alpha=\frac{1}{2},1,2$ are displayed from left to right.
Across all panels, we find good agreement between our calculation, supplemented with non-perturbative corrections, and the particle-level simulation, particularly at small $\lambda^1_\alpha$. This confirms the presence of a peak-like structure in the first bin. 

\begin{figure}
\centering
\includegraphics[width=0.32\linewidth]{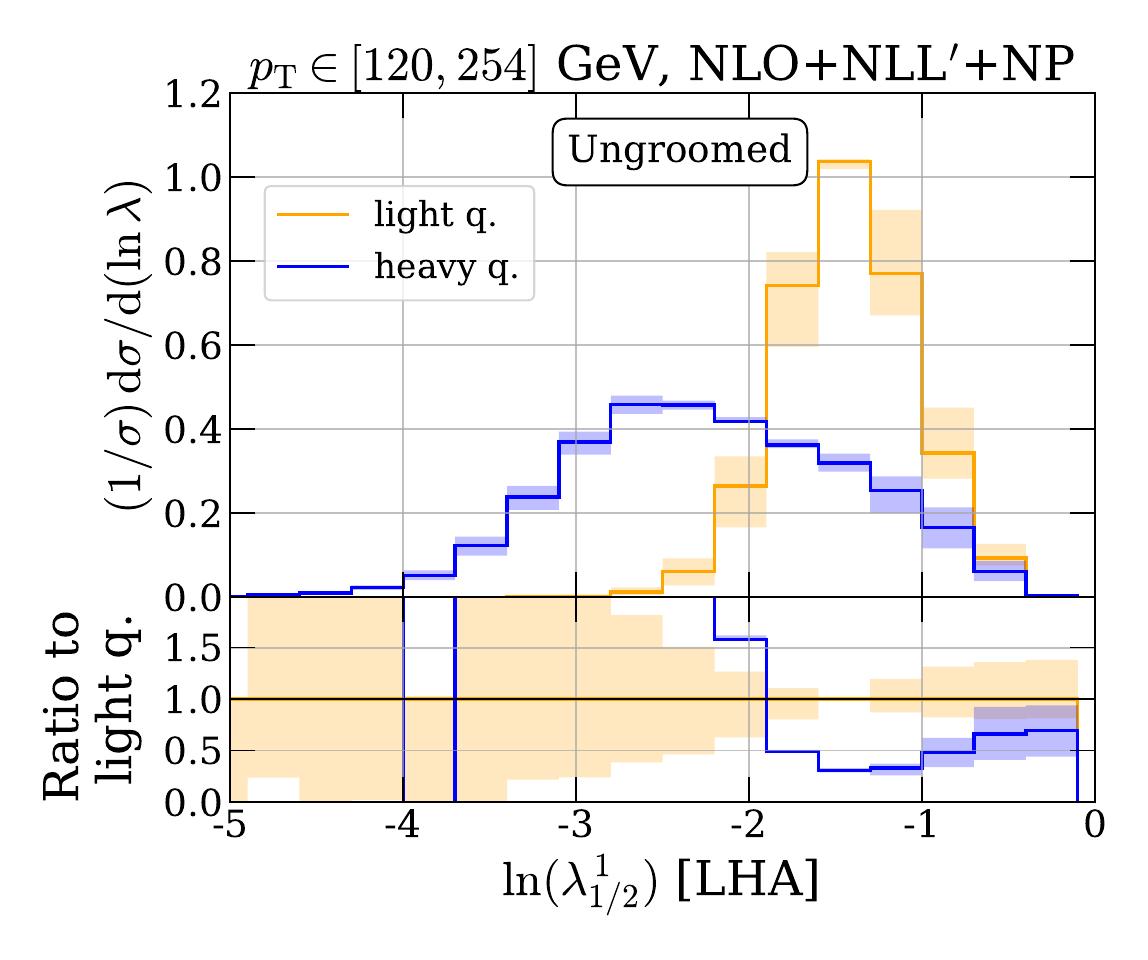}
\includegraphics[width=0.32\linewidth]{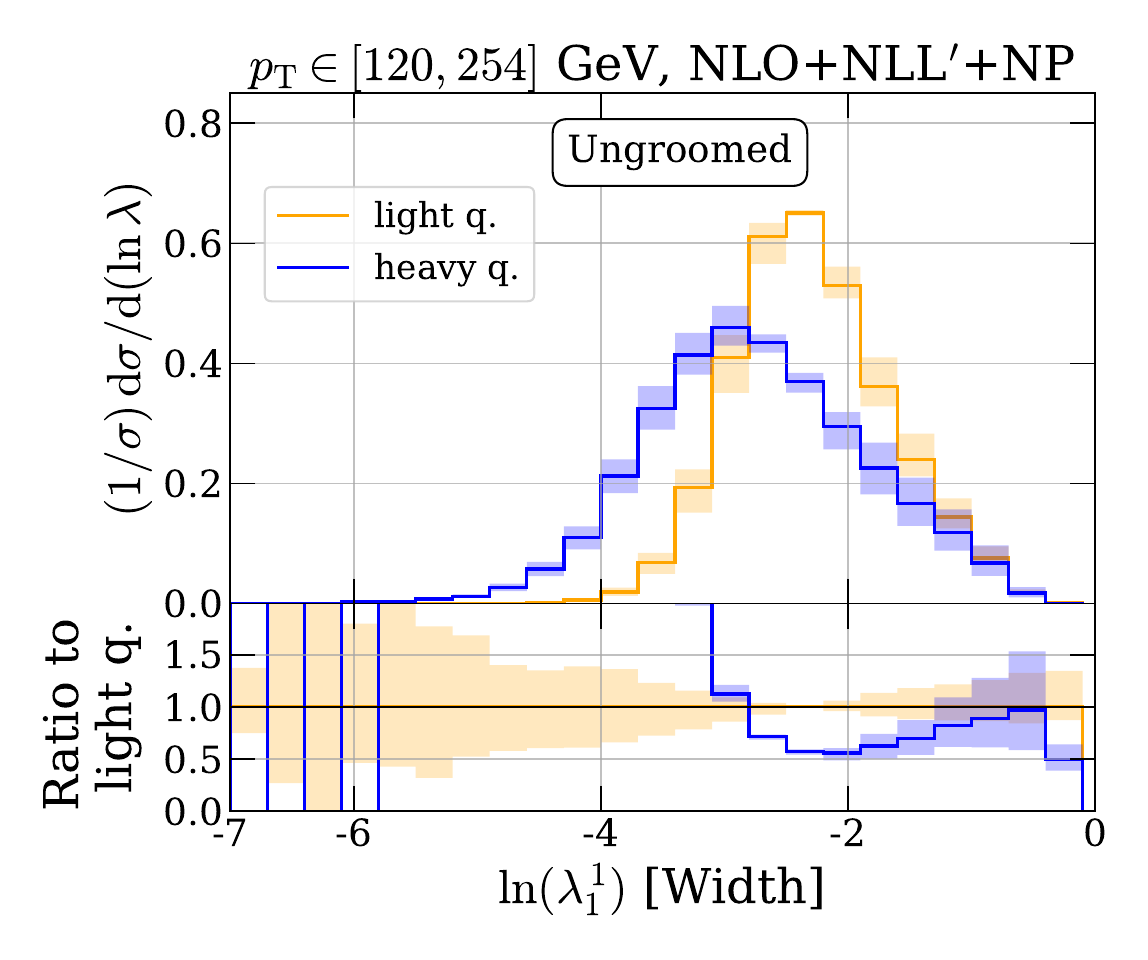}
\includegraphics[width=0.32\linewidth]{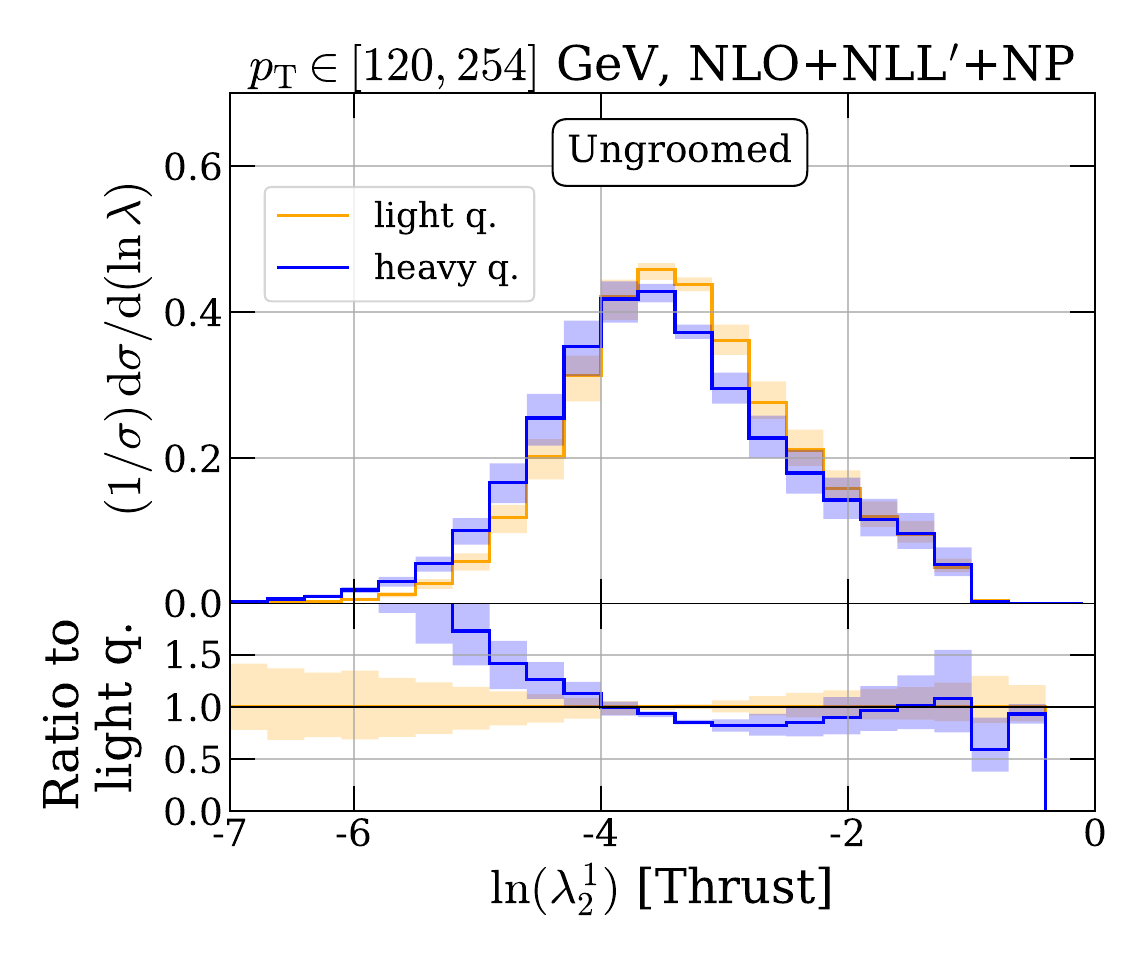}
\includegraphics[width=0.32\linewidth]{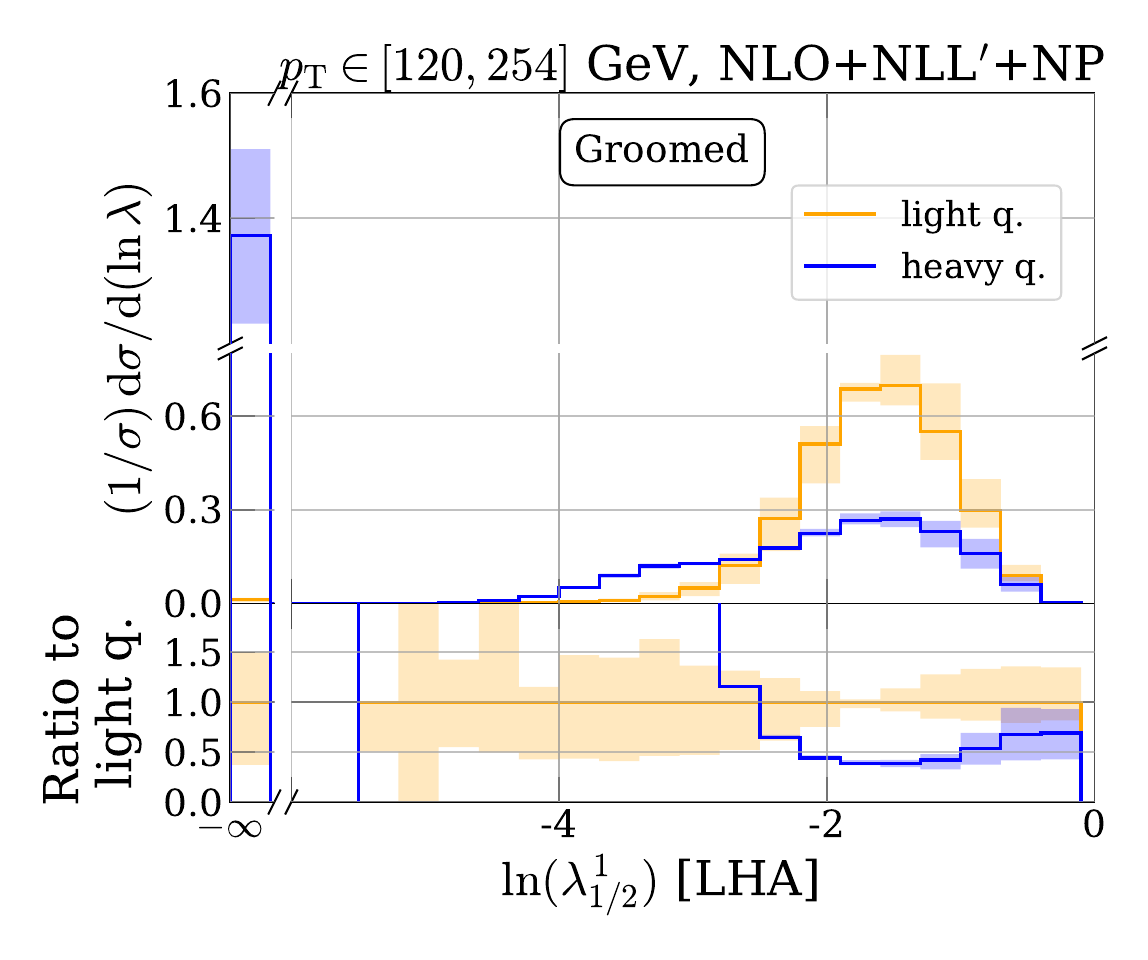}
\includegraphics[width=0.32\linewidth]{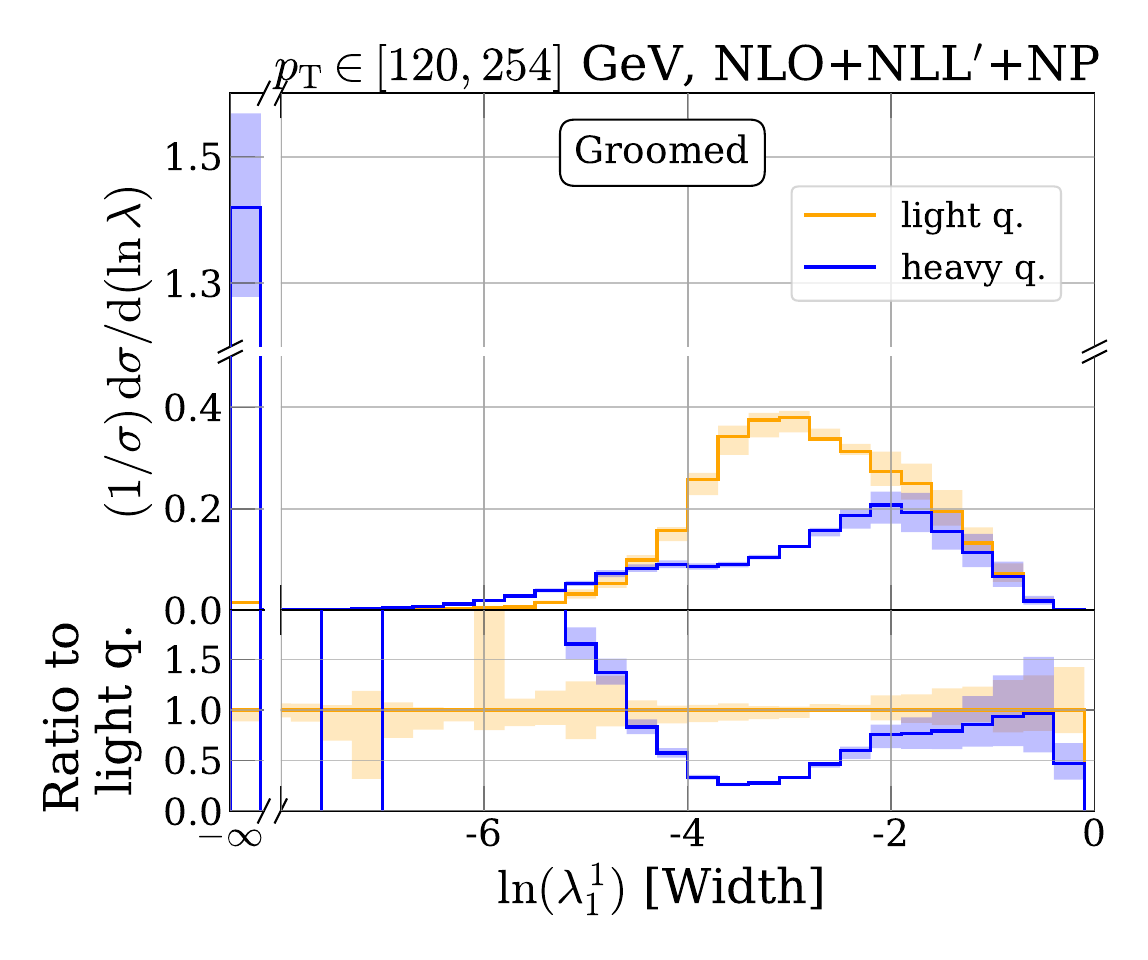}
\includegraphics[width=0.32\linewidth]{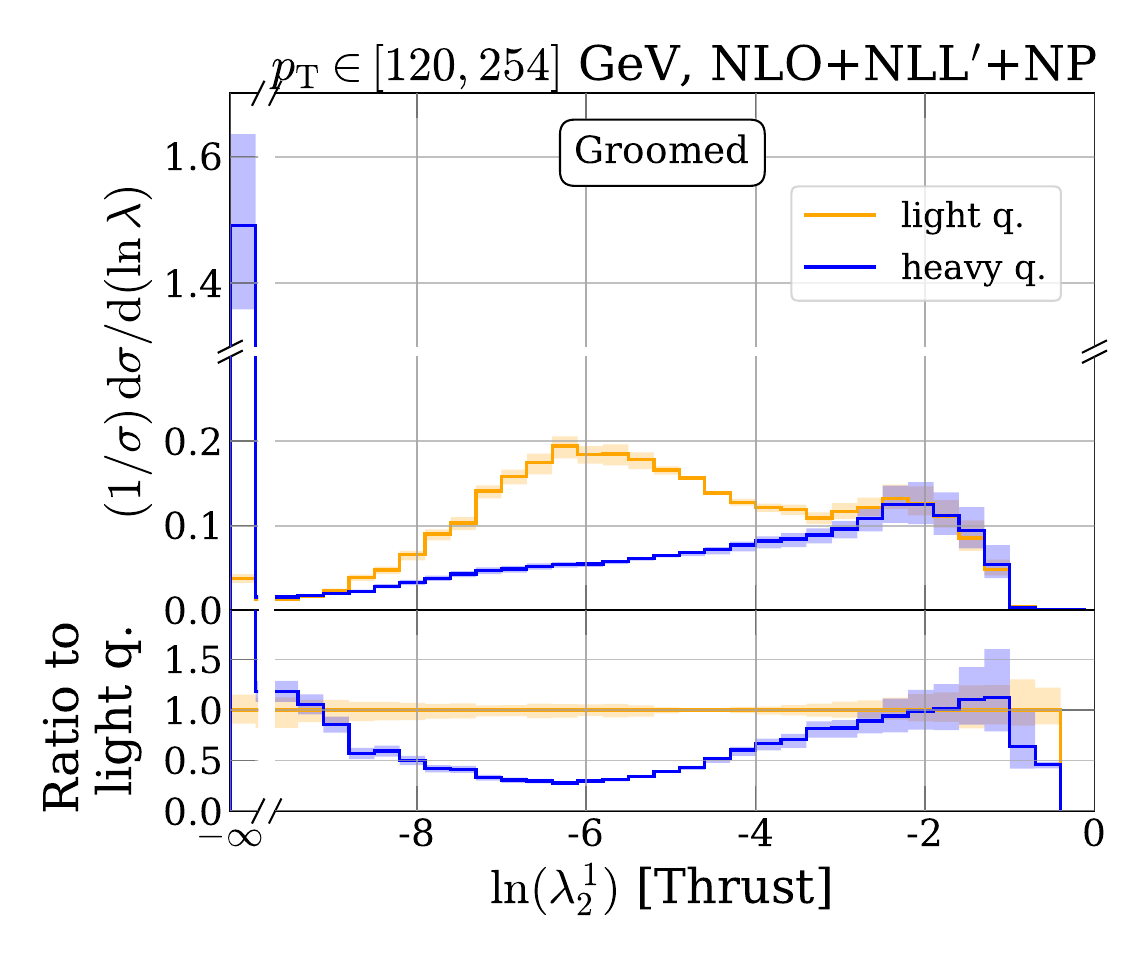}
\caption{Comparison between jet angularities measured on $b$-jets (with undecayed $B$-hadrons) and on a light-quark jet sample using fine bins. The plots at the top are for standard jets, while the ones at the bottom for \softdrop-groomed jets. From left to right, we show LHA, Width and Thrust angularities.}
    \label{fig:heavy_vs_light_fine}
\end{figure}

After validating our predictions for $b$-jet angularity distributions against detailed Monte Carlo simulations, it is instructive to return to the comparison between $b$-quark and light-quark jets using a finer binning, in order to better highlight the differences between the two spectra. This is done in Fig.~\ref{fig:heavy_vs_light_fine}, where we present the same results as in Figs.~\ref{fig:heavy_vs_light_coarse} and~\ref{fig:heavy_vs_light_coarse_SD}, but with increased bin resolution. We focus on the intermediate $\ptjet$ slice, although similar conclusions hold in the low- and high-$\ptjet$ regions.
The upper panels correspond to standard jets. The finer binning allows us to clearly resolve the transition at the value of the angularity given in eq.~(\ref{eq:transpoint}). Above this point, the $b$-jet and light-jet spectra are very similar, while they begin to diverge as one moves across the transition region. As noted in Ref.~\cite{Dhani:2024gtx}, the differential distribution at \NLL accuracy exhibits a discontinuity at this point; however, matching to fixed order (together with finite bin widths) smoothes out this feature.

The lower panels of Fig.~\ref{fig:heavy_vs_light_fine} show the corresponding results for groomed jets. At moderate and large values of $\lambda^1_\alpha$, the behaviour closely mirrors that observed for standard jets, with a clear transition between regions where mass effects are negligible and where they significantly impact the spectrum. However, an additional striking feature appears: a pronounced spike in the first bin, which helps explain the structure already observed with coarser binning.
As discussed in Sec.~\ref{sec: NLL SD}, this effect originates already at parton level because the combined effect of grooming and the dead cone closes up the (logarithmic) phase space. As a result, even at resummed level, we find a delta-function contribution $\delta(\lambda^1_\alpha)$, corresponding to jets composed of a single parton. Non-perturbative effects smear this contribution to some extent, but configurations remain in which, after \softdrop grooming, the jet contains only the $B$ hadron, leading to the observed enhancement near zero. This effect could be removed by applying \softdrop in tagging mode, i.e.\ rejecting jets that fail the \softdrop condition, as proposed in Ref.~\cite{Caletti:2023spr}. In such a case, it would then be natural to consider ratios of $b$- and light-jet distributions, which would better emphasize shape differences at finite $\lambda^1_\alpha$ while reducing theoretical uncertainties.

We conclude this section by noting that it is worth revisiting our earlier predictions~\cite{Caletti:2021oor,Reichelt:2021svh} for inclusive jets in light of the framework developed here for jets initiated by massive quarks. In particular, we can now provide predictions for inclusive jets that consistently incorporate contributions from massive $b$ quarks. In addition, heavy-quark thresholds can be included in the evaluation of the running-coupling integrals entering the radiator. A comparison of these improved predictions with CMS data~\cite{CMS:2021iwu} is presented in App.~\ref{app:CMS}.
\\

\section{Conclusions and Outlook}  \label{sec:conclusion}

We have presented a phenomenological study of jet angularities for heavy-flavour jets in hadronic $Z+$jet production, considering both ungroomed and \softdrop-groomed jets. Our predictions, which are obtained with the \sherpa~resummation plugin, combine  fixed-order calculations at NLO with resummation at NLL$'$ accuracy, supplemented by non-perturbative corrections obtained through a transfer-matrix approach. 
A central aspect of this work is the inclusion of finite quark-mass effects, both in the resummation and in the fixed-order matching, allowing for a direct assessment of their impact on experimentally relevant observables. In particular, the resummation is performed within a variable-flavour-number scheme for the running coupling. 

Our findings provide a phenomenological validation of the picture developed in Ref.~\cite{Dhani:2024gtx}, where jet angularities and related observables were shown to be sensitive probes of heavy-quark mass effects and, in particular, of the dead-cone suppression of collinear radiation. In the present work, these features are confirmed in a realistic hadron-collider environment, including fiducial cuts, matching to fixed order, and non-perturbative corrections.

When considering standard, i.e.\ ungroomed, jets, the $b$-quark mass reduces the logarithmic enhancement of the spectrum, replacing the double-logarithmic behaviour of the massless case with single-logarithmic contributions. For groomed observables, the interplay between the dead-cone effect and the \softdrop\ condition results in a partial closure of the phase space for emissions. This gives rise to a non-vanishing probability for configurations with very small angularity, which is reflected in a pronounced enhancement in the first bin of the distribution once non-perturbative effects are included. These features are qualitatively reproduced by parton-shower simulations, providing a useful cross-check of the analytical framework.

At the particle level, we find that the separation between heavy- and light-flavour jets depends on both the angularity exponent and the jet transverse momentum. The discrimination is most pronounced for smaller angular exponents, e.g.\ $\alpha=\frac{1}{2}$ and, to some extent, $\alpha=1$, and moderate transverse momenta, where mass effects remain sizeable, while it gradually diminishes at higher transverse momenta, where the spectra approach the massless limit. Grooming tends to enhance sensitivity to the collinear region and therefore retains some discriminating power even in regimes where ungroomed observables become less sensitive.

As a byproduct of our study on $b$-tagged jets, we have been able to revisit and improve our previous predictions for inclusive jets by including the $b$-jet component and the treatment of quark-mass thresholds in the running coupling. Our updated comparison with the existing CMS measurement~\cite{CMS:2021iwu} indicates that heavy-flavour contributions remain subleading in inclusive jet samples, and therefore do not significantly modify the overall description of the data. Nevertheless, the framework developed here provides a consistent basis for studying heavy-flavour jet substructure in more differential or flavour-tagged measurements.

Several directions for future work suggest themselves. On the theoretical side, extending the accuracy of the resummation and improving the control over matching corrections would help to reduce residual uncertainties, particularly in transition regions. A more systematic investigation of power corrections, including their dependence on the heavy-quark mass, would also be of interest. From a phenomenological perspective, it would be natural to extend the present analysis to more complex final states, such as $Z+b\bar{b}$ production, as well as to jets originating from top-quark decays. 
For this purpose, we have collected the necessary theoretical ingredients for these analyses in an appendix. 
In addition, alternative observable definitions or grooming/tagging strategies could be explored to reduce the contribution from single-particle configurations and enhance sensitivity to shape differences at finite angularity.

We look forward to upcoming measurements of heavy-flavour jet substructure at the LHC, which will provide an opportunity to test our predictions more directly.
Such comparisons will help identify aspects of the theoretical description of heavy-quark fragmentation that may require further refinement. This has broader implications for a range of studies within the LHC physics programme, including precision tests of the Yukawa sector through Higgs-boson decays into heavy-flavour (sub)jets.

\paragraph{Acknowledgements:} 
First and foremost, we would like to thank our long-standing collaborator Gregory Soyez for many useful discussions. 
We also acknowledge fruitful exchanges with colleagues who are currently measuring $b$-jet substructure at the LHC: Leticia Cunqueiro Mendez, Matt Leblanc, Ezra Lesser, Alberto Rescia, Jennifer Roloff, and Federico Sforza.

The work of AG is supported by the Excellence Cluster ORIGINS, funded
by the Deutsche Forschungsgemeinschaft (DFG, German Research Foundation) under Germany’s
Excellence Strategy — EXC-2094-390783311 and by a TUM Global Postdoc Fellowship.
The work of LM and SM was supported by the Italian Ministry of Research (MUR) under grant PRIN 2022SNA23K funded by the European Union -- Next Generation EU, Mission 4, Component 2, CUP D53D23002880006. SS and LS are grateful for financial support from the German Federal Ministry of Research, Technology, and Space (projects 05D23MG1 and 05H24MGA).
DR is supported by the European Union under the HORIZON program in Marie Sk{\l}odowska-Curie Project No. 101153541.
AG and LS are grateful to the COMETA COST Action CA22130 for the financial support provided during the course of this project. 

\appendix 
\section{Kinematics of the emission}
\label{app: kinematics}
In this brief appendix we derive the kinematic constraints for a single-gluon emission off a quark with mass $m$ in the soft and collinear limits, which enter the expressions of the radiators in eq.~\eqref{eq:radiator_ungromeed_massive} for the ungroomed case and eq.~\eqref{eq:radiator_gromeed_massive} for \softdrop\ grooming. Given that the jet is initiated by a $b$-quark, we work in the quasi-collinear limit to obtain the kinematics of the gluon emission. Within this approximation, the azimuth–rapidity distance and the momentum fraction of the emission are given by
\begin{align}
    \Delta R_k^2 = 4e^{-2 \eta}\cosh \yjet + \frac{m^2}{\ptjet^2}, 
    \qquad 
    z = \frac{p_{T_k}}{\ptjet} = \frac{\kt e^{\eta}}{2 \ptjet \cosh \yjet},
\end{align}
where $\eta$ and $\kt$ denote, respectively, the rapidity and transverse momentum of the emission with respect to the emitting leg, and $p_{T_k}$ the transverse momentum of the emission with respect to the beam.  
$\Delta R_k$ is the standard azimuth--rapidity distance between the jet axis and the emission.
From the expression of $\Delta R_k^2$ one can extract a lower bound on the rapidity:
\begin{align}
    \eta_{\text{min}} = \ln\!\left(\frac{2 \cosh \yjet}{R_0}\right) 
    + \order{\frac{m^2}{\ptjet^2 R_0^2}}.
\end{align}

Within this parametrization, the definition of the angularity, cf. eqs.~\eqref{eq:jet_ang} and \eqref{eq: CAESAR parametrization}, becomes
\begin{align}
\label{eq: angularity def}
    \frac{p_{T_k}}{\ptjet}\left(\frac{\Delta R_k}{R_0}\right)^\alpha 
    = d(\mu_Q)g(\phi)\,\frac{\kt}{\mu_Q}\,
    e^{-(\alpha-1)\eta}\,
    \left(\frac{\kt^2+ z^2 m^2}{\kt^2}\right)^{\frac{\alpha}{2}},
\end{align}
with
\begin{align}
    d(\mu_Q) g(\phi) = \frac{\mu_Q}{\ptjet R_0}\, e^{(\alpha-1)\eta_{\text{min}}}.
\end{align}
Consequently, the radiator associated with a single gluon emission takes the form
\begin{align}
\label{eq: ungroomed radiator}
    \mathcal{R}_{b} =&  \int^1_0 \de z\int^{\mu^2_Q}_0 \frac{\de \kt^2}{\kt^2+z^2 m^2}\;
    \frac{\as^{\text{CMW}}(\kt^2)}{2\pi}\,P_{gb}(z, \kt^2)\,
    \Theta(\eta-\eta_{\text{min}})\times \nonumber \\
    &\Theta\!\left(
    d(\mu_Q)g(\phi)\frac{\kt}{\mu_Q} 
    e^{-(\alpha-1)\eta}
    \left(\frac{\kt^2+ z^2 m^2}{\kt^2}\right)^{\alpha/2}
    - \lambda^1_\alpha\right).
\end{align}
However, as shown in~\cite{Ghira:2023bxr, Caletti:2023spr, Dhani:2024gtx}, 
at NLL accuracy the radiator in eq.~\eqref{eq: ungroomed radiator} reduces to
\begin{align}
    \mathcal{R}_{b} \xrightarrow[]{\text{NLL}} R_{b},
\end{align}
with
\begin{align}
    R_b = \int^1_0 \de z& \int^{\mu^2_Q}_{z^2 m^2}\frac{\de \kt^2}{\kt^2} 
   \; \frac{\as^{\text{CMW}}(\kt^2)}{2\pi}\,
    P_{gb}(z,\kt^2-z^2m^2) \Theta(\eta-\eta_{\text{min}})\times\nonumber\\
    &\Theta\!\left(
    d(\mu_Q)g(\phi)\frac{\kt}{\mu_Q} e^{-(\alpha-1)\eta}-\lambda^1_{\alpha}\right) 
    .
\end{align}
Thus, at NLL accuracy one can still employ the parametrization of eq.~\eqref{eq: CAESAR parametrization} even for heavy-flavour jets.  

For the groomed radiator, the \softdrop\ condition imposes an additional constraint,
\begin{align}
    z \geq \zc \left(\frac{\Delta R_k}{R_0}\right)^\beta,
\end{align}
which can be recast as
\begin{align}
    \frac{\kt e^{(1+\beta)\eta}}{2 \ptjet \cosh\yjet}
    \geq \zc'\,
    \left(\frac{\kt^2+z^2 m^2}{\kt^2}\right)^{\beta/2},
\end{align}
with $\zc'$ defined in eq.~\eqref{eq: zcut prime}.  
The same reasoning as in the ungroomed case applies here, following the analysis of~\cite{Dhani:2024gtx}, the groomed radiator defined as
\begin{align}
    &\bar{\mathcal{R}}_b= \int^1_0 \de z\int^{\mu^2_Q}_0 \frac{\de \kt^2}{\kt^2+z^2 m^2} \;
    \frac{\as^{\text{CMW}}(\kt^2)}{2\pi}\,P_{gb}(z, \kt^2)\,
    \Theta(\eta-\eta_{\text{min}}) \times\nonumber \\
    & \Theta\!\left(
    d(\mu_Q)g(\phi)\frac{\kt}{\mu_Q} 
    e^{-(\alpha-1)\eta}
    \left(\frac{\kt^2+ z^2 m^2}{\kt^2}\right)^{\alpha/2}
    - \lambda^1_\alpha\right)\Theta\left(\frac{\kt e^{(1+\beta)\eta}}{2 \ptjet \cosh\yjet}-\zc'\,
    \left(\frac{\kt^2+z^2 m^2}{\kt^2}\right)^{\beta/2}\right),
\end{align}
can be rewritten at NLL accuracy as
\begin{align}
    \bar{R}_b =\int^1_0 \de z &\int^{\mu^2_Q}_{z^2 m^2}\frac{\de \kt^2}{\kt^2} 
    \;\frac{\as^{\text{CMW}}(\kt^2)}{2\pi}\,
    P_{gb}(z,\kt^2-z^2m^2)\,\Theta(\eta-\eta_{\text{min}}) \times\nonumber  \\
    &\Theta\!\left(
d(\mu_Q)g(\phi)\frac{\kt}{\mu_Q} e^{-(\alpha-1)\eta}-\lambda^1_{\alpha}\right) \Theta\left(\frac{\kt e^{(1+\beta)\eta}}{2 \ptjet \cosh\yjet}-\zc'\right)
    .
\end{align}
\section{Dipole calculation for the soft function}
\label{app:dipoles}

In this appendix, we collate the results needed to determine the global part of the soft function $\mathcal{S}$. For 
the resummed predictions of $b$-jet angularities in $pp\to Z+b$ production we need to consider the $bg\to Z+b$ partonic channel. However, in view of future phenomenological studies of $b$-jets,  in Sec.~\ref{app:dipoles-Zbb} we also 
compile results for the $pp\to Z+b\bar{b}$ process, i.e. the $gg\to Z+b\bar{b}$ and $q\bar{q}\to Z+b\bar{b}$ partonic channels. 

\subsection{$bg\to Z+b$}
\label{app:dipoles-Zb}
We parametrize the momenta of the hard QCD legs as follows:
\begin{align}
    &p_1= \frac{\sqrt{s}}{2} x_1\left(1,0,0,v_1\right), \nonumber \\
    &p_2= \frac{\sqrt{s}}{2} x_2\left(1,0,0,-v_2\right), \nonumber \\
    & p_3= \left(m_{T_3} \cosh y_3, p_{T_3},0,m_{T_3}\sinh y_3\right),
\end{align}
with $x_{1,2}$ denoting the fractions of the proton-beam energies carried by the initial-state partons, $s$ the centre-of-mass energy squared of the colliding beams, and $y_3$ the rapidity of the observed jet. The transverse momentum of the jet is denoted with $p_{T_3}$ and $m_{T_3} = \sqrt{m_3^2+p_{T_3}^2}$ is its transverse mass. Additionally, we introduced the velocities $v_1, v_2$ of the incoming particles, which read
\begin{align}
    v_i= \sqrt{1- \frac{4 m_i^2}{s x_i^2}}, \quad i=1,2.
\end{align}
For the process of interest, we have $m_1=m_3=m$ and $m_2=0$.
The soft-gluon momentum is instead parametrized as
\begin{align}
    \quad k = p_{T_k} \left(\cosh \eta_{k}, \cos \phi_{k}, \sin \phi_{k}, \sinh\eta_{k}\right),
\end{align}
with $p_{T_k}, \eta_k , \phi_k$ measured with respect to the beam direction.

We start by considering the eikonal factor
\begin{equation}\label{eq:w-massive}
    w_{ij}= \frac{p_i\cdot p_j}{p_i\cdot k \, p_j\cdot k}-\frac{p_i^2}{2 (p_i \cdot k)^2}-\frac{p_j^2}{2 (p_j \cdot k)^2},
\end{equation}
where $p_i,p_j$ are the momenta of the dipole constituents.
The contribution of the dipoles to the differential distribution of the angularity variable $\lambda^1_\alpha$ reads
\begin{align}
    \frac{\de \sigma_{ij}}{\de \ln \lambda^1_\alpha}=\frac{\as C_{ij}}{2\pi}\; \lambda^1_{\alpha}\int^{p_{T_3}}_0 \de p_{T_k} \, p_{T_k}\int^{\infty}_{\eta_{\text{min}}}\de \eta_k \int^{2\pi}_0\frac{\de \phi_k}{2\pi} w_{ij}\, \delta\left(\frac{p_{T_k}}{p_{T_3}} \left(\frac{\Delta R_k}{R_0}\right)^\alpha-\lambda^1_{\alpha}\right),
\end{align}
with $\Delta R_k =\sqrt{(\eta_k-y_3)^2+ \phi_k^2} $ and $C_{ij}$ the colour factor for the dipole $(ij)$.

\begin{figure}[tb]
    \centering
    \includegraphics[width=0.45\linewidth]{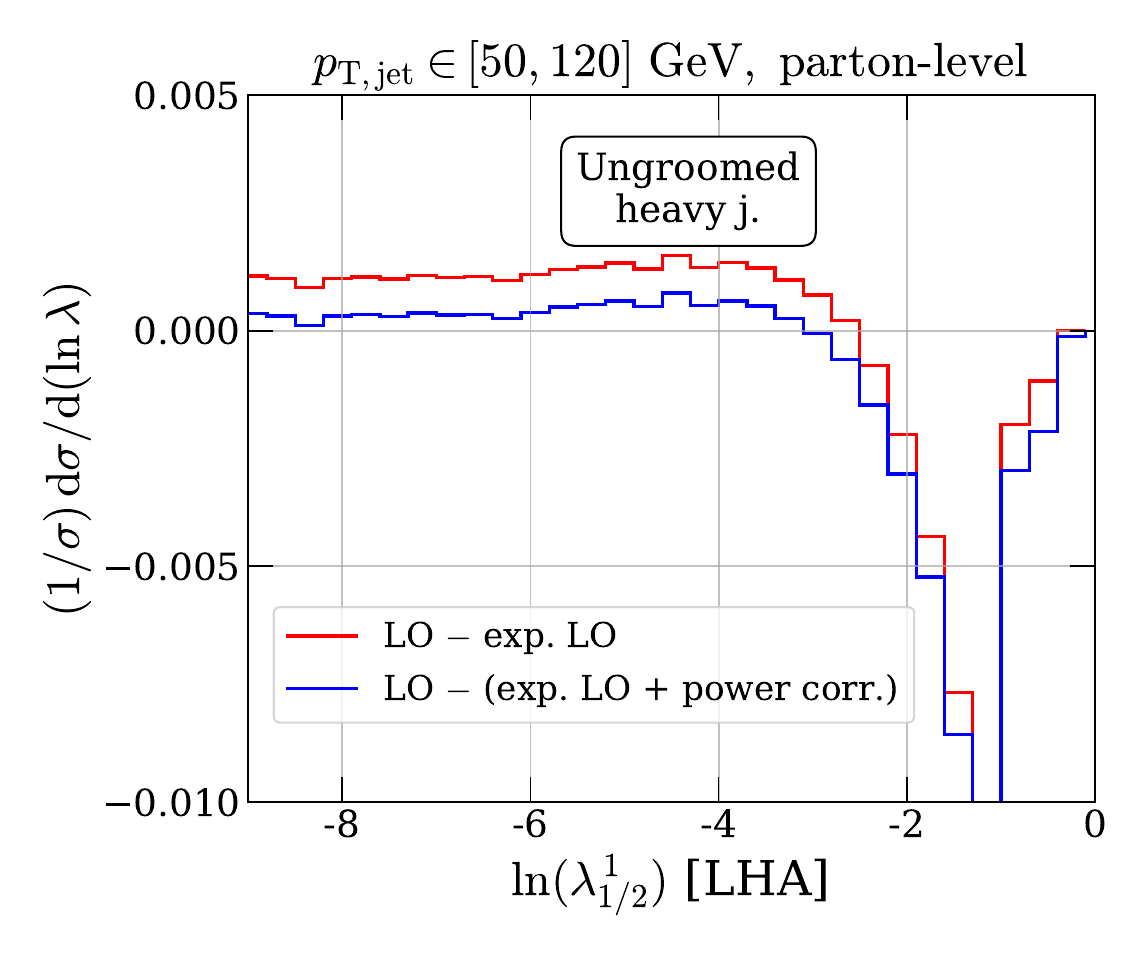}      
    \includegraphics[width=0.45\linewidth]{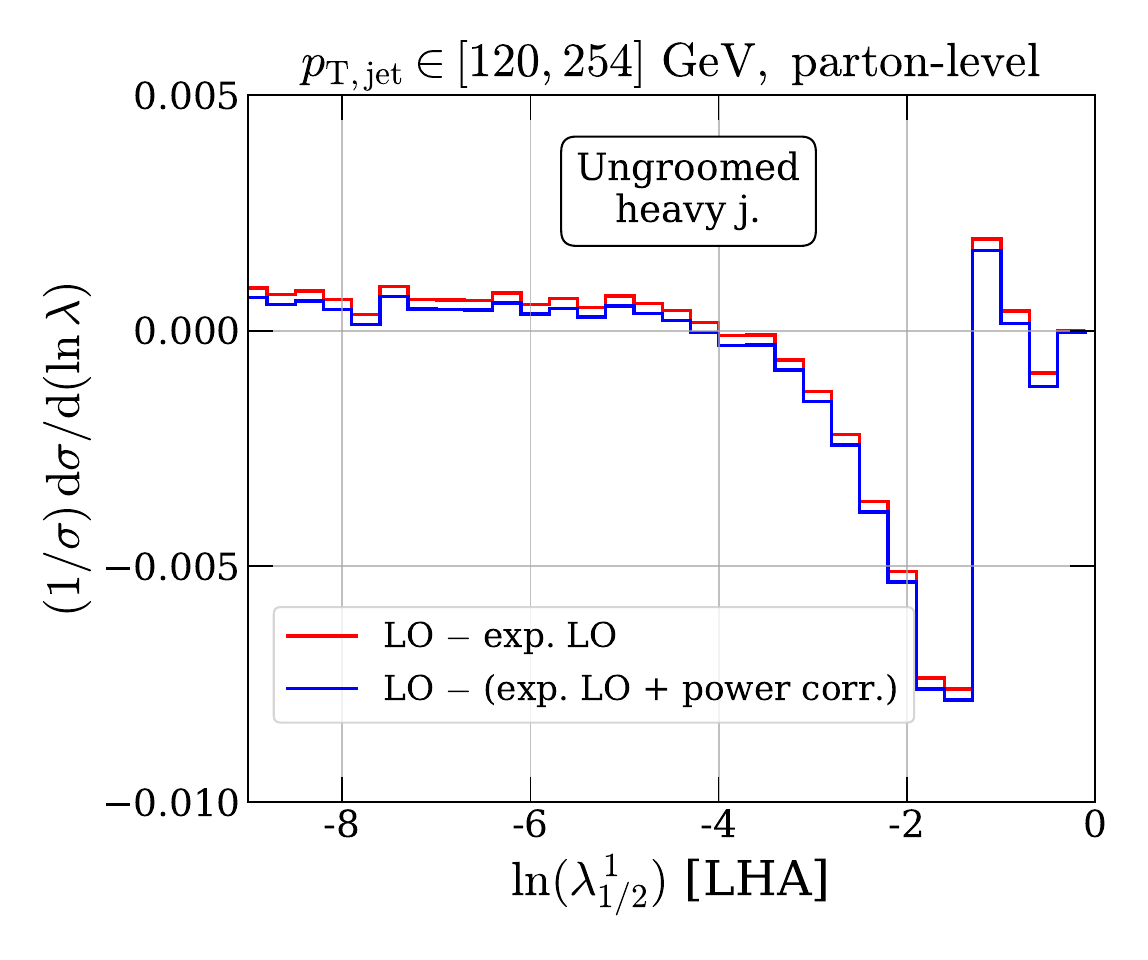}   \\ 
    \includegraphics[width=0.45\linewidth]{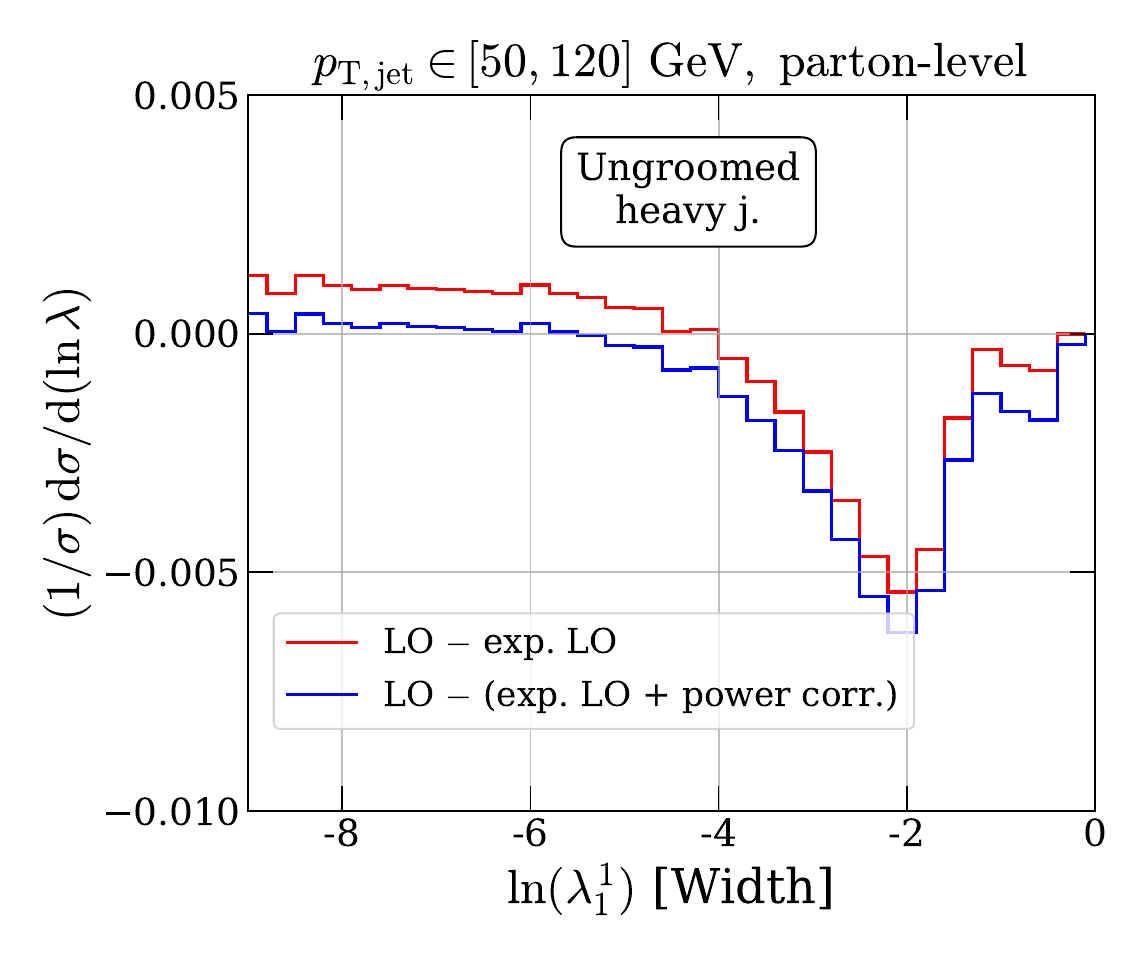}      
    \includegraphics[width=0.45\linewidth]{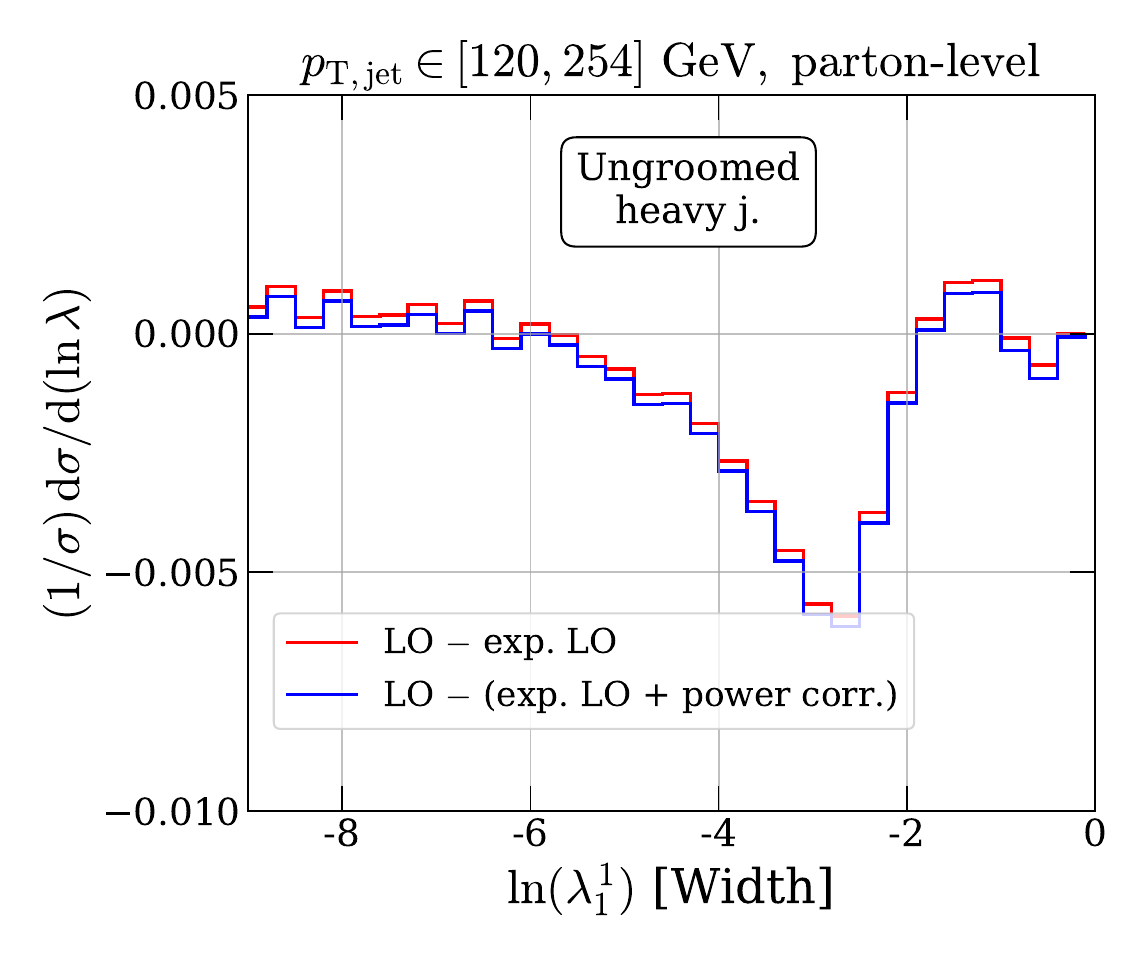}   \\
    \includegraphics[width=0.45\linewidth]{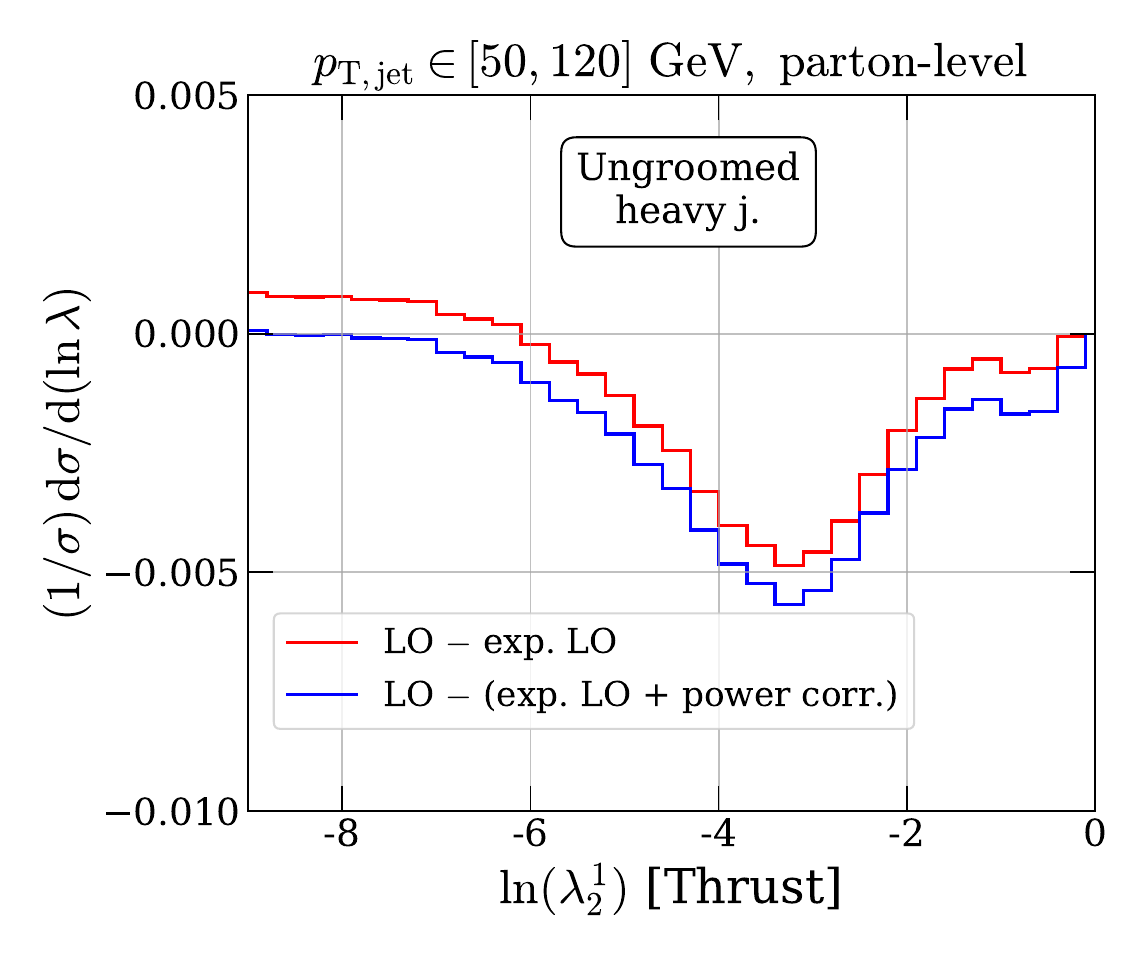}      
    \includegraphics[width=0.45\linewidth]{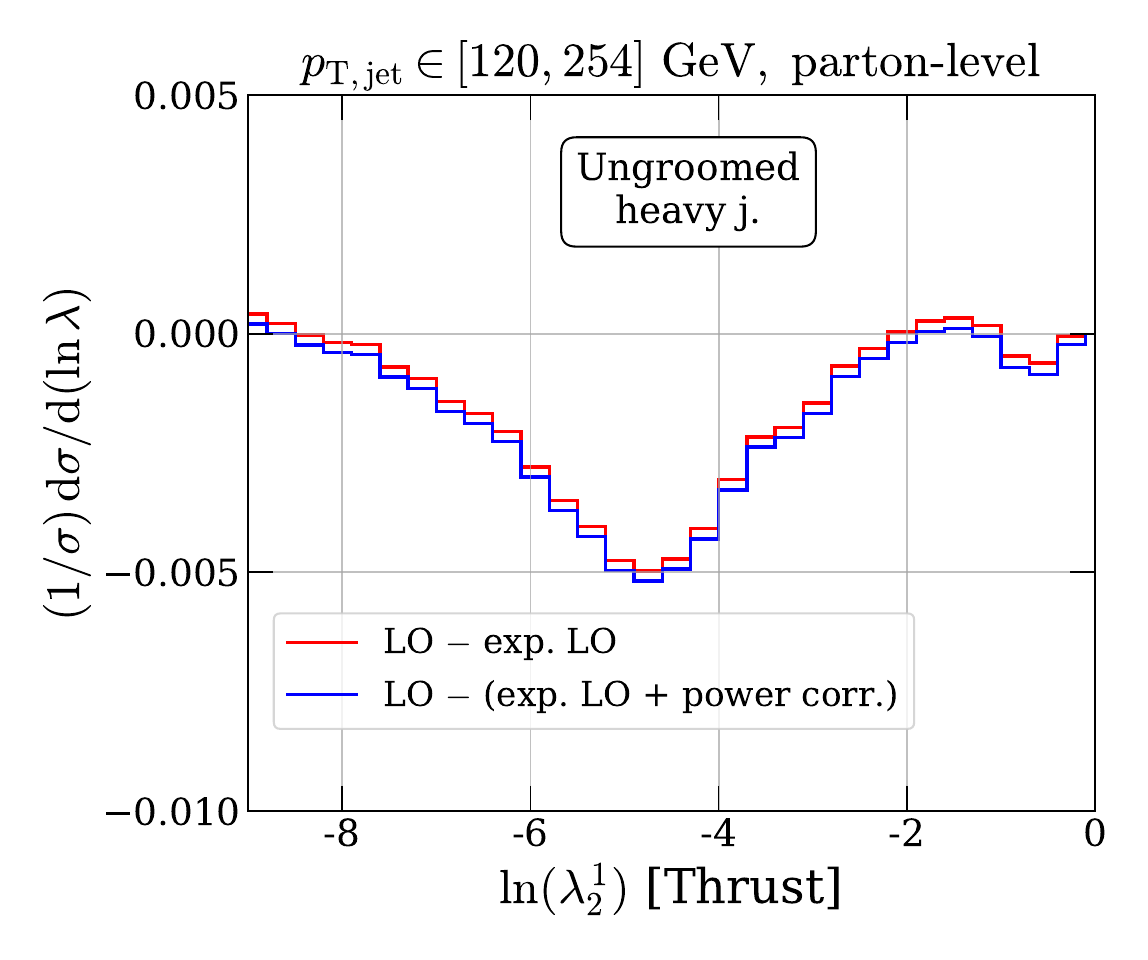}   \\
    \caption{Difference of expanded resummation to fixed order, with and without the inclusion of power corrections at LO. 
    }
    \label{fig:PowerCorrections}
\end{figure}

For each dipole configuration, we perform the following change of variables
\begin{align}
    \eta_k-y_3= \Delta R_k  \cos\psi\,,\quad  \phi_k= \Delta R_k  \sin\psi\,.
\end{align}
The integral over $p_{T_k}$ can be performed trivially using the $\delta$-function. The $\Delta R_k $ and $\psi$ integration cannot be done in closed form. However, we can consider the quasi-collinear limit, i.e.\ we consider $\Delta R_k^2, \frac{m^2}{p^2_{T_3}}, \frac{m^2}{s x_1^2}$ to be small, but of the same order. Within this approximation and expanding the result in powers of the jet radius, we obtain

\begin{subequations}
\begin{align}
\frac{\de \sigma_{12}}{\de \ln \lambda^1_\alpha}=&\;\frac{\as}{2\pi}\ca R_0^2\Bigg[1- 2e^{2y_3} \frac{m^2}{s x_1^2} +\dots\Bigg],\\
    \frac{\de \sigma_{13}}{\de \ln \lambda^1_\alpha}=&\;\frac{\as}{2\pi}\left(2 \cf-\ca\right)\Bigg[-\ln \xi_3 -1 + \frac{R_0^2}{288}\left(72+R_0^2-216 e^{2y_3} \frac{m^2}{s x_1^2}\right)\nonumber \\
    &\qquad\qquad\qquad\qquad+\frac{ \xi_3}{36}\left(72+18 R_0^2-72 e^{2 y_3} \frac{m^2 R_0^2}{s x_1^2}\ln \xi_3 \right) +\dots\Bigg],  \\
     \frac{\de \sigma_{23}}{\de \ln \lambda^1_\alpha}=&\;\frac{\as}{2\pi}\ca\Bigg[-\ln \xi_3 -1 + \frac{R_0^2}{288}\left(72+R_0^2\right)
    +\frac{ \xi_3}{36}\left(72+18 R_0^2\right)+\dots \Bigg],
\end{align}
\end{subequations}
where the dots indicate higher-order contributions $\mathcal{O}\left(\xi_3 R_0^4,\,\xi_3 \frac{m^4}{s^2},\, \xi_3^2 R_0^2,\, \xi_3^2 \frac{m^2}{s} \right)$ with $\xi_3= \frac{m^2}{p^2_{T_3} R_0^2}$.
We observe that the leading contribution coincides with the expansion of the Sudakov form factor calculation in the small-$\lambda^1_\alpha$ limit. The massless calculation cannot be recovered by taking the small-mass limit due to the non-commutativity between massless and massive results, discussed at length in \cite{Gaggero:2022hmv,Ghira:2023bxr}. The impact of the power corrections in the mass at $\order{\as}$ is shown in Fig.~\ref{fig:PowerCorrections}. We examine two different $\ptjet$ windows, namely $\ptjet \in [50,120]~\mathrm{GeV}$ and $\ptjet \in [120,254]~\mathrm{GeV}$.
For each value of $\alpha$ considered in the text, we show two curves. The red curve represents the difference between the fixed-order leading-order calculation and the $\mathcal{O}(\as)$ expansion of the resummed result. The blue curve instead corresponds to the difference between the fixed-order calculation and the $\mathcal{O}(\as)$ expansion of the resummation including mass power corrections.
In the small-observable limit at $\mathcal{O}(\as)$, the blue curves lie closer to zero than the red ones, as expected. This indicates that the inclusion of subleading power corrections in $m^2$ improves the agreement between the resummed expansion and the fixed-order result. We further observe that the agreement between the resummed expansion and the leading-order calculation worsens in the small $\ptjet$ range for larger values of $\alpha$, while the inclusion of power corrections does not alter this trend.
In the high-$\ptjet$ window, mass power corrections become essentially negligible: the red and blue curves are almost overlapped.

\subsection{$jj\to Z+b\bar b$}
\label{app:dipoles-Zbb}

Although our analysis focuses exclusively on $Z+b$ final states, the formalism presented here can be readily extended to the process $pp \to Z+b\bar{b}$. This channel is particularly relevant, as it plays an important role in upcoming measurements at the LHC, see e.g.~\cite{Rescia:2026ozo}.
The main difference between the two processes lies in the structure of the soft function: in the $Z+b\bar{b}$ case, a larger number of dipoles contribute. In particular, for dijet production recoiling against a colour singlet, one must account for six dipoles, with the additional complication that the final-state legs do not have back-to-back kinematics. 

We parametrize the momenta of the hard QCD legs as follows:
\begin{align}
    p_1 &=x_1 \frac{\sqrt{s}}{2}(1,0,0,v_1),\nonumber\\
    p_2 &=x_2 \frac{\sqrt{s}}{2}(1,0,0,-v_2),\nonumber\\
    p_3 &= \left(m_{T_3}\cosh y_3,p_{T_3},0,m_{T_3}\sinh y_3\right),\nonumber\\
    p_4 &=\left(m_{T_4}\cosh y_4,p_{T_4}\cos \phi_4,p_{T_4}\sin \phi_4,m_{T_4}\sinh y_4\right),\nonumber\\
\end{align}
where  $p_3$ is the momentum of the measured jet and $p_4$ is the momentum of the recoil jet with transverse momentum $p_{T_4}$, rapidity $y_4$, azimuth $\phi_4$, and transverse mass $m_{T_4}= \sqrt{m^2_4+ p^2_{T_4}}$.
The calculation of dipoles $(1,2)$, $(1,3)$ and $(2,3)$ is very similar to the case of $Z+b$ production.
Using the same assumptions as before, we obtain\clearpage

\begin{subequations}
\begin{align}
\frac{\de \sigma_{12}}{\de \ln \lambda^1_\alpha}=\;&\frac{\as}{2\pi} C_{12} R_0^2\Bigg[1- 2e^{2y_3} \frac{m_1^2}{s x_1^2} -2e^{-2y_3} \frac{m_2^2}{s x_2^2}+\dots\Bigg],\\
    \frac{\de \sigma_{13}}{\de \ln\lambda^1_\alpha}=&\;\frac{\as}{2\pi}C_{13}\Bigg[-\ln \xi_3 -1 + \frac{R_0^2}{288}\left(72+R_0^2-216 e^{2y_3} \frac{m_1^2}{s x_1^2}\right)\nonumber \\
    &\qquad\qquad+\frac{ \xi_3}{36}\left(72+18 R_0^2-72 e^{2 y_3} \frac{m_1^2 R_0^2}{s x_1^2}\ln \xi_3 \right) +\dots\Bigg],  \\
     \frac{\de \sigma_{23}}{\de \ln\lambda^1_\alpha}=&\;\frac{\as}{2\pi}C_{23}\Bigg[-\ln \xi_3 -1 + \frac{R_0^2}{288}\left(72+R_0^2-216 e^{-2y_3} \frac{m_2^2}{s x_2^2}\right)\nonumber \\
    &\qquad\qquad+\frac{ \xi_3}{36}\left(72+18 R_0^2-72 e^{-2 y_3} \frac{m_2^2 R_0^2}{s x_2^2}\ln \xi_3 \right) +\dots\Bigg],
\end{align}
\end{subequations}
where in the case at hand $m_{1,2}=0$ for gluon or light-quark induced channels or $m_{1,2}=m$ for the $b \bar b$ initiated channel, while $C_{ij}$ are the appropriate colour factors. As before, the dots denote power suppressed contributions. 
On the other hand, the dipoles involving the recoiling leg read
\begin{subequations}
\begin{align}
\frac{\de \sigma_{14}}{\de \ln\lambda^1_\alpha}= &\;\frac{\as}{2\pi}C_{14}\frac{R_0^2}{4 \left(\cosh \Delta y - \cos\Delta \phi\right)}  \Bigg[   \frac{4 e^{\Delta y}\left(\cosh \Delta y- \cos \Delta \phi\right)+ R_0^2 }{2\left(\cosh \Delta y - \cos\Delta \phi\right)} \nonumber \\
 &\qquad\qquad\qquad\qquad\qquad\qquad\qquad-\frac{4 m_1^2}{s x_1^2} e^{2y_3} \left(e^{\Delta y} -\cos \Delta \phi \right)+\dots\Bigg]\,,\\
 \frac{\de \sigma_{24}}{\de \ln\lambda^1_\alpha}= &\;\frac{\as}{2\pi}C_{24}\frac{R_0^2}{4 \left(\cosh \Delta y - \cos\Delta \phi\right)}  \Bigg[   \frac{4 e^{-\Delta y}\left(\cosh \Delta y- \cos \Delta \phi\right)+ R_0^2 }{2\left(\cosh \Delta y - \cos\Delta \phi\right)} \nonumber \\
 &\qquad\qquad\qquad\qquad\qquad\qquad\qquad- \frac{4 m_2^2}{s x_2^2} e^{-2y_3} \left(e^{-\Delta y} -\cos \Delta \phi \right)  +\dots\Bigg]\,,\\
 \frac{\de \sigma_{34}}{\de \ln\lambda^1_\alpha}=& \;\frac{\as}{2\pi} C_{34} \Bigg[-1-\ln\xi_3 + \frac{R_0^2}{4} \frac{\cosh\Delta y+ \cos\Delta \phi}{\left(\cosh\Delta y- \cos\Delta \phi\right)} +\frac{1}{2} \xi_3 \left(4+R_0^2\right)+\dots\Bigg].
\end{align}
\end{subequations}
We emphasize that the final result depends on the rapidity and azimuthal separation, $\Delta y = y_3 - y_4$ and $\Delta \phi = -\phi_4$, between the jet and the recoiling leg. In addition, we find that the mass of the recoiling leg enters only as a power-suppressed correction, as expected.
Finally, we briefly comment on the non-global configuration. For $Z+b\bar b$ production, our result agrees with that of Ref.~\cite{Dhani:2024gtx}, up to power-suppressed corrections in the jet radius and the heavy-quark mass. In our implementation of the soft function, the dependence on $\Delta y$ and $\Delta \phi$ has been incorporated within the power expansion in the jet radius.

Clearly, the soft function also receives contributions from non-global logarithms. The necessary dipoles are the same as the ones computed for the jet-mass distribution in dijets~\cite{Dasgupta:2012hg}, see also~\cite{Larkoski:2025afg} for the inclusion of subleading quark effects. However, as for the global contribution, the calculation needs to be performed for a generic angular separation $\Delta \phi$. This can be straightforwardly accounted for in the large-$N_C$ dipole shower that we use for the resummation of non-global logarithms.

\section{Comparison to CMS data for untagged jets}\label{app:CMS}
In this section, we compare a selection of our predictions with the CMS measurement of jet angularities reported in \cite{CMS:2021iwu}. We refine the theoretical framework of \cite{Caletti:2021oor} by incorporating contributions from $b$-initiated jets. In addition, the angularity spectra for quark- and gluon-initiated jets are computed using the decoupling scheme for the running of $\as$, as described in Sections~\ref{sec: NLL} and \ref{sec: NLL SD}. Furthermore, with respect to the predictions presented in Ref.~\cite{Reichelt:2021svh}, the non-perturbative modelling in \sherpa\ has been updated and tuned to 
data~\cite{Knobbe:2023njd,Knobbe:2023ehi}. 

\begin{figure}[tb]
    \centering
    \includegraphics[width=0.45\linewidth]{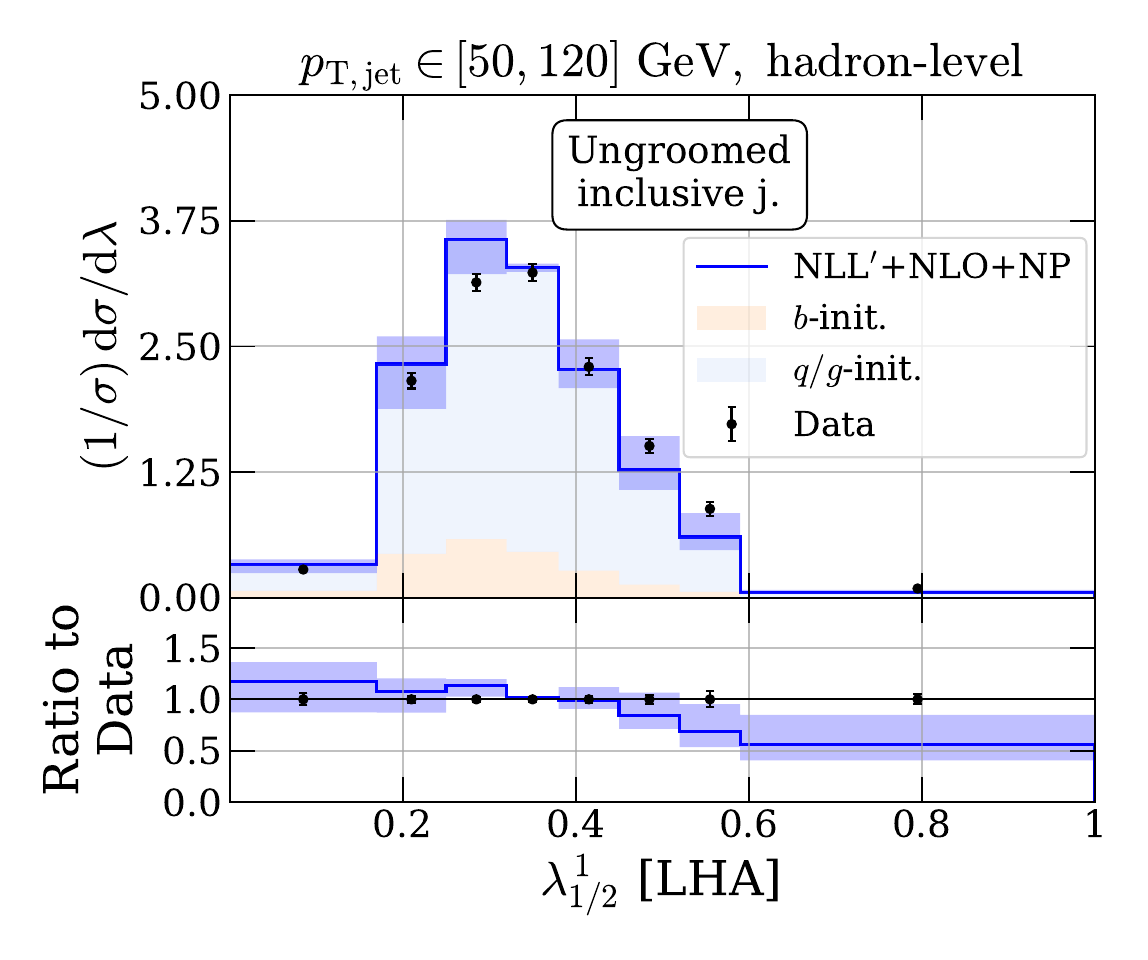}      
    \includegraphics[width=0.45\linewidth]{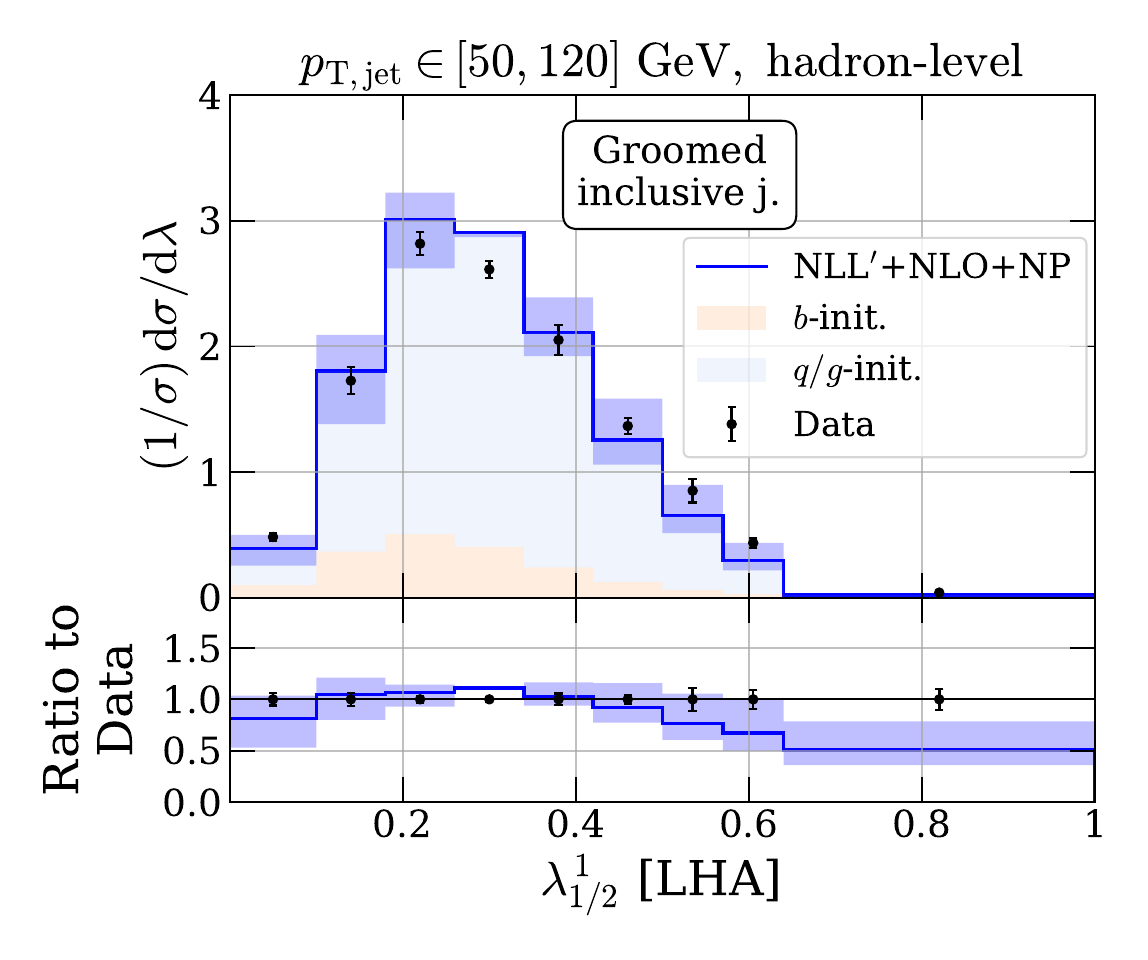}   \\ 
   \includegraphics[width=0.45\linewidth]{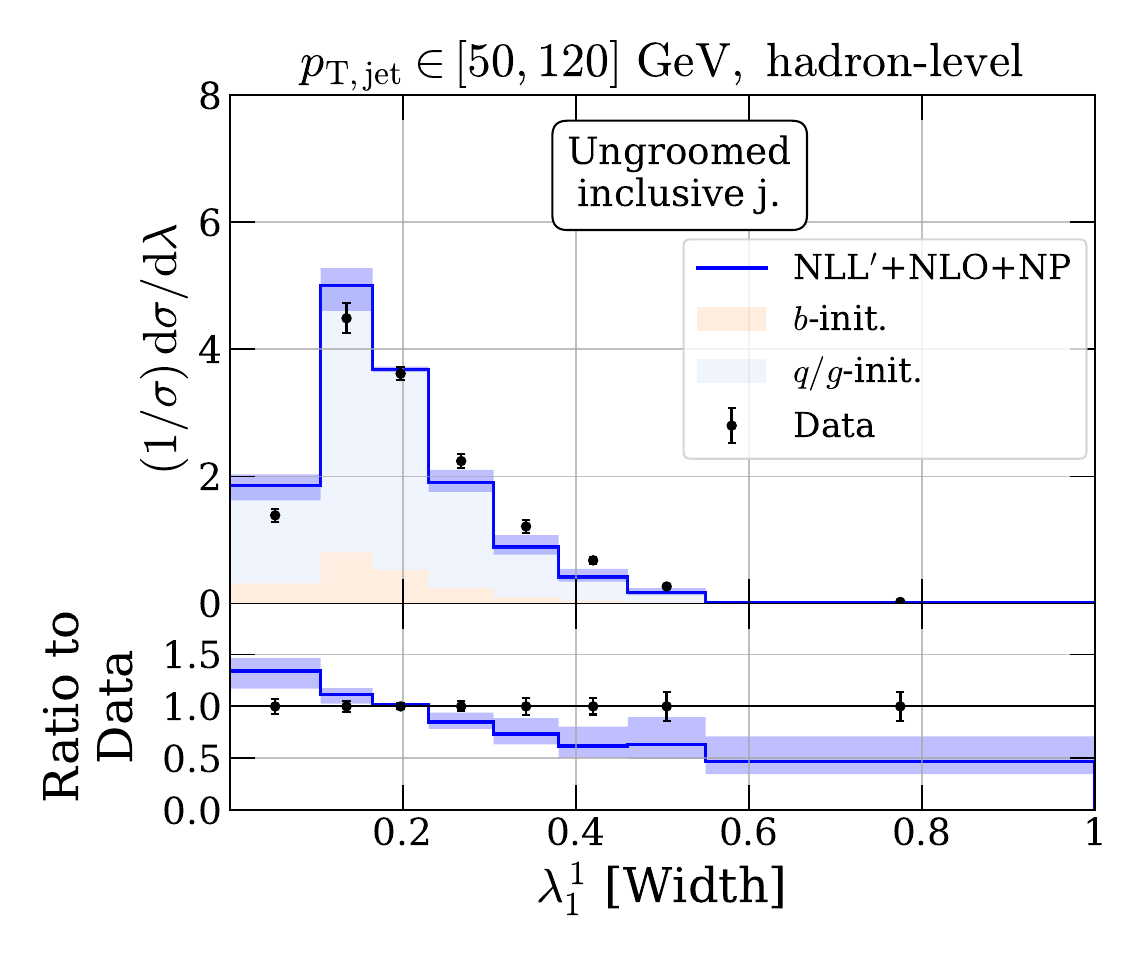}      
    \includegraphics[width=0.45\linewidth]{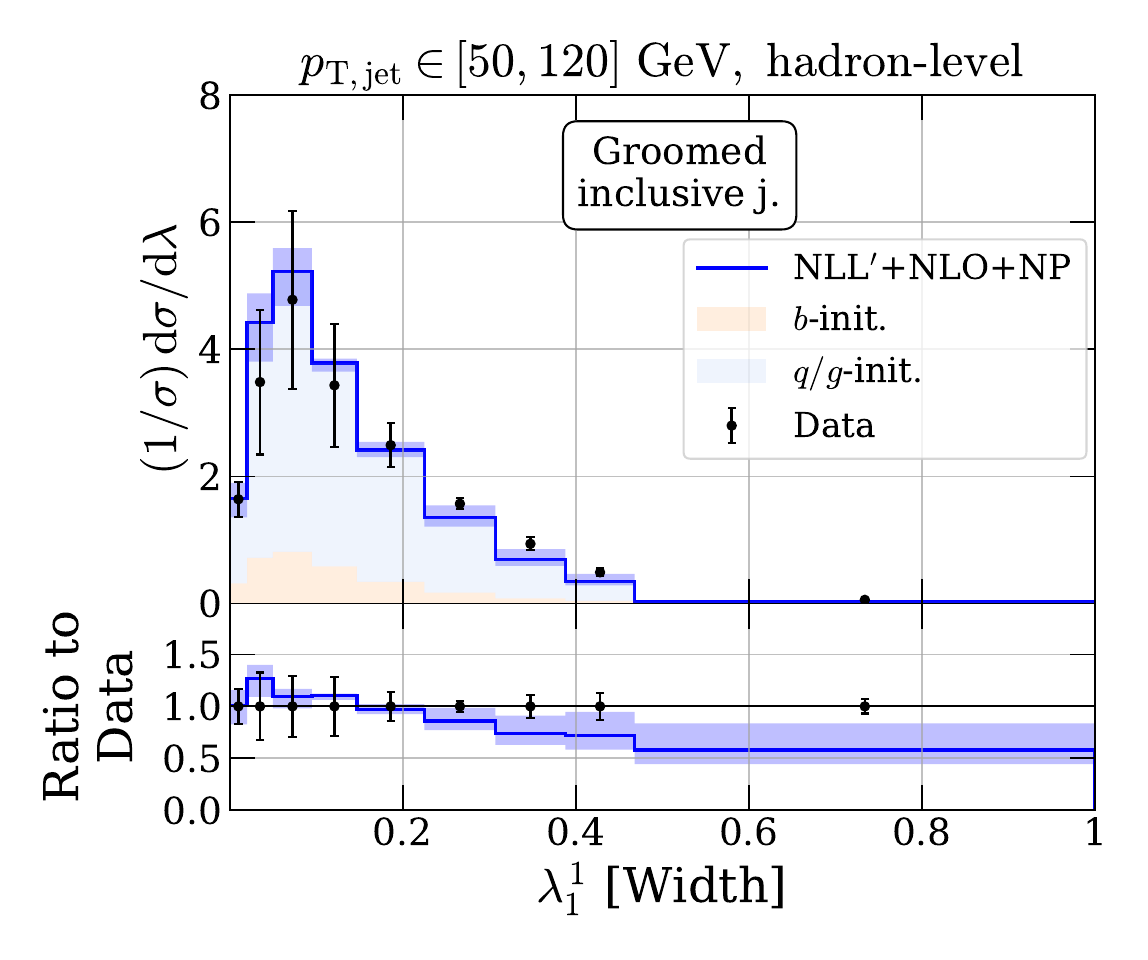}   \\
    \includegraphics[width=0.45\linewidth]{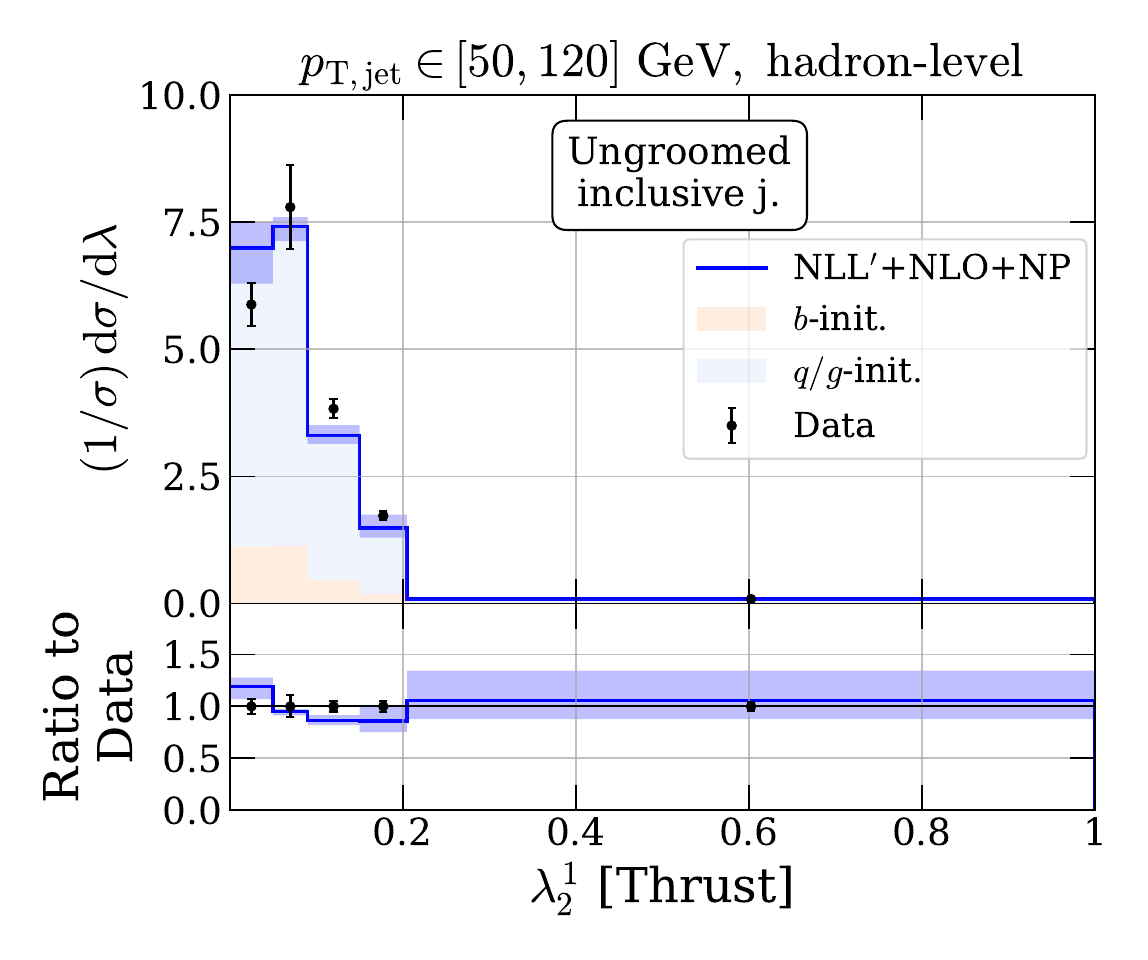}      
    \includegraphics[width=0.45\linewidth]{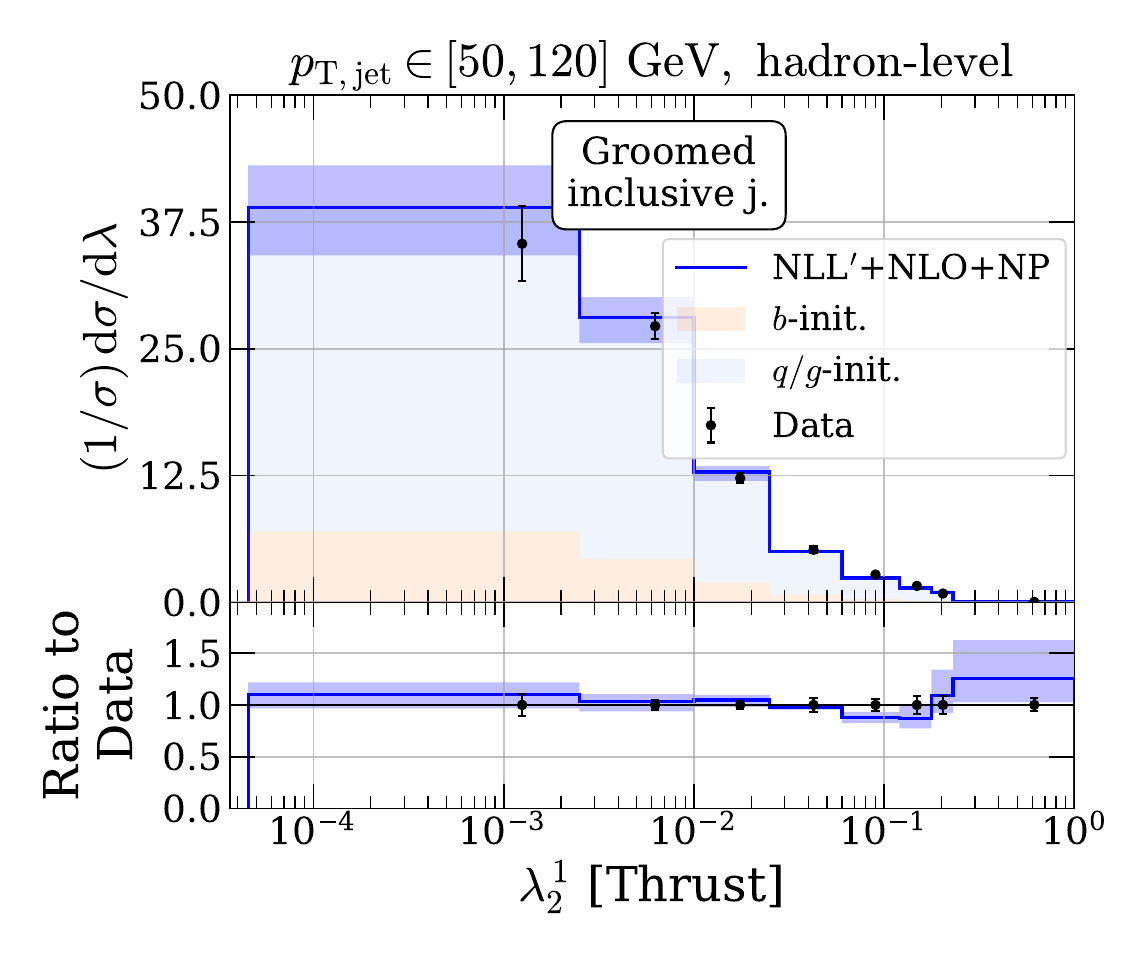}   \\
    \caption{Comparison of NLL predictions including NLO matching supplemented with non-perturbative corrections against CMS data~\cite{CMS:2021iwu} for ungroomed (left column) and groomed (right column) angularities $\lambda^1_\alpha$, for $\alpha \in \{1/2,1,2\}$ (top to bottom). }
    \label{fig:comparison CMS MEPS}
\end{figure}

This comparison is presented in Fig. \ref{fig:comparison CMS MEPS}. Here we show distributions for ungroomed (left) and groomed (right) angularities $\lambda^1_\alpha$, with $\alpha \in \{1/2,1,2\}$.
The blue curves correspond to the NLL calculation matched to the next-to-leading order (NLO) fixed-order prediction supplemented with non-perturbative corrections obtained via the transfer-matrix approach as described in Section~\ref{sec:transfer_matrices}.

In each plot the combined contribution from light-quark and gluon jets to the angularity spectrum is depicted with the blue band, whereas the orange region displays the contribution from $b$-jets. Finally, in the sub-panels we show the ratio of each theoretical prediction to the experimental data. 

We find good agreement with the data across all values of $\alpha$, consistent with the results of \cite{Caletti:2021oor}.

\newpage
\bibliography{Ref}

\end{document}